\documentclass[11pt]{article}
\usepackage{amsmath}
\usepackage{graphicx,psfrag,epsf}
\usepackage{enumerate}
\usepackage{natbib}
\usepackage{url} 

\usepackage{xr-hyper}

\usepackage{setspace}
\usepackage{authblk}
\usepackage[utf8]{inputenc}
\usepackage{color,soul,yhmath}
\usepackage{xcolor,titlesec}
\usepackage{algpseudocode}
\usepackage{bm}
\usepackage{graphicx}
\graphicspath{ {.} }
\usepackage{hyperref}
\usepackage{bbm}
\usepackage{amssymb}
\usepackage[makeroom]{cancel}

\DeclareMathOperator*{\argmin}{arg\,min}
\hypersetup{colorlinks,linkcolor=red,citecolor=blue,urlcolor=blue,breaklinks}

\newtheorem{theorem}{Theorem}

\newtheorem{corollary}{Corollary}

\newtheorem{lemma}{Lemma}

\newtheorem{remark}{Remark}

\numberwithin{equation}{section}
\def\c#1{\mathcal{#1}}
\def\b#1{\mathbbm{#1}}
\def\bf#1{\mathbf{#1}}

\def\wt#1{\widetilde{#1}}
\def\wh#1{\widehat{#1}}
\def\vec#1{\text{vec}\big(#1\big)}

\def\cA{\mathcal{A}}

\def\cM{\mathcal{M}}

\def\cF{\mathcal{F}}

\def\cN{\mathcal{N}}
\def\cH{\mathcal{H}}
\def\cG{\mathcal{G}}

\def\cU{\mathcal{U}}

\def\cP{\mathcal{P}}
\def\cL{\mathcal{L}}

\def\A{\mathbf{A}}
\def\B{\mathbf{B}}
\def\C{\mathbf{C}}
\def\D{\mathbf{D}}

\def\G{\mathbf{G}}
\def\H{\mathbf{H}}
\def\I{\mathbf{I}}
\def\Q{\mathbf{Q}}
\def\S{\mathbf{S}}
\def\Y{\mathbf{Y}}
\def\M{\mathbf{M}}

\def\R{\mathbf{R}}
\def\W{\mathbf{W}}
\def\X{\mathbf{X}}

\def\0{\mathbf{0}}
\def\1{\mathbf{1}}
\def\V{\mathbf{V}}
\def\U{\mathbf{U}}
\def\diag{\text{diag}}

\def\bLambda{\boldsymbol{\Lambda}}
\def\bSigma{\boldsymbol{\Sigma}}
\def\balpha{\boldsymbol{\alpha}}
\def\bgamma{\boldsymbol{\gamma}}
\def\bGamma{\boldsymbol{\Gamma}}
\def\bmu{\boldsymbol{\mu}}
\def\bepsilon{\boldsymbol{\epsilon}}
\def\tr{\text{tr}}

\def\a{\bf{a}}
\def\bb{\bf{b}} 
\def\e{\bf{e}}
\def\h{\bf{h}}
\def\f{\bf{f}}
\def\u{\bf{u}}

\def\x{\bf{x}}
\def\y{\bf{y}}
\def\z{\bf{z}}
\def\bbeta{\bm{\beta}}
\def\bphi{\bm{\phi}}
\def\bPhi{\bm{\Phi}}
\def\bPi{\bm{\Pi}}

\def\bXi{\bm{\Xi}}

\allowdisplaybreaks

\title{Inference on Dynamic Spatial Autoregressive Models with Change Point Detection}
\author{Zetai Cen$^*$, Yudong Chen$^\dagger$, Clifford Lam$^*$\\
		$^*$ Department of Statistics, London School of Economics\\
		$^\dagger$ Department of Statistics, University of Warwick}
\date{(\today)}

\begin{document}
\maketitle

\begin{abstract}
      We analyze a varying-coefficient dynamic spatial autoregressive model with spatial fixed effects. One salient feature of the model is the incorporation of multiple spatial weight matrices through their linear combinations with varying coefficients, which help solve the problem of choosing the most ``correct'' one for applied econometricians who often face the availability of multiple expert spatial weight matrices. We estimate and make inferences on the model coefficients and coefficients in basis expansions of the varying coefficients through penalized estimations, establishing the oracle properties of the estimators and the consistency of the overall estimated spatial weight matrix, which can be time-dependent. We further consider two applications of our model in change point detections in dynamic spatial autoregressive models, providing theoretical justifications in consistent change point locations estimation and practical implementations. Simulation experiments demonstrate the performance of our proposed methodology, and real data analyses are also carried out.
\end{abstract}

\addtocontents{toc}{\protect\setcounter{tocdepth}{0}}

\section{Introduction}\label{sec: introduction}
The study of spatial dependence in regional science gives rise to the techniques in spatial econometrics that we commonly use nowadays. Restricting to cross-sectional data only, a very general form of a model describing spatial dependence can be $\y = f(\y) + \bepsilon$ \citep{Anselin1988}, where $\y$ denotes a vector of $d$ observed units, and $\bepsilon$ denotes an error term. A prominent and widely used candidate model is the spatial autoregressive model (see for example \cite{LeSagePace2009}), which assumes a known {\em spatial weight matrix} $\W$ with zero diagonal and $f(\y)$ of the form $f(\y) = \rho\W\y$ (or $f(\y) = \rho\W\y + \X\bbeta$ for a model with matrix of covariates $\X$), where $\rho$ is called the spatial correlation coefficient.

Users of these models need to specify the $d\times d$ spatial weight matrix $\W$, which can be a contiguity matrix of 0 and 1, a matrix of inverse distances between two cities/regions, relative amount of import export etc. An obvious shortcoming for practitioners is to specify an ``accurate'' spatial weight matrix for use, often in the face of too many potential choices. This leads to a series of attempts to estimate the spatial weight matrix itself from data. For instance, see \cite{Pinkseetal2002} and \cite{Sun2016} for models dealing with cross-sectional data only, both allowing for nonlinear spatial weight matrix estimation.
\cite{BeenstockFelsenstein2012}, \cite{BhattacharjeeJensen-Butler2013}, \cite{LamSouza2020} and \cite{HigginsMartellosio2023} use spatial panel data for spatial weight matrix estimation, with \cite{LamSouza2020} and \cite{HigginsMartellosio2023} allowing for multiple specified spatial weight matrices by a linear combination of them with constant coefficients.

Recent advances in spatial econometrics allow researchers to specify more complex models with an observed panel $\{\y_t\}$. \cite{ZhangShen2015} consider partially linear covariate effects and constant spatial interactions using a sieve method to estimate a nonlinear function, while \cite{SunMalikov2018} consider varying coefficients in both the spatial correlation coefficient (with underlying variables differ over observed units) and the covariate effects, assuming the nonlinear functions are smooth for kernel estimations. \cite{Liangetal2022} use kernel estimation on a model with constant spatial interactions but deterministic time-varying coefficient functions for the covariates, while \cite{Changetal2025} generalize the model to include an unknown random time trend and deterministic time-varying spatial correlation coefficient, still using kernel estimation. \cite{Hongetal2024} investigate a model similar to \cite{SunMalikov2018}, but added dynamic terms involving $\y_{t-1}$.

However, all the above allow for one specified spatial weight matrix only. As mentioned before, practitioners often face with too many potential choices for a spatial weight matrix. Combining the flexibility of allowing for multiple specified spatial weight matrices as input in \cite{LamSouza2020} and varying effects in spatial interactions over observed variables or time directly, we propose a model akin to that in \cite{LamSouza2020}, but with varying coefficients in the linear combination of spatial weight matrices. The varying coefficients can be varying over some observed variables (stochastic) or time directly (non-stochastic). 

Our contributions in this paper are three-folds. Firstly, using basis representations, we allow for the varying coefficients to be either stochastic or directly time-varying, without the need for any smoothness conditions. Hence, the final estimated spatial weight matrix can be either stochastic or deterministic, e.g., directly time dependent. Secondly, our adaptive LASSO estimators are proved to have the oracle properties, so that ill-specified spatial weight matrices which are irrelevant in the end will be dropped with probability going to 1 as the dimension $d$ and the sample size $T$ go to infinity. Meanwhile, the effects of relevant spatial weight matrices
can be seen to be truly varying or not, again with probability going to 1 as $d,T \rightarrow \infty$. This greatly facilitates the interpretability of the spillover effects over time. Last but not least, our framework includes special cases such as spatial autoregressive panel data models with structural changes \citep{Li2018} or threshold variables \citep{Deng2018, LiLin2024}. Section \ref{sec:Change_Point_Detection} explores the applications to multiple change points detection in both spatial autoregressive models with structural changes or threshold variables, suggesting an applicable algorithm for consistent change points detection in both cases.

The rest of the paper is organized as follows. Section~\ref{sec: model_estimation} introduces the dynamic spatial autoregressive model and presents a procedure using adaptive LASSO to estimate the spatial fixed effect, spatial autoregressive parameters in a basis expansion, and the regression coefficients. Section~\ref{sec: assumption_theorem} includes the required assumptions and the theoretical guarantees on the parameter estimators. Section~\ref{sec: practical_implementation} covers the algorithm for practical implementations including model selection and covariance matrix estimation for our estimators. Section \ref{sec:Change_Point_Detection} focuses on applying our framework for change point detection for a spatial autoregressive model with threshold variables or structural changes. Finally, numerical results are presented in Section~\ref{sec: numerical}, with a case study of stock returns on the NYSE. Section~\ref{sec: conclusion} concludes. Technical proofs, additional lemmas, extra simulations and real data analysis are provided in the online supplementary material.

Throughout this paper and unless otherwise specified, we use a lower-case letter $a$, a bold lower-case letter $\bf{a}$, and a bold capital letter $\bf{A}$ to denote a scalar, a vector, and a matrix, respectively. We also use $a_i, A_{ij}, \bf{A}_{i\cdot}, \bf{A}_{\cdot j}$ to denote, respectively, the $i$-th element of $\bf{a}$, the $(i,j)$-th element of $\bf{A}$, the $i$-th row vector (as a column vector) of $\bf{A}$, and the $j$-th column vector of $\bf{A}$. Hereafter, given a positive integer $m$, define $[m]:=\{1,2,\dots,m\}$, and we use $\otimes$ to represent the Kronecker product.

For a given set, we denote by $|\cdot|$ its cardinality, and $\b{1}\{\cdot\}$ the indicator function. For a vector $\bf{v}$, we denote its $L_2$-norm by $\|\bf{v}\|$. For a given matrix $\bf{A}$, we use $\|\bf{A}\|$ to denote its spectral norm, $\|\bf{A}\|_F$ its Frobenius norm, $\|\A\|_\text{max} := \max_{i,j}|A_{ij}|$ its max norm, $\|\A\|_{1} := \max_{j}\sum_{i}|A_{ij}|$ its $L_1$-norm, and $\|\A\|_{\infty} := \max_{i}\sum_{j}|A_{ij}|$ its $L_{\infty}$-norm. For $q>0$, we define the $L_q$-norm of a given real-valued random variable $x$ as $\|x\|_q := (\b{E}|x|^q)^{1/q}$. Without loss of generality, we always assume that the eigenvalues of a matrix are arranged in descending order, with their eigenvectors ordered correspondingly.

\section{Model and estimation}\label{sec: model_estimation}

\subsection{Dynamic spatial autoregressive model}\label{subsec: DSL_model}

We propose a framework of dynamic spatial autoregressive models with fixed effects such that for each time $t\in[T]$,
\begin{equation}
\label{eqn: spatiallag}
    \y_t = \bmu^\ast + \sum_{j=1}^{p} \Big(\phi_{j,0}^\ast + \sum_{k=1}^{l_j}\phi_{j,k}^\ast z_{j,k,t} \Big) \W_j\y_t
    + \X_t\bbeta^\ast + \bepsilon_t,
\end{equation}
where $\y_t\in\b{R}^d$ is the observed vector at time $t$, and $\bmu^\ast$ is a constant vector of spatial fixed effects. Each $\W_j\in\b{R}^{d\times d}$ is a pre-specified spatial weight matrix provided by researchers to feature the spillover effects of cross-sectional units from their neighbors. Each $\W_j$ has zero entries on its main diagonal with no restrictions on the signs of off-diagonal entries, and can be asymmetric. Each term $(\phi_{j,0}^\ast+ \sum_{k=1}^{l_j}\phi_{j,k}^\ast z_{j,k,t} )$ is essentially a spatial correlation coefficient for the spatial weight matrix $\W_j$ (see also \cite{LamSouza2020}), which can be time-varying by being presented as either a basis expansion using some non-random pre-specified set of basis $\{z_{j,k,t}\}$, or an affine combination of random variables $\{z_{j,k,t}\}$. In either case, we call the $\{z_{j,k,t}\}$'s the dynamic variables hereafter. For $j\in[p], \, k\in[l_j]$, the parameters $\phi_{j,0}^\ast, \phi_{j,k}^\ast$ are unknown and need to be estimated. The covariate matrix $\X_t$ has size $d \times r$, with $\bbeta^\ast$ the corresponding unknown regression coefficients of length $r$. Finally, $\bepsilon_t$ is the idiosyncratic noise with zero mean.

Without loss of generality, we assume $\X_t$ to have zero mean. Otherwise, we read $
\bmu^\ast + \X_t\bbeta^\ast = (\bmu^\ast + \b{E}[\X_t]\bbeta^\ast) + (\X_t -  \b{E}[\X_t]) \bbeta^\ast$,
which leads to estimating $(\bmu^\ast+ \b{E}[\X_t] \bbeta^\ast)$ as the spatial fixed effects instead. We can rewrite (\ref{eqn: spatiallag}) as a traditional spatial autoregressive model $\y_t = \bmu^\ast + \W_t^\ast \y_t + \X_t\bbeta^\ast + \bepsilon_t$ by defining the true spatial weight matrix at time $t$ as
\begin{equation}
\label{eqn: true_W}
\W_t^\ast := \sum_{j=1}^{p} \Big(\phi_{j,0}^\ast + \sum_{k=1}^{l_j}\phi_{j,k}^\ast z_{j,k,t} \Big) \W_j,
\;\;\; \text{with }
-1 < \rho_t^\ast := \sum_{j=1}^{p} \Big(\phi_{j,0}^\ast + \sum_{k=1}^{l_j}\phi_{j,k}^\ast z_{j,k,t} \Big) < 1.
\end{equation}
The restrictions on $\rho_t^\ast$ for all $t\in[T]$  ensure the model is stationary. See Assumptions (M2) and (M2') for technical details. We define $L:=p+ \sum_{j=1}^p l_j$ and reformulate \eqref{eqn: spatiallag} as
\begin{align}
    \y_t &= \bmu^\ast + (\bLambda_t \bPhi^\ast) \y_t + \X_t \bbeta^\ast + \bepsilon_t, \;\;\;
    \text{where}
    \label{eqn: reformulate} \\
    \bLambda_t &:=
    (\bLambda_{1,t}, \bLambda_{2,t}, \dots, \bLambda_{p,t}) \in \mathbb{R}^{d\times dL},
    \;\; \text{with }
    \bLambda_{j,t} := (\W_j, z_{j,1,t}\W_j, \dots, z_{j,l_j,t}\W_j) \in \mathbb{R}^{d\times (d+dl_j)} ,
    \notag \\
    \bPhi^\ast &:= \big( \bPhi_1^{\ast \top}, \bPhi_2^{\ast \top}, \dots, \bPhi_p^{\ast \top} \big)^\top \in \mathbb{R}^{dL \times d},
    \;\;\; \text{with }
    \bPhi_j^\ast := \big( \phi_{j,0}^\ast \I_d, \phi_{j,1}^\ast \I_d, \dots, \phi_{j,l_j}^\ast \I_d \big)^\top . \notag
\end{align}

Due to the endogeneity in $\y_t$ and potentially $\X_t$, we assume that a set of valid instrumental variables $\U_t$ are available for $t\in[T]$. More specifically, each $\U_t$ is independent of $\bepsilon_t$ but is correlated with $\y_t$ and the endogenous $\X_t$. Note that if $\X_t$ is exogenous, we may simply have $\U_t = \X_t$. Following \cite{KelejianPrucha1998}, we can construct instruments $\B_t$ as a $d \times v$ matrix with $v\geq r$ by interacting each given spatial weight matrix with $\U_t$ such that $\B_t$ is composed of at least a subset of linearly independent columns in\footnote{Ideally, we should have $\B_t$ of the form $[\sum_{j=1}^p (\phi_{j,0}+ \sum_{k=1}^{l_j}\phi_{j,k} z_{j,k,t} )\W_j]^m\U_t$ for $m=0,1,2,\dots$. However, similar to \cite{LamSouza2020}, each $\phi_{j,0}$ and $\phi_{j,k}$ is unknown and hence we exclude any cross-terms with more than one $\W_j$.}
\begin{equation}
\label{eqn: instrument}
\Big\{
\U_t, \W_1\U_t,\W_1^2\U_t, \dots, \W_1\U_t,
\W_1^2\U_t,\dots, \W_p\U_t, \W_p^2\U_t, \dots,
\W_p\U_t, \W_p^2\U_t, \dots
\Big\} .
\end{equation}

To enhance the interpretability of the true spatial weight matrix $\W_t^\ast$, we assume that the dynamic feature of model (\ref{eqn: spatiallag}) is driven only by a few $\{z_{j,k,t}\}$. In other words, the vector of coefficients $\bphi^\ast := ( \bphi_1^{\ast \top}, \bphi_2^{\ast \top}, \dots, \bphi_p^{\ast \top} )^\top$ (with $\bphi_j^\ast := (\phi_{j,0}^\ast, \phi_{j,1}^\ast, \dots, \phi_{j,l_j}^\ast )^\top$) is assumed to be sparse. Using LASSO \citep{Tibshirani1996}, an $L_1$ penalty $\lambda \|\bphi\|_1$ can be included in the regression problem to shrink the estimators towards zero and some of them to exactly zero,  where $\lambda>0$ is a tuning parameter.
However, this form of regularization penalizes uniformly on each entry, which may lead to over- or under-penalization. The former induces bias while the latter fails sign-consistency, i.e., zeros are estimated exactly as zeros and non-zeros are estimated with the correct signs.

To ensure zero-consistency in variable selection, a necessary ``irrepresentable condition'' is often imposed \citep{ZhaoYu2006}. Subsequently, \cite{Zou2006} reweighs the regularization to be $\lambda \u^\top |\bphi|$ where $|\cdot|$ is applied entrywise and $\u$ contains the inverse of some initial estimators of $\bphi^\ast$. Now, the sign-consistency can be ensured even without the irrepresentable condition if the estimators in $\u$ are $\sqrt{T}$-consistent. Such a framework adaptively penalizes the magnitude of the estimators and is hence called ``adaptive LASSO''. To this end, we start by profiling out $\bbeta$. To make use of the instruments, define $\bar{\B}:= T^{-1}\sum_{t=1}^T \B_t$. If $\bphi$ (and hence $\bPhi$) is given, by multiplying $(\B_t-\bar{\B})^\top$ and summing over all $t\in [T]$ on both sides of (\ref{eqn: reformulate}), we then have
\[
\sum_{t=1}^T(\B_t-\bar{\B})^\top(\I_d - \bLambda_t \bPhi) \y_t = \sum_{t=1}^T(\B_t-\bar{\B})^\top\X_t \bbeta + \sum_{t=1}^T(\B_t-\bar{\B})^\top\bepsilon_t ,
\]
where the true values $\bPhi^\ast$ and $\bbeta^\ast$ are replaced by $\bPhi$ and $\bbeta$ respectively. Note the spatial fixed effect $\bmu^\ast$ vanishes as $\sum_{t=1}^T(\B_t-\bar{\B})^\top\bmu^\ast = \0$. Thus, the least squares estimator of $\bbeta^\ast$ given $\bphi$ can be denoted as
\begin{equation}
\label{eqn: profiled_beta}
    \bbeta(\bphi) = \Big\{\sum_{s=1}^T \X_s^\top(\B_s-\bar{\B}) \sum_{t=1}^T(\B_t-\bar{\B})^\top \X_t\Big\}^{-1} \sum_{s=1}^T \X_s^\top( \B_s-\bar{\B}) \sum_{t=1}^T(\B_t-\bar{\B})^\top(\I_d - \bLambda_t \bPhi) \y_t .
\end{equation}

To facilitate formulating the adaptive LASSO problem by accommodating the instrumental variables, we write $\bgamma := v^{-1}\1_v$ and the $i$-th row of $\B_t$ and $\bar{\B}$ by $\B_{t,i\cdot}$ and $\bar{\B}_{i\cdot}$, respectively. Define the outcome and covariates filtered through instrumental variables as
\[
\y_{B,i} := \sum_{t=1}^T (\B_{t,i\cdot} - \bar{\B}_{i\cdot})^\top \bgamma \y_t, \;\;\;
\X_{B,i} := \sum_{t=1}^T (\B_{t,i\cdot} - \bar{\B}_{i\cdot})^\top \bgamma \X_t, \;\;\; \text{for }
i\in [d].
\]
The least squares problem is then
\begin{align}
    \wt{\bphi} &= \argmin_{\bphi} \;\; \frac{1}{2T} \;\;
    \sum_{i=1}^d \Big\| (\I_d - \bLambda_t \bPhi)\y_{B,i} - \X_{B,i} \bbeta(\bphi ) \Big\|^2 .
    \label{eqn: phi_ls}
\end{align}
With the solution as an initial estimator, the adaptive LASSO problem now solves for
\begin{align}
    \wh{\bphi} &= \argmin_{\bphi} \;\; \frac{1}{2T} \;\;
    \sum_{i=1}^d \Big\| (\I_d - \bLambda_t \bPhi)\y_{B,i} - \X_{B,i} \bbeta(\bphi ) \Big\|^2
    + \lambda \u^\top |\bphi| ,
    \label{eqn: phi_ada_lasso} \\
    & \;\;\;\;\; \text{subj. to} \;\;\;
    \|\bLambda_t \bPhi \|_{\infty} < 1, \;\;\; \text{with} \;\;\;
    |\z_t^\top \bphi| < 1, \notag
\end{align}
where $\z_t := (\z_{1,t}^\top, \z_{2,t}^\top, \dots, \z_{p,t}^\top)^\top$, $\z_{j,t} := (1, z_{j,1,t}, \dots, z_{j,l_j,t})^\top$, $\u := (|\wt{\phi}_{1,0}|^{-1}, \dots, |\wt{\phi}_{p,l_p} |^{-1})^\top$, $|\bphi|:= (|\phi_{1,0}|, \dots, |\phi_{p,l_p}| )^\top$ and $\lambda$ is a tuning parameter. With $\wh{\bphi}$ (and hence $\wh\bPhi$), the adaptive LASSO estimators for $\bbeta^\ast$ can be obtained by $\wh{\bbeta} := \bbeta(\wh{\bphi})$ and the fixed effect estimator by
\begin{equation}
\label{eqn: def_mu_hat}
\wh\bmu := \frac{1}{T} \sum_{t=1}^T \Big\{ (\I_d - \bLambda_t \wh\bPhi) \y_t - \X_t \wh\bbeta \Big\}.
\end{equation}

\subsection{Full matrix notations}\label{sec: full_matrix_notations}
To facilitate both the theoretical results and  practical implementation, we present the least squares and the adaptive LASSO problems in matrix notations here. First, we introduce
\begin{equation}
\label{eqn: def_B}
\B := T^{-1/2}d^{-a/2}(\B_{\bgamma}- \bar{\B}_{\bgamma}) := T^{-1/2}d^{-a/2} \I_d \otimes \big\{(\I_T \otimes \bgamma^\top ) (\B_1 - \bar{\B}, \dots, \B_T - \bar{\B})^\top \big\} ,
\end{equation}
where $a$ is a constant that gauges the correlation between $\B_t$ and $\X_t$ so that a larger $a$ generally means that $\B_t$ is correlated with more covariates in $\X_t$. See Assumption (R4) for more technical details. As in \cite{LamSouza2020}, in practice we can set $a=1$ to compute $\B$, which would not change the optimal values of any tuning parameters or estimators in the adaptive LASSO problem below.

For ease of notation, denote $z_{j,0,t}=1$ for all $j\in[p], t\in[T]$. Now rewrite~\eqref{eqn: spatiallag} as
\begin{align}
    &\y = \bmu^\ast \otimes \1_T + \V \bphi^\ast + \X_{\bbeta^\ast} \Vec{\I_d} + \bepsilon,
    \;\;\; \text{where}
    \label{eqn: spatiallag_rewrite1} \\
    & \y := \Vec{(\y_1, \dots, \y_T)^\top}, \;\;\;
    \bepsilon := \Vec{(\bepsilon_1, \dots, \bepsilon_T)^\top}, \;\;\;
    \V:= (\V_1, \dots, \V_p), \notag \\
    & \V_j := \Big[ \bGamma_{j,0} \Vec{\W_j^\top}, \bGamma_{j,1}\Vec{\W_j^\top}, \dots,
    \bGamma_{j,l_j} \Vec{\W_j^\top} \Big], \notag \\
    & \bGamma_{j,k} := \I_d \otimes (z_{j,k,1}\y_1, \dots, z_{j,k,T}\y_T)^\top , \;\;\;
    \X_{\bbeta^\ast} := \I_d \otimes \big\{(\I_T \otimes (\bbeta^\ast)^\top ) (\X_1, \dots, \X_T)^\top \big\} . \notag
\end{align}
In this form, the model now has design matrix $\V$ in a classical linear regression setting, except that the endogenous variables $\y_t$ are present in $\V$. We thus obtain the augmented model by left-multiplying both sides of~\eqref{eqn: spatiallag_rewrite1} by $\B^\top$:
\begin{equation}
\label{eqn: spatiallag_rewrite1_augmented}
    \B^\top\y = \B^\top\V \bphi^\ast + \B^\top\X_{\bbeta^\ast} \Vec{\I_d} + \B^\top\bepsilon ,
\end{equation}
where the augmented spatial fixed effect vanishes since $\B^\top(\bmu^\ast \otimes \1_T) = T^{-1/2}d^{-a/2} \bmu^\ast \otimes \big\{ (\B_1 - \bar{\B}, \dots, \B_T - \bar{\B}) (\I_T\otimes \bgamma)\1_T\big\}  =\0$. For any matrix $\C$, denote $\C^\otimes := \I_T\otimes \C$. We can also rewrite (\ref{eqn: spatiallag}) as
\begin{align}
    \y^\nu &= \1_T \otimes \bmu^\ast  + \sum_{j=1}^p \sum_{k=0}^{l_j} \phi_{j,k}^\ast \W_j^\otimes \y_{j,k}^\nu + \X \bbeta^\ast + \bepsilon^\nu,
    \;\;\; \text{where} \;\;  \y^\nu := (\y_1^\top, \dots, \y_T^\top)^\top, 
    \label{eqn: spatiallag_rewrite2} \\
    \y_{j,k}^\nu &:= (z_{j,k,1} \y_1^\top, \dots, z_{j,k,T} \y_T^\top)^\top, \;\;\;
    \bepsilon^\nu := (\bepsilon_1^\top, \dots, \bepsilon_T^\top)^\top, \;\;\; \X := (\X_1^\top, \dots, \X_T^\top)^\top. \notag
\end{align}
Thus with $\B^\nu := (\B_1^\top - \bar{\B}^\top, \dots, \B_T^\top - \bar{\B}^\top)^\top$, we may write (\ref{eqn: profiled_beta}) in matrix form as
\begin{align}
    \bbeta(\bphi) &= \Big(\X^\top \B^\nu (\B^\nu)^\top \X \Big)^{-1} \X^\top \B^\nu (\B^\nu)^\top \Big(\y^\nu - \sum_{j=1}^p \sum_{k=0}^{l_j} \phi_{j,k} \W_j^\otimes \y_{j,k}^\nu \Big) .
    \label{eqn: profiled_beta_matrix}
\end{align}

To rewrite the least squares and adaptive LASSO problem for $\wt\bphi$ and $\wh\bphi$, define
\begin{align*}
    \Y_W &:= (\W_1^\otimes \y_{1,0}^\nu, \dots, \W_1^\otimes \y_{1,l_1}^\nu, \dots, \W_p^\otimes \y_{p,0}^\nu, \dots, \W_p^\otimes \y_{p,l_p}^\nu) ,\\
    \bXi &:= T^{-1/2}d^{-a/2} \Big(\sum_{t=1}^T \X_t \otimes (\B_t - \bar{\B}) \bgamma \Big) \big[\X^\top \B^\nu (\B^\nu)^\top \X \big]^{-1} \X^\top \B^\nu (\B^\nu)^\top.
\end{align*}
Then with all simplification steps relegated to the online supplement, (\ref{eqn: phi_ls}) can be rewritten as
\begin{equation}
\label{eqn: simplify_phi_ls_matrix}
\begin{split}
    \wt{\bphi} &= \argmin_{\bphi} \frac{1}{2T}
    \Big\| \B^\top\y - \bXi \y^\nu - ( \B^\top\V - \bXi \Y_W ) \bphi \Big\|^2 \\
    &= \big[( \B^\top\V - \bXi \Y_W )^\top ( \B^\top\V - \bXi \Y_W ) \big]^{-1} ( \B^\top\V - \bXi \Y_W )^\top (\B^\top\y - \bXi \y^\nu) .
\end{split}
\end{equation}
The adaptive LASSO problem in (\ref{eqn: phi_ada_lasso}) can be written as
\begin{align}
    \wh{\bphi} &= \argmin_{\bphi} \frac{1}{2T}
    \Big\| \B^\top\y - \bXi \y^\nu - ( \B^\top\V - \bXi \Y_W ) \bphi \Big\|^2 + \lambda \u^\top |\bphi| ,
    \label{eqn: simplify_phi_ada_lasso_matrix} \\
    & \;\;\;\;\; \text{subj. to} \;\;\;
    \|\bLambda_t \bPhi\|_{\infty} < 1, \;\;\; \text{with} \;\;\;
    |\z_t^\top \bphi| < 1. \notag
\end{align}

\section{Assumptions and theoretical results}\label{sec: assumption_theorem}

We first present some notations involving the measure of serial dependence of all time series variables, which is gauged by the functional dependence measure introduced by \cite{Wu2005}. We state all the assumptions in this paper in Section \ref{subsec: assumption}. Denote $\{\x_t\} := \big\{\hspace{-1pt} \vec{\X_t} \hspace{-1pt} \big\}$ and $\{\bb_t\} := \big\{\hspace{-1pt} \vec{\B_t} \hspace{-1pt} \big\}$ to be the vectorized processes for $\{\X_t\}$ and $\{\B_t\}$ with length $dr$ and $dv$, respectively. For $t\in[T]$, assume that
\begin{equation}
\label{eqn: functional_x_b_epsilon}
\x_t = \big[ f_i(\cF_t) \big]_{i\in[dr]}, \,\,\,\,\,\;\;\;
\bb_t = \big[ g_i(\cG_t) \big]_{i\in[dv]}, \,\,\,\,\,\;\;\;
\bepsilon_t = \big[ h_i(\cH_t) \big]_{i\in[d]},
\end{equation}
where $f_i(\cdot)$'s, $g_i(\cdot)$'s, $h_i(\cdot)$'s are measurable functions defined on the real line, and $\cF_t = (\dots, \e_{x,t-1}, \e_{x,t})$, $\cG_t = (\dots, \e_{b,t-1}, \e_{b,t})$, $\cH_t = (\dots, \e_{\epsilon,t-1}, \e_{\epsilon,t})$ are defined by \text{i.i.d.} processes $\{\e_{x,t}\}$, $\{\e_{b,t}\}$ and $\{\e_{\epsilon,t}\}$ respectively, with $\{\e_{b,t}\}$ independent of $\{\e_{\epsilon,t}\}$ but correlated with $\{\e_{x,t}\}$. For $q>0$, we define
\begin{equation}
\label{eqn: def_functional_dependence}
\begin{split}
    & \theta_{t,q,i}^x := \big\|x_{t,i} - x_{t,i}'\big\|_q = (\b{E}| x_{t,i} - x_{t,i}' |^q)^{1/q},
    \;\;\; i\in[dr], \\
    & \theta_{t,q,i}^b := \big\|b_{t,i} - b_{t,i}'\big\|_q = (\b{E}| b_{t,i} - b_{t,i}' |^q)^{1/q},
    \;\;\; i\in[dv], \\
    & \theta_{t,q,i}^\epsilon := \big\|\epsilon_{t,i} - \epsilon_{t,i}'\big\|_q = (\b{E}| \epsilon_{t,i} - \epsilon_{t,i}' |^q)^{1/q},
    \;\;\; i\in[d],
\end{split}
\end{equation}
where $x_{t,i}' = f_i(\cF_t')$, $\cF_t' = (\dots, \e_{x,\text{-}1}, \e_{x,0}', \e_{x,1}, \dots, \e_{x,t})$, with $\e_{x,0}'$ independent of all other $\e_{x,j}$'s. Hence $x_{t,i}'$ is a coupled version of $x_{t,i}$ with $\e_{x,0}$ replaced by an \text{i.i.d.} copy $\e_{x,0}'$. We define $b_{t,i}'$ and $\epsilon_{t,i}'$ similarly.

\subsection{Assumptions}\label{subsec: assumption}

We present here the set of assumptions for our model. In summary, (I1) helps to identify the model; assumptions prefixed ``M'' renders the model framework; those prefixed ``R'' are more technical.

\begin{itemize}
\item[(I1)] (Identification)
\em
The matrix $\Q^\top\Q$ has all its eigenvalues uniformly bounded away from $0$, where
\[
\Q = [\b{E}(\B^\top\V), \b{E}(\B^\top\wt{\X})], \;\;\;
    \wt{\X} = (\x_{1,1\cdot},\dots, \x_{T,1\cdot}, \dots, \x_{1,d\cdot}, \dots, \x_{T,d\cdot})^\top .
\]
\end{itemize}

\begin{itemize}
\item[(M1)] (Time series in $\X_t$, $\B_t$ and $\bepsilon_t$)
\em
The processes $\{\X_t\}$, $\{\B_t\}$ and $\{\bepsilon_t\}$ are second-order stationary and satisfy (\ref{eqn: functional_x_b_epsilon}), with $\{\X_t\}$ and $\{\bepsilon_t\}$ having mean zero. The tail condition $\b{P}(|Z| > z)\leq D_1 \exp(-D_2 z^\ell)$ is satisfied by the variables $B_{t,ij}$, $X_{t,ij}$, $\epsilon_{t,i}$ for the same constants $D_1$, $D_2$ and $\ell$.

With (\ref{eqn: def_functional_dependence}), define the tail sums
\begin{align*}
    \Theta_{m,q}^x &= \sum_{t=m}^\infty \max_{i\in[dr]} \; \theta_{t,q,i}^x, \;\;\;
    \Theta_{m,q}^b = \sum_{t=m}^\infty \max_{i\in[dv]} \; \theta_{t,q,i}^b, \;\;\;
    \Theta_{m,q}^\epsilon = \sum_{t=m}^\infty \max_{i\in[d]} \; \theta_{t,q,i}^\epsilon .
\end{align*}
For some $w> 2$, assume $\Theta_{m,2w}^x,\; \Theta_{m,2w}^b,\; \Theta_{m,2w}^\epsilon \leq C m^{-\alpha}$ with constants $\alpha, \, C > 0$ that can depend on $w$.
\end{itemize}

\begin{itemize}
\item[(M2)] (True spatial weight matrix $\W_t^\ast$ with non-random $z_{j,k,t}$)
\em
$\W_t^\ast$ defined in (\ref{eqn: true_W}) uses a uniformly bounded non-stochastic basis $\{z_{j,k,t}\}$ for $j\in[p], \, k\in[l_j]$. There exists a constant $\eta >0$ such that for all $t\in[T]$, $\|\W_t^\ast\|_\infty < \eta < 1$ uniformly as $d\to \infty$. The elements in $\W_t^\ast$ can be negative, and $\W_t^\ast$ can be asymmetric. Furthermore, $\rho_t^\ast$ defined in (\ref{eqn: true_W}) satisfies $|\rho_t^\ast| < 1$.
\end{itemize}

\begin{itemize}
\item[(M2')] (True spatial weight matrix $\W_t^\ast$ with random $z_{j,k,t}$)
\em
Same as Assumption (M2), except that $\{z_{j,k,t}\}$ is a zero mean stochastic process with support $[-1,1]$, such that $z_{j,k,t} = u_{j,k}(\cU_t)$, similar to (\ref{eqn: functional_x_b_epsilon}), satisfies $\b{E}(z_{j,k,t} \X_t) =\0$, $\b{E}(z_{j,k,t} \bepsilon_t) =\0$, and $\Theta_{m,2w}^z \leq Cm^{-\alpha}$ as in Assumption (M1). Furthermore:
\begin{itemize}
    \item[1.] there exists $\eta >0$ such that $\sum_{j=1}^{p} \sum_{k=0}^{l_j} \|\phi_{j,k}^\ast \W_j\|_\infty < \eta < 1$ uniformly as $d\to \infty$;
    \item[2.] $\sum_{j=1}^{p} \sum_{k=0}^{l_j} |\phi_{j,k}^\ast| < 1$.
\end{itemize}
\end{itemize}

\begin{itemize}
\item[(R1)]
\em
Denote the $d^2 L\times L$ block diagonal matrix $\D_W := \textnormal{diag}\big\{\I_{1+l_1} \otimes\, \textnormal{vec}(\W_1^\top), \dots, \I_{1+l_p} \otimes\, \textnormal{vec}(\W_p^\top) \big\}$. Then, there exists a constant $u>0$ such that the $L$-th largest singular value of $\D_W$ satisfies $\sigma_L^2(\D_W) \geq du >0$ uniformly as $d\to \infty$.

Moreover, $\max_j \big\{ \|\W_j\|_1, \|\W_j\|_\infty \big\} \leq c < \infty$ uniformly as $d\to \infty$ for some constant $c>0$.
\end{itemize}

\begin{itemize}
\item[(R2)]
\em
Write $\bepsilon_t = \bSigma_\epsilon^{1/2} \bepsilon_t^\ast$ with $\bSigma_\epsilon$ being the covariance matrix of $\bepsilon_t$. Assume $\big\|\bSigma_\epsilon \big\|_\textnormal{max} \leq \sigma_\textnormal{max}^2 < \infty$ uniformly as $d\to \infty$. The same applies to the variance of the elements in $\B_t$.

Assume also $\big\|\bSigma_\epsilon^{1/2} \big\|_\infty \leq S_\epsilon < \infty$ uniformly as $d\to \infty$, with $\{\epsilon_{t,i}^\ast\}_{i\in[d]}$ being a martingale difference with respect to the filtration generated by $\sigma(\epsilon_{t,1}, \dots, \epsilon_{t,i})$. Furthermore, $\{\bepsilon_t^\ast \}_{t\in[T]}$ satisfies the tail condition and the functional dependence in Assumption (M1).
\end{itemize}

\begin{itemize}
\item[(R3)]
\em
All singular values of $\b{E}(\X_t^\top \B_t)$ are uniformly larger than $du$ for some constant $u>0$, while the maximum singular value is of order $d$. Individual entries in the matrix $\b{E}(\bb_t \x_t^\top)$ are uniformly bounded away from infinity, with $\x_t$ and $\bb_t$ defined in (\ref{eqn: functional_x_b_epsilon}).
\end{itemize}

\begin{itemize}
\item[(R4)]
\em
With the same $a \in [0,1]$ introduced in (\ref{eqn: def_B}), we define
\begin{align*}
    & \G := d^{-a}\, \I_d \otimes \{\b{E}(\check\G) \, \b{E}(\check\G)^\top \} , \;\;\;
    \check\G := (\check\G_{1,0},\dots, \check\G_{1,l_1},\dots, \check\G_{p,0}, \dots, \check\G_{p,l_p}) ,\\
    & \check\G_{j,k}:= \frac{1}{T}\sum_{t=1}^T z_{j,k,t} (\B_t - \bar{\B})\bgamma \bbeta^{\ast \top} \X_t^\top \bPi_t^{\ast \top} , \;\;\;
    \bPi_t^\ast := (\I_d - \W_t^\ast)^{-1}.
\end{align*}
We assume that $\G$ has full rank and that there exists a constant $u>0$ such that $\lambda_\textnormal{min}(\G) \geq u>0$ and $\lambda_\textnormal{max}(\G) <\infty$ uniformly as $d\to \infty$.
\end{itemize}

\begin{itemize}
\item[(R5)]
\em
For the same constant $a$ as in Assumption (R4), we have for each $d$,
\[
\max_{i\in[d]} \sum_{j=1}^d \big\| \b{E}(\B_{t,i\cdot} \X_{t,j\cdot}^\top) \big\|_\textnormal{max}, \;\;\;
\max_{j\in[d]} \sum_{i=1}^d \big\| \b{E}(\B_{t,i\cdot} \X_{t,j\cdot}^\top) \big\|_\textnormal{max} = O(d^a) .
\]
At the same time, assume $\b{E}(\X_t \otimes \B_t\bgamma)$ has all singular values of order $d^{1+a}$.
\end{itemize}

\begin{itemize}
\item[(R6)]
\em
With $\bPi_t^\ast$ in Assumption (R4), define
\[
\ddot\G := (\ddot\G_{1,0},\dots, \ddot\G_{1,l_1},\dots, \ddot\G_{p,0}, \dots, \ddot\G_{p,l_p}) , \;\;\;
\ddot\G_{j,k}:= \frac{1}{T}\sum_{t=1}^T z_{j,k,t} (\B_t - \bar{\B})^\top \big(\I_d \otimes \bbeta^{\ast \top} \X_t^\top \bPi_t^{\ast \top}\big) .
\]
Assume that $\b{E}(\ddot\G) \b{E}(\ddot\G)^\top$ has full rank and that there exists a constant $u>0$ such that $\lambda_v\big( \b{E}(\ddot\G) \b{E}(\ddot\G)^\top\big) \geq u>0$ and $\lambda_\textnormal{max}\big( \b{E}(\ddot\G) \b{E}(\ddot\G)^\top\big) < \infty$ uniformly as $d\to \infty$.
\end{itemize}

\begin{itemize}
\item[(R7)]
\em
Define the predictive dependence measures
\[
P_0^b(B_{t,jk}) := \b{E}(B_{t,jk}\mid \cG_0) -\b{E}(B_{t,jk}\mid \cG_{-1}), \;\;\;
P_0^\epsilon(\epsilon_{t,j}) := \b{E}(\epsilon_{t,j} \mid \cH_0) -\b{E}(\epsilon_{t,j} \mid \cH_{-1}),
\]
with $\cG_t$ and $\cH_t$ specified after (\ref{eqn: functional_x_b_epsilon}). Assume
\[
\sum_{t\geq 0} \max_{j\in[d]}\; \max_{k\in[v]} \big\| P_0^b(B_{t,jk}) \big\|_2 <\infty, \;\;\;
\sum_{t\geq 0} \max_{j\in[d]} \big\| P_0^\epsilon(\epsilon_{t,j}) \big\|_2 <\infty .
\]
\end{itemize}

\begin{itemize}
\item[(R8)]
\em
For $b\in [0,1]$, the eigenvalues of $\textnormal{Var}(d^{-b/2} \B_{t,\cdot k})$ and $\textnormal{Var}(\bepsilon_t)$ are uniformly bounded away from both zero and infinity, and respectively dominate the singular values of $d^{-b} \b{E}\big\{ [\B_{t, \cdot k} -\b{E}(\B_{t,\cdot k})] [\B_{t+\tau, \cdot k} -\b{E}(\B_{t,\cdot k})]^\top\big\}$ and $\b{E}(\bepsilon_t \bepsilon_{t +\tau}^\top)$ for any $\tau\neq 0$. The sum of the $i$-th largest singular values over all lags $\tau \in \b{Z}$ for each $i\in[d]$ is assumed to be finite for both autocovariance matrices of $\{d^{-b/2} \B_t\}$ and $\{\bepsilon_t\}$.
\end{itemize}

\begin{itemize}
\item[(R9)]
\em
Define $c_T :=g T^{-1/2} \log^{1/2}( T\vee d)$ for some constant $g>0$. The tuning parameter for the adaptive LASSO problem (\ref{eqn: phi_ada_lasso}) is $\lambda = C c_T$ for some constant $C>0$.
\end{itemize}

\begin{itemize}
\item[(R10)] (Rate assumptions)
\em
We assume that as $L,\, d,\, T\to \infty$,
\begin{align*}
    c_T L^{3/2} d^{1-a}, \;\;\;
    L d^{-1}, \;\;\;
    L^2 d^3 T^{2-w}, \;\;\;
    d^{b +2a +1/w} T^{-1} = o(1), &\\
    d^{-1/w}\log(T \vee d), \;\;\;
    d^{b-a -1/w} \log^{-1}( T\vee d) = O(1). &
\end{align*}
\end{itemize}

Assumption (I1) facilitates identifying the model (\ref{eqn: spatiallag_rewrite1_augmented}); see the online supplement for more details. In the sequel, we discuss in detail the structure of the true spatial weight matrix $\W_t^\ast$, and some technical assumptions made above.

The time series components in the model are described in Assumption (M1), allowing for weak serial dependence in addition to strict stationarity, as commonly seen in the spatial panel literature (e.g.~\cite{Giordanoetal2024} and \cite{LiLin2024}). While mixing conditions are imposed in previous works, we adopt the notion of functional or physical dependence, following earlier studies such as \cite{Shao2010} and \cite{Wuetal2024}. We further assume exponential tails in (M1), enabling the application of a Nagaev-type inequality for functionally dependent data \citep{Liuetal2013}.

Assumptions (M2) and (M2') describe the structure of $\W_t^\ast$ under two different settings for $\{z_{j,k,t}\}$. The row sum condition for $\W_t^\ast$ ensures that the model (\ref{eqn: spatiallag}) is uniformly stationary. Assumption (M2) treats $\{z_{j,k,t}\}$ as a non-random series, while (M2') allows $\{z_{j,k,t}\}$ to be stochastic. Note that in (M2'), the stationarity is guaranteed if $\|\W_t^\ast\|_\infty < \eta < 1$ and $|\rho_t^\ast| < 1$ hold with probability 1. The assumption of zero mean and support $[\text{-}1,1]$ is equivalent to the general assumption of bounded support, as we can always rewrite each $\phi_{j,k}^\ast z_{j,k,t}$ as
\begin{align*}
    \phi_{j,k}^\ast z_{j,k,t} &= \phi_{j,k}^\ast \b{E}(z_{j,k,t}) + \phi_{j,k}^\ast \big[z_{j,k,t} - \b{E}(z_{j,k,t})\big] = \phi_{j,k}^\ast \b{E}(z_{j,k,t}) + \big(\phi_{j,k}^\ast z_{j,k}^\ast \big) \big\{\big[ z_{j,k,t} - \b{E}(z_{j,k,t})\big] /z_{j,k}^\ast \big\},
\end{align*}
where $[\text{-}z_{j,k}^\ast, z_{j,k}^\ast]$ is the support of $z_{j,k,t}$, and $\phi_{j,k}^\ast \b{E}(z_{j,k,t})$, a constant term, can be absorbed into $z_{j,0,t}$. We note that, in fact, we may allow some $\{z_{j,k,t}\}$ for $j\in\cP \subseteq [p], k\in\cL_j \subseteq [l_j]$ satisfying (M2), while all other $\{z_{j,k,t}\}$ satisfying (M2'), as all theoretical results hold under either assumption.

Assumption (R1) describes how sparse each spatial weight matrix candidate is. It is worthwhile pointing out that although each $\W_j$ is not necessarily linearly independent with each other by (R1), we actually implicitly impose such linear independence condition from (I1) through the combination of $\{z_{j,k,t}\}$ and $\W_j$ in $\B^\top \V$. (R2) is included as a technical addition to (M2).

Assumptions (R3) to (R6) all draw on the relation between $\B_t$ and $\X_t$. Their dependence structure is non-trivial due to the extra complication from spatial weight matrix and the time-varying components here. Naturally, $\B_t$ needs to be correlated with $\X_t$ to a certain extent, captured by an unknown constant $a$ which facilitates the presentation of theoretical results. For instance, as an immediate consequence of (R5), we may derive $\big\| \b{E}(\X_t \otimes \B_t\bgamma) \big\|_1 = O(d^{1+a})$, which is a key ingredient in obtaining the rates for $\big\|\wt{\bphi} -\bphi^\ast \big\|_1$ in Theorem~\ref{thm: LASSO_rate}.

The predictive dependence measure defined in (R7) allows us to apply the central limit theorem for data with functional dependence and the Assumption (R7) can be satisfied by, for example, causal AR processes. Assumption (R8) further restricts the serial correlation in the noise. It also introduces another constant $b$ that characterizes how elements in $\B_t$ are contemporarily and temporally correlated with each other. From the comparison of rates derived in the proofs, we conclude that $b$ is actually bounded above by $1/w$, which is intuitive since a large $w$ in (M1) generally implies light-tails and hence a small $b$. Lastly, (R9) sets the rate for $\lambda$ and (R10) characterizes the relation between $T$, $d$ and $L$. As an example, (R10) holds when $w=6$, $a=1/2$, $L =O(d^{1/3})$ and $T \asymp d^2$. More generally, (R10) can be satisfied as long as $T$ grows at a polynomial rate in $d$.

\subsection{Main results}\label{sec: main_theorem}

In this subsection, we formally present the main results for the estimators of our model.

\begin{theorem}\label{thm: LASSO_rate}
Let all assumptions in Section \ref{subsec: assumption} hold ((M2) or (M2')). Given any $\bphi$ as an estimator of $\bphi^\ast$, with $c_T$ defined in Assumption (R9), $\bbeta(\bphi)$ according to (\ref{eqn: profiled_beta_matrix}) satisfies
\[
\big\|\bbeta(\bphi) - \bbeta^\ast\big\|_1 =
O_p\Big( \big\|\bphi - \bphi^\ast \big\|_1 + c_T d^{-\frac{1}{2} + \frac{1}{2w}} \Big).
\]
In particular, the least squares estimator $\wt{\bphi}$ in (\ref{eqn: simplify_phi_ls_matrix}) and $\wt{\bbeta} := \bbeta(\wt{\bphi})$ under $L=O(1)$ satisfy
\[
\big\|\wt{\bphi} -\bphi^\ast \big\|_1 = O_P\big( c_T d^{-\frac{1}{2} +\frac{1}{2w}}\big) = \big\|\wt{\bbeta} -\bbeta^\ast \big\|_1 .
\]
\end{theorem}

Theorem~\ref{thm: LASSO_rate} serves as a foundational step for the results hereafter. From Theorem \ref{thm: LASSO_rate}, the error of our least squares estimator $\bbeta(\bphi)$ might be inflated by the plugged-in estimator for $\bphi^\ast$. With a dense estimator $\wt\bphi$, we arrive at the rate $c_T d^{-\frac{1}{2} +\frac{1}{2w}}$. The dependence of the rate on $w$ confirms that weaker temporal dependence in the data results in better estimation, as expected. We now present the sign-consistency of our adaptive LASSO estimator.

\begin{theorem}\label{thm: ada_LASSO_phi_asymp}
(Oracle property for $\wh\bphi$) Let all assumptions in Section \ref{subsec: assumption} hold ((M2) or (M2')), except that (R4) and (R6) are satisfied with $\check{\G}$ and $\ddot{\G}$ respectively replaced by
$\check{\G}_{H}$ and $\ddot{\G}_{H}$, where $H := \big\{i: (\bphi^\ast)_i \neq 0 \big\}$ and for any matrix with $L$ columns, $(\cdot)_H$ denotes the corresponding submatrix with columns restricted on $H$. Then, as $T, d \to \infty$, with probability approaching 1, $\textnormal{sign}(\wh\bphi_H) = \textnormal{sign}(\bphi_H^\ast)$ and $\wh\bphi_{H^c} = \0$.

If we further assume the smallest eigenvalue of $\R_H \S_{\bgamma} \S_{\bgamma}^\top \R_H^\top$ is of constant order, where $\R_H$ and $\S_{\bgamma}$ are defined below, then $\wh\bphi_H$ is asymptotically normal with rate $T^{-1/2} d^{-(1-b)/2}$ such that
\begin{align*}
& T^{1/2} (\R_H \S_{\bgamma} \R_{\beta} \bSigma_{\beta} \R_{\beta}^\top \S_{\bgamma}^\top \R_H^\top)^{-1/2} (\wh\bphi_H - \bphi_H^\ast) \xrightarrow{D} \cN(\0, \I_{|H|}) ,
\quad \text{where}
\end{align*}
\begin{align*}
    & \R_{\beta} = \big[\b{E}(\X_t^\top \B_t) \b{E}(\B_t^\top \X_t)\big]^{-1} \b{E}(\X_t^\top \B_t),
    \;\;\;
    \bSigma_\beta = \sum_{\tau} \b{E}\big\{(\B_t- \b{E}(\B_t))^\top \bepsilon_t \bepsilon_{t+\tau}^\top (\B_{t+\tau}- \b{E}(\B_t))\big\} , \\
    & \S_{\bgamma} = \Big( \b{E}\big[ \X_{t,1\cdot} (\B_{t,1\cdot} -\b{E}(\B_{t,1\cdot}))^\top \big] \bgamma, \;\dots,\; \b{E}\big[ \X_{t,1\cdot} (\B_{t,d\cdot} -\b{E}(\B_{t,d\cdot}))^\top \big] \bgamma, \\
    &\hspace{12pt}
    \dots,\; \b{E}\big[ \X_{t,d\cdot} (\B_{t,1\cdot} -\b{E}(\B_{t,1\cdot}))^\top \big] \bgamma, \;\dots,\; \b{E}\big[ \X_{t,d\cdot} (\B_{t,d\cdot} -\b{E}(\B_{t,d\cdot}))^\top \big] \bgamma \Big)^\top ,\\
    & \R_H = [(\H_{20} -\H_{10})_H^\top (\H_{20} -\H_{10})_H]^{-1} (\H_{20} -\H_{10})_H^\top ,\\
    & \H_{10} =
    \hspace{-2pt} \Big\{
    \hspace{-2pt} \Big(\I_d \otimes  (\bgamma^\top
    \hspace{-2pt} \otimes \I_d) \b{E}\big(\U_{\x, 1,0} (\bbeta^\ast \otimes \I_d) \bPi_t^{\ast \top} \big)
    \Big) \textnormal{vec}(\W_1^\top) +
    \hspace{-2pt} \Big(
    \I_d \otimes (\bgamma^\top
    \hspace{-2pt} \otimes \I_d) \b{E}\big( \U_{\bmu, 1,0} \bPi_t^{\ast \top} \big)
    \Big) \textnormal{vec}(\W_1^\top), \\
    &\hspace{12pt}
    \dots, \Big(\I_d \otimes (\bgamma^\top
    \hspace{-2pt} \otimes \I_d) \b{E}\big( \U_{\x, p,l_p} (\bbeta^\ast \otimes \I_d) \bPi_t^{\ast \top} \big)
    \Big) \textnormal{vec}(\W_p^\top) + 
    \hspace{-2pt} \Big(
    \I_d \otimes (\bgamma^\top
    \hspace{-2pt} \otimes \I_d) \b{E}\big( \U_{\bmu, p,l_p} \bPi_t^{\ast \top} \big) 
    \Big) \textnormal{vec}(\W_p^\top)
    \hspace{-2pt} \Big\} , \\
    & \H_{20} = \b{E}(\X_t \otimes \B_t\bgamma) \big[\b{E}(\X_t^\top \B_t) \b{E}(\B_t^\top \X_t)\big]^{-1} \b{E}(\X_t^\top \B_t) \\
    &\hspace{12pt}
    \cdot \Big\{\V_{\W_1^\top, v}^\top \b{E}\big[( \I_d \otimes \U_{\x, 1,0}) \V_{\bPi_t^\ast, r} \big] \bbeta^\ast + \V_{\W_1^\top, v}^\top \b{E}\big[ (\I_d \otimes \U_{\bmu, 1,0}) \textnormal{vec}(\bPi_t^{\ast \top}) \big], \\
    &\hspace{12pt}
    \dots, \V_{\W_p^\top, v}^\top  \b{E}\big[ (\I_d \otimes \U_{\x, p,l_p}) \V_{\bPi_t^\ast, r} \big] \bbeta^\ast + \V_{\W_p^\top, v}^\top \b{E}\big[ (\I_d \otimes \U_{\bmu, p,l_p}) \textnormal{vec}(\bPi_t^{\ast \top}) \big] \Big\} ,\\
    & \U_{\x, j,k} = \frac{1}{T}\sum_{t=1}^T z_{j,k,t} \textnormal{vec}(\B_t -\bar{\B}) \x_t^\top ,
    \;\;\;
    \U_{\bmu, j,k} = \frac{1}{T}\sum_{t=1}^T z_{j,k,t} \textnormal{vec}(\B_t -\bar{\B}) \bmu^{\ast \top} ,\\
    & \hspace{12pt}
    \text{with } \V_{\H, K} = \big(
    \I_K \otimes \h_1^\top, 
    \;\;\; \hdots , \;\;\;
    \I_K \otimes \h_n^\top
    \big)^\top \;\;\; \text{for any given $n\times d$ matrix $\H = (\h_1, \dots, \h_n)^\top$} .
\end{align*}
\end{theorem}

Regarding the new assumptions in Theorem \ref{thm: ada_LASSO_phi_asymp}, the term $\check{\G}$ (\text{resp.} $\ddot{\G}$) in (R4) (\text{resp.} (R6)) is simply replaced by its $H$-restricted version. Moreover, we show in the proof that $\R_H \S_{\bgamma} \S_{\bgamma}^\top \R_H^\top$ has its largest eigenvalue of constant order, and hence the requirement on its smallest eigenvalue is not particularly strong. Nonetheless, two key results are obtained: $\wh\bphi$ consistently estimates the zeros in $\bphi^\ast$ as exact zeros, and is asymptotically normal on the nonzero entries in $\bphi^\ast$. The convergence rate is $T^{-1/2} d^{-(1-b)/2}$, which is slower if more variables in $\B_t$ are correlated. In the special case where $\B_t$ is cross-sectionally and serially uncorrelated with $b=0$, the convergence rate simplifies to $(Td)^{-1/2}$.

Theorem~\ref{thm: ada_LASSO_phi_asymp} enables us to perform inference on $\bphi_H$ in practice, with the covariance matrix replaced by the plug-in estimator; see Section~\ref{sec: practical_implementation} for more details. If $\{z_{j,k,t}\}$'s are non-stochastic, inference on $\rho_t^\ast$ by $\wh{\rho}_t := (\z_t)_H^\top \wh{\bphi}_H$ is also feasible, since
\[
T^{1/2} ((\z_t)_H^\top \R_H \S_{\bgamma} \R_{\beta} \bSigma_{\beta} \R_{\beta}^\top \S_{\bgamma}^\top \R_H^\top (\z_t)_H)^{-1/2} (\wh{\rho}_t -\rho_t^\ast) \xrightarrow{D} \cN(0,1).
\]

Lastly, we present the consistency of the spatial weight matrix estimator and the spatial fixed effect estimator. 

\begin{theorem}\label{thm: spatial_weight_rate}
Let all assumptions in Theorems \ref{thm: LASSO_rate} and \ref{thm: ada_LASSO_phi_asymp} hold. Then, for $\wh\W_t := \sum_{j=1}^{p} \Big(\wh{\phi}_{j,0} + \sum_{k=1}^{l_j} \wh{\phi}_{j,k} z_{j,k,t} \Big) \W_j$ and the spatial fixed effect estimator $\wh\bmu$ defined in (\ref{eqn: def_mu_hat}), we have
\[
\big\|\wh\W_t - \W_t^\ast \big\|_\infty =O_P\big( T^{-1/2} d^{-(1-b)/2} \big) =\big\|\wh\W_t - \W_t^\ast \big\|_1, \;\;\;
\big\|\wh\bmu -\bmu^\ast \big\|_\textnormal{max} =O_P(c_T).
\]
\end{theorem}
Note that this result implies that the spectral norm error of the spatial weight matrix estimator $\wh\W_t$ also satisfies $\big\|\wh\W_t - \W_t^\ast \big\| = O_P\big( T^{-1/2} d^{-(1-b)/2} \big)$.

\section{Practical implementation}\label{sec: practical_implementation}

From Section \ref{sec: full_matrix_notations}, we estimate $(\bmu^\ast, \bphi^\ast, \bbeta^\ast)$ by first obtaining a penalized estimator for $\bphi^\ast$, followed by the least squares estimator for $\bbeta^\ast$ and $\bmu^\ast$. We now present the step-by-step algorithm.
\bigskip

\noindent \ul{Algorithm for $(\bmu^\ast, \bphi^\ast, \bbeta^\ast)$ Estimation}
\begin{itemize}
  \item[1.] Compute the least squares estimator stated in (\ref{eqn: simplify_phi_ls_matrix}) and denote it as $\wt\bphi$.
  \item[2.] Construct $\u$ using $\wt\bphi$. Using the Least Angle Regressions (LARS) \citep{Efronetal2004}, solve the adaptive LASSO problem stated in (\ref{eqn: simplify_phi_ada_lasso_matrix}), and denote the solution by $\wh\bphi$.
  \item[3.] Using (\ref{eqn: profiled_beta_matrix}), obtain the least squares estimator for $\bbeta^\ast$ as $\wh\bbeta = \bbeta(\wh\bphi)$.
  \item[4.] According to (\ref{eqn: reformulate}), construct $\wh\bPhi$ using $\wh\bphi$ and obtain the least squares estimator for $\bmu^\ast$ as $\wh\bmu = T^{-1} \sum_{t=1}^T \big\{ (\I_d - \bLambda_t \wh\bPhi) \y_t - \X_t \wh\bbeta \big\}$.
\end{itemize}
\bigskip

The tuning parameter $\lambda$ in step 2 can be determined via minimizing the following BIC:
\begin{equation}
\label{eqn: BIC_lambda}
\text{BIC}(\lambda) = \log\Big( \frac{1}{T}\; \Big\| \B^\top\y - \B^\top\V \wh\bphi - \B^\top\X_{\bbeta(\wh\bphi)} \Vec{\I_d} \Big\|^2 \Big) + |\wh{H}|\; \frac{\log(T)}{T} \;\log(\log(L)),
\end{equation}
as inspired by \cite{Wangetal2009} (\text{cf.} equation (2.7) within), where $\wh\bphi$ is the adaptive LASSO solution with parameter $\lambda$ and $\wh{H}$ is the set of indices on which $\wh\bphi$ is nonzero. Note that although $\B$ contains the unknown constant $a$, the optimal $\lambda$ is independent of it. A procedure for assessing the goodness of fit of a given set of dynamic variables $\{z_{j,k,t}\}$ can also be facilitated by \eqref{eqn: BIC_lambda}. In detail, we can compare \eqref{eqn: spatiallag} with the null model $\c{H}_0: \y_t = \bmu^\ast + \sum_{j=1}^{p} \phi_{j,0}^\ast \W_j\y_t + \X_t\bbeta^\ast + \bepsilon_t$ using BIC. For $\c{H}_0$, we compute the following BIC, borrowed from equation (10) in \cite{WangLeng2007}:
\begin{equation}
\label{eqn: BIC_lambda_null}
\text{BIC}(\c{H}_0) = \log\Big( \frac{1}{T}\; \Big\| \B^\top\y - \B^\top\V \wh\bphi - \B^\top\X_{\bbeta(\wh\bphi)} \Vec{\I_d} \Big\|^2 \Big) + |\wh{H}|\; \frac{\log(T)}{T},
\end{equation}
where the variables are constructed under $\c{H}_0$ and the above algorithm is implemented accordingly. Compared to \eqref{eqn: BIC_lambda_null}, the criterion in \eqref{eqn: BIC_lambda} accounts for potentially growing $L$. Hence, if $L$ is small, we may also use \eqref{eqn: BIC_lambda_null} (with variables constructed under the model of interest) to select $\lambda$. More discussion is relegated to the online supplement, with extensive numerical results therein.

Finally, to utilize the asymptotic normality of $\wh\bphi_H$ for feasible inference, we require estimators for $\R_H$, $\R_\beta$, $\bSigma_\beta$ and $\S_{\bgamma}$ in Theorem \ref{thm: ada_LASSO_phi_asymp}. By replacing the expected values by their sample estimates, using $\wh{H} = \{i:(\wh\bphi)_i \neq 0\}$ to estimate the set $H$ and leveraging all the consistency results for $\bbeta(\wh\bphi)$, $\wh\bmu$ and $\wh{H}$, we obtain estimators $\wh\R_{\wh{H}}$, $\wh\R_\beta$ and $\wh\S_{\bgamma}$. For $\bSigma_\beta$, we use a consistent estimator of $\bepsilon_t$ denoted by $\wh\bepsilon_t := \y_t -\wh\bmu -\wh\W_t \y_t - \X_t \bbeta(\wh\bphi)$. As $\bSigma_\beta$ involves an infinite sum, we can sum up to a cut-off $\tau^*$ after which the sum changes little, and denote the constructed estimator $\wh\bSigma_\beta$. Putting everything together, the covariance matrix of $\wh\bphi_H$ can be estimated by $T^{-1} \wh\R_{\wh{H}} \wh\S_{\bgamma} \wh\R_{\beta} \wh\bSigma_{\beta} \wh\R_{\beta}^\top \wh\S_{\bgamma}^\top \wh\R_{\wh{H}}^\top$.

Establishing the theoretical validity of inference based on plug-in estimators would require additional assumptions to rigorously justify each step outlined above. As this is not the primary focus of our work, we do not pursue a formal proof of consistency for the plug-in estimator. Instead, we provide empirical evidence supporting its adequacy through simulation studies in Section~\ref{subsec:simulations}.

\section{Change point detection and estimation in dynamic spatial autoregressive models}\label{sec:Change_Point_Detection}

\subsection{Threshold spatial autoregressive models}\label{subsec: example_threshold}
The early work by \cite{Tong1978} proposes a regime switching mechanism via the threshold autoregressive model. Since then, it has been studied extensively in the past few decades. More recently, such threshold structure is used by researchers in spatial econometrics; see, for example, threshold spatial autoregressive models for purely cross-sectional data by \cite{Deng2018} and \cite{LiLin2024}. One benefit of the framework introduced in this work is that threshold variables can be directly adapted into (\ref{eqn: spatiallag}) under a panel data setup. As a simple example, we consider
\begin{equation}
\label{eqn: threshold_spatial}
    \y_t =
    \bmu^\ast + \Big(\phi_1^\ast \W_1 \b{1}\{q_t \leq \gamma^\ast\} + \phi_2^\ast \W_2 \b{1}\{q_t > \gamma^\ast\} \Big)\y_t +\X_t\bbeta^\ast + \bepsilon_t .
\end{equation}
This is a spatial autoregressive model with regime switching on the spatial weight matrix, where $q_t$ is some observed threshold variable with an unknown threshold value $\gamma^\ast$. By rewriting (\ref{eqn: threshold_spatial}) in the form of (\ref{eqn: spatiallag}), we have
\begin{equation}
\label{eqn: threshold_spatial_rewrite}
\y_t = \bmu^\ast + z_{1,1,t} \phi_1^\ast \W_1 \y_t +  z_{2,1,t} \phi_2^\ast \W_2 \y_t +\X_t\bbeta^\ast + \bepsilon_t ,
\;\;\; \text{where }
z_{1,1,t}:= \b{1}\{q_t \leq \gamma^\ast\}, \;
z_{2,1,t}:= \b{1}\{q_t > \gamma^\ast\}.
\end{equation}

We consider the estimation of the threshold value $\gamma^\ast$. Suppose there is a domain of possible threshold values $\Gamma = [\gamma_\text{min}, \gamma_\text{max}]$. A standard approach in threshold models is to search the minimum regression error over the intersection $\Gamma \cap \{q_1, \dots, q_T\}$; see, for example, \cite{Deng2018}. Our framework provides an alternative approach. Denote the elements in the intersection by $\gamma_1\leq \gamma_2\leq \dots \leq \gamma_L$, with $L \equiv |\Gamma \cap \{q_1, \dots, q_T\}|$, and let $\gamma^\ast$ be identified as one of them. Let $z_{1,l,t} := \b{1}\{q_t \leq \gamma_l\}$ and $z_{2,l,t} := \b{1}\{q_t > \gamma_l\}$ for all $l\in[L]$. Then, we can consider a spatial autoregressive model:
\begin{equation}
\label{eqn: threshold_spatial_rewrite_saturated}
\y_t = \bmu^\ast + \sum_{l=1}^{L} z_{1,l,t} \phi_{1,l}^\ast \W_1 \y_t + \sum_{l=1}^{L} z_{2,l,t} \phi_{2,l}^\ast \W_2 \y_t +\X_t\bbeta^\ast + \bepsilon_t ,
\end{equation}
where $\phi_{1,l}^\ast$ and $\phi_{2,l}^\ast$ would be nonzero\footnote{
In practice, we often have no prior information on which spatial weight matrix corresponds to the regime $q_t \leq \gamma^\ast$. This can be resolved in \eqref{eqn: threshold_spatial_rewrite_saturated} by writing $\big( \sum_{l=1}^{L} z_{1,l,t} \phi_{1,l}^\ast \big)$ and $\big( \sum_{l=1}^{L} z_{2,l,t} \phi_{2,l}^\ast \big)$ as $\big( \phi_{1,0}^\ast + \sum_{l=1}^{L} z_{1,l,t} \phi_{1,l}^\ast \big)$ and $\big( \phi_{2,0}^\ast + \sum_{l=1}^{L} z_{2,l,t} \phi_{2,l}^\ast \big)$, respectively.
}
only for $\gamma_l = \gamma^\ast$. The threshold value can be selected consistently in one step by the oracle property of our adaptive LASSO estimator in Theorem~\ref{thm: ada_LASSO_phi_asymp}. We present this result in Corollary \ref{corollary: threshold_consistency}. Given the sparse solution $(\wh\phi_{1,l}, \wh\phi_{2,l})_{l \in [L]}$ of (\ref{eqn: threshold_spatial_rewrite_saturated}), we can re-estimate all model parameters. Note that such an approach remains applicable for $L$ growing with $T$.
In practice, the order of $L$ may not fulfill Assumption (R10), but we can circumvent this issue by a two-step procedure, as we will discuss later in Section~\ref{subsec: example_change_point}.

Our framework also allows us to consider a spatial autoregressive model with regime switching on the spatial correlation coefficients, similar to \cite{LiLin2024} except that their regimes change over spatial units while ours over timestamps as follows,
\begin{equation}
\label{eqn: threshold_spatial_coef}
\y_t =
\bmu^\ast + \Big( \phi_1^\ast \b{1}\{q_t \leq \gamma^\ast\} + \phi_2^\ast \b{1}\{q_t > \gamma^\ast\} \Big) \W_1 \y_t +\X_t\bbeta^\ast + \bepsilon_t .
\end{equation}
To estimate the parameters, \cite{LiLin2024} uses quasi maximum likelihood estimators (QMLE) and traverses over a finite parameter space for the threshold value $\gamma^\ast$. In contrast, a one-step estimation is again feasible by our framework. To this end, we may read (\ref{eqn: threshold_spatial_coef}) in the form,
\begin{equation}
\label{eqn: threshold_spatial_coef_rewrite_saturated}
\y_t =\bmu^\ast + \Big( \phi_{1,0}^\ast + \sum_{l=1}^{L} z_{1,l,t} \phi_{1,l}^\ast \Big) \W_1 \y_t +\X_t\bbeta^\ast + \bepsilon_t ,
\end{equation}
where $z_{1,l,t}$ for $l\in[L]$ is as previously defined. With our adaptive LASSO estimators, only $\wh\phi_{1,0}$ and one $\wh\phi_{1,l}$ such that $\gamma_l = \gamma^\ast$ are expected to be nonzero. The consistency of such estimator for $\gamma^\ast$ is included in Corollary \ref{corollary: threshold_consistency}.

\begin{corollary}
\label{corollary: threshold_consistency}
(Consistency on threshold value estimation)
Let all assumptions in Theorem \ref{thm: ada_LASSO_phi_asymp} hold.

\noindent (a) For model \eqref{eqn: threshold_spatial}, the threshold value $\gamma^\ast$ can be consistently estimated by the set of estimators $\{\gamma_l : \wh\phi_{1,l} \neq 0, \, \wh\phi_{2,l} \neq 0\}$, where $\{ \wh\phi_{1,l}, \wh\phi_{2,l} \}_{l\in[L]}$ is the adaptive LASSO solution for $\{ \phi_{1,l}^\ast, \phi_{2,l}^\ast \}_{l\in[L]}$ in \eqref{eqn: threshold_spatial_rewrite_saturated}.

\noindent (b) For model \eqref{eqn: threshold_spatial_coef}, the threshold value $\gamma^\ast$ can be consistently estimated by the set of estimators $\{\gamma_l : \wh\phi_{1,l} \neq 0\}$, where $\{ \wh\phi_{1,l} \}_{l\in[L]}$ is the adaptive LASSO solution for $\{ \phi_{1,l}^\ast \}_{l\in[L]}$ in \eqref{eqn: threshold_spatial_coef_rewrite_saturated}.
\end{corollary}

In fact, our framework~\eqref{eqn: spatiallag} can be applied to spatial autoregressive models with more complicated threshold structures in the spatial weight matrix $\W_t^\ast$, e.g., regimes from multiple threshold variables with multiple threshold values. We leave this discussion to the online supplement.

\subsection{Spatial autoregressive models with structural change points}\label{subsec: example_change_point}
Structural changes in the relationship of variables in econometric models have been studied extensively in the literature; see, for example, \cite{Sengupta2017} and \cite{BarigozziTrapani2020}. As a second example demonstrating the applicability of our framework, we consider the spatial autoregressive model with a structural change:
\begin{equation}
\label{eqn: change_spatial}
\y_t =
\bmu^\ast + \Big(\phi_1^\ast \W_1 \b{1}\{t \leq t^*\} + \phi_2^\ast \W_2 \b{1}\{t > t^*\} \Big) \y_t +\X_t\bbeta^\ast + \bepsilon_t .
\end{equation}
where $t^*$ is some unknown change location. Similar to the threshold model example, this can also be expressed in the form of~\eqref{eqn: threshold_spatial_rewrite}, now with $z_{1,1,t}:= \b{1}\{t \leq t^*\}$ and $z_{2,1,t}:= \b{1}\{t > t^*\}$. It is important to note that, despite the models taking the same form, the dynamic variables $z_{1,1,t}$ and $z_{2,1,t}$ are random in the threshold model but non-random in the change point model here. Recall that our main results hold for both types of $\{z_{j,k,t}\}$; see Section~\ref{subsec: assumption} for a more detailed discussion.

To estimate the change location $t^*$, \cite{Li2018} calculates the quasi maximum likelihood for each possible change location and sets the maximizer as the estimator. When the set of possible change locations is large, this approach requires a significant number of model fittings. Using our framework, it is again possible to estimate $t^*$ in one go. Let $\c{T}$ denote the set of all candidate change point locations such that $\c{T} = \{t_1, \ldots, t_{|\c{T}|}\}$. Then, we rewrite model \eqref{eqn: change_spatial} as
\begin{equation}
\label{eqn: change_rewrite_saturated}
\y_t = \bmu^\ast + \sum_{l=1}^{|\mathcal{T}|} z_{1,l,t} \phi_{1,l}^\ast \W_1 \y_t + \sum_{l=1}^{|\mathcal{T}|} z_{2,l,t} \phi_{2,l}^\ast \W_2 \y_t +\X_t\bbeta^\ast + \bepsilon_t ,
\end{equation}
where $z_{1,l,t} := \mathbbm{1}\{t \leq t_l\}$ and $z_{2,l,t} := \mathbbm{1}\{t > t_l\}$. We then follow the same argument used in the threshold model below \eqref{eqn: threshold_spatial_rewrite_saturated} and the consistency of our change location estimate is again guaranteed.

\begin{corollary}
\label{corollary: change_consistency}
(Consistency on change location estimation)
Let all assumptions in Theorem \ref{thm: ada_LASSO_phi_asymp} hold, with $L$ replaced by $|\c{T}|$. Consider model \eqref{eqn: change_spatial} and assume that $t^* \in \c{T}$. Then, the change location $t^*$ can be consistently estimated by the set of estimators $\bigl\{l \in [|\c{T}|] : \wh\phi_{1,l} \neq 0, \, \wh\phi_{2,l} \neq 0\bigr\}$, where $\bigl\{ \wh\phi_{1,l}, \wh\phi_{2,l} \bigr\}_{l\in [|\c{T}|]}$ is the adaptive LASSO solution for $\{ \phi_{1,l}^\ast, \phi_{2,l}^\ast \}_{l\in [|\c{T}|]}$ in \eqref{eqn: change_rewrite_saturated}.
\end{corollary}

We can also consider a multiple change model:
\begin{equation}
\label{eqn: change_spatial_multiple}
\y_t =
\bmu^\ast + \sum_{j=1}^p \Big(\phi_{j,1}^\ast \b{1}\{t \leq t_1^*\} + \phi_{j,2}^\ast \b{1}\{t_1^* < t \leq t_2^*\} + \dots + \phi_{j,k}^\ast \b{1}\{t > t_k^*\}\Big) \W_j \y_t +\X_t\bbeta^\ast + \bepsilon_t .
\end{equation}
This model allows for $k$ change points in $\W_t^\ast$ consisting of $p$ spatial weight candidates, with the number of change points $k$ unknown. The result below confirms the consistency of the estimations in both change point numbers and locations.
\begin{corollary}
\label{corollary: change_consistency_multiple}
(Consistency on the estimations for the number of changes and the change locations)
Given a set of all candidate change point locations $\c{T}$ and assume that $t_i^* \in \mathcal{T}$ for all $i\in[k]$. Let all assumptions in Theorem~\ref{thm: ada_LASSO_phi_asymp} hold, with $L$ replaced by $|\c{T}|$. For model~\eqref{eqn: change_spatial_multiple}, let $\{ \wh\phi_{j,l} \}_{j\in [p], l\in [|\c{T}|]}$ denote the adaptive LASSO solution for $\{ \phi_{j,l}^\ast \}_{j\in [p], l\in [|\c{T}|]}$ in
\begin{equation*}
\y_t =\bmu^\ast + \sum_{j=1}^p \Big(\phi_{j,0}^\ast + \sum_{l=1}^{|\c{T}|} z_{j,l,t} \phi_{j,l}^\ast \Big) \W_j \y_t +\X_t\bbeta^\ast + \bepsilon_t ,
\end{equation*}
where $z_{j,l,t} := \mathbbm{1}\{t \leq t_l\}$ with $t_l$ being the $l$-th element of $\c{T}$ for $j\in[p], l \in \big[|\c{T} |\big]$. Write $\wh{\c{T}}:= \{l\neq 0: \wh\phi_{j,l} \neq 0 \text{ for some } j\in [p]\}$. Then, $\wh{k} := |\wh{\c{T}}|$ estimates $k$ consistently and, for every $i\in[\wh{k}]$, the $i$-th smallest element in $\wh{\c{T}}$ estimates $t_i^\ast$ consistently.
\end{corollary}
\begin{remark}\label{remark: sequential_procedure}
Suppose the true number of changes is of order $L$ satisfying Assumption (R10), but throughout Section~\ref{subsec: example_change_point}, the size of the set $\c{T}$ is restricted by (R10). This means that applying our procedure to a single set $\mathcal{T}$ containing many candidate change point locations may violate this assumption. A similar concern arises with the order of $L$ in threshold spatial autoregressive models in Section~\ref{subsec: example_threshold}.

To address this issue in practice, we may resort to a divide-and-conquer scheme to work with smaller candidate sets instead. Consider model \eqref{eqn: change_spatial} for instance. We first form some subsets $\c{T}_1, \c{T}_2,\ldots$ such that $\c{T} = \cup_{j} \c{T}_j$ and that each $\c{T}_j$ satisfies (R10) (with $L$ replaced by $|\c{T}_j|$). Note that we may allow these subsets to overlap. On each subset, we run the estimation algorithm and obtain all identified potential change locations within the subset. Then, we aggregate all those locations into a set $\wt{\c{T}}$, which can be shown, using Corollary~\ref{corollary: change_consistency} on each $\c{T}_j$, to satisfy (R10) (with $L$ replaced by $|\wt{\c{T}}|$). Finally, we estimate \eqref{eqn: change_rewrite_saturated} with $\c{T}$ replaced by $\wt{\c{T}}$ to determine the change point location. Simulations in the online supplement confirms the effectiveness of this scheme.

\end{remark}

\section{Numericla studies}\label{sec: numerical}
\subsection{Simulations} \label{subsec:simulations}
In this subsection, we conduct Monte Carlo simulations to demonstrate the performance of our estimators. For the general setting, we consider
\begin{equation}
\label{eqn: sim_dgp_general}
\y_t = \Big\{\I_d- \big(0.2 +0.2\, z_{1,1,t} +0\, z_{1,2,t}\big) \W_1 - \big(0 +0\, z_{2,1,t} +0.3\, z_{2,2,t} \big) \W_2 \Big\}^{-1} \big(\bmu^\ast + \X_t\bbeta^\ast + \bepsilon_t \big) ,
\end{equation}
where $\bmu^\ast$ and $\bbeta^\ast$ are vectors of $1$'s, $\W_1$ is generated to have two neighbors ahead and two behind as in \cite{KelejianPrucha1998}, and $\W_2$ is a contiguity matrix with off-diagonal entries being i.i.d. $\text{Bernoulli}\,(0.2)$. The true parameter is $\bphi^\ast = (0.2,\, 0.2,\, 0,\, 0,\, 0,\, 0.3)^\top$. The disturbance $\bepsilon_t$ is jointly Gaussian with its variance-covariance matrix having 1 on the diagonal and each upper triangular entries 0.1 with probability 0.2 and 0 otherwise. For any row of $\W_1$ or $\W_2$ with row sum exceeding one, we divide each entry by the $L_1$ norm of the row. We use independent standard normal random variables for the dynamic variables $\{z_{1,1,t}\}$, $\{z_{1,2,t}\}$, $\{z_{2,1,t}\}$ and $\{z_{2,2,t}\}$. The covariate matrix $\X_t$ has three columns, with each entry generated as independent standard normal, except that the third column is endogenous by adding $0.5\; \bepsilon_t$. Let $\X_{\text{exo},t}$ be $\X_t$ with the disturbance part removed, then the instruments can be set as $\B_t = \big[ \X_{\text{exo},t}, \W_1 \X_{\text{exo},t}, \W_2 \X_{\text{exo},t} \big]$. The tuning parameter for the adaptive LASSO is selected by minimizing the BIC in \eqref{eqn: BIC_lambda}.

We experiment $d=25, 50, 75$ and $T=50, 100, 150$, with each setting repeated 1000 times. Results are presented in Table \ref{tab: sim_general_setting}. In there, MSE is the mean squared error; specificity is the proportion of true zeros estimated as zeros; sensitivity is the proportion of nonzeros estimated as nonzeros. The MSE results corroborate the consistency of parameter estimation in Theorem \ref{thm: LASSO_rate} and \ref{thm: spatial_weight_rate}, while the specificity and sensitivity results corroborate the sparsity consistency in Theorem \ref{thm: ada_LASSO_phi_asymp}. Zeros in $\bphi^\ast$ can be selected with high accuracy, yet the sensitivity results suggest a mild over-identification of zeros. Both increasing the spatial dimension $d$ and time span $T$ improve the performance of our estimators in general, except that when $T$ increases from 100 to 150 and $d=25$, all performance measures slightly deteriorate, which is similarly observed in Table 1 of \cite{LamSouza2020}. This might suggest the issue of a data set with unbalanced dimensions in practice.

\begin{table}[ht!]
\setlength{\tabcolsep}{5pt}
\small
\begin{center}
\begin{tabular}{l||ccc|ccc|ccc}
\hline &  & $T=50$ &  &  & $T=100$ &  &  & $T=150$ &  \\ \cline{2-10}
& $d=25$ & $d=50$ & $d=75$ & $d=25$ & $d=50$ & $d=75$ & $d=25$ & $d=50$ & $d=75$ \\
\hline & & & & & & & & &  \\[-1em]
$\wh\bphi$ MSE & .002 & .001 & .002 & .000 & .000 & .000 & .003 & .001 & .000  \\
 & {\small (.008)} & {\small (.001)} & {\small (.001)} & {\small (.000)} & {\small (.000)} & {\small (.000)} & {\small (.007)} & {\small (.000)} & {\small (.000)} \\
 $\wh\bbeta$ MSE & .087 & .029 & .080 & .005 & .008 & .001 & .088 & .013 & .005  \\
 & {\small (.434)} & {\small (.011)} & {\small (.025)} & {\small (.004)} & {\small (.002)} & {\small (.001)} & {\small (.278)} & {\small (.005)} & {\small (.002)} \\
 $\wh\bmu$ MSE & .039 & .038 & .044 & .013 & .010 & .011 & .036 & .010 & .008  \\
 & {\small (.124)} & {\small (.010)} & {\small (.010)} & {\small (.006)} & {\small (.002)} & {\small (.002)} & {\small (.083)} & {\small (.003)} & {\small (.002)} \\
 $\wh\bphi$ Specificity & .992 & 1.000 & 1.000 & 1.000 & 1.000 & 1.000 & .994 & 1.000 & 1.000  \\
 & {\small (.064)} & {\small(.000)} & {\small (.000)} & {\small (.000)} & {\small (.000)} & {\small (.000)} & {\small (.048)} & {\small (.000)} & {\small (.000)} \\
 $\wh\bphi$ Sensitivity & .921 & .983 & .992 & .910 & .961 & .991 & .873 & .983 & .956  \\
 & {\small (.133)} & {\small (.069)} & {\small (.048)} & {\small (.132)} & {\small (.099)} & {\small (.051)} & {\small (.148)} & {\small (.070)} & {\small (.103)} \\
\hline
\end{tabular}
\end{center}
\caption{Simulation results for the general setting~\eqref{eqn: sim_dgp_general}. The mean and standard deviation (in brackets) of the corresponding error measures over 1000 repetitions are reported.}
\label{tab: sim_general_setting}
\end{table}

\begin{figure}[ht!]
\begin{center}
\centerline{\includegraphics[width=0.8\textwidth]{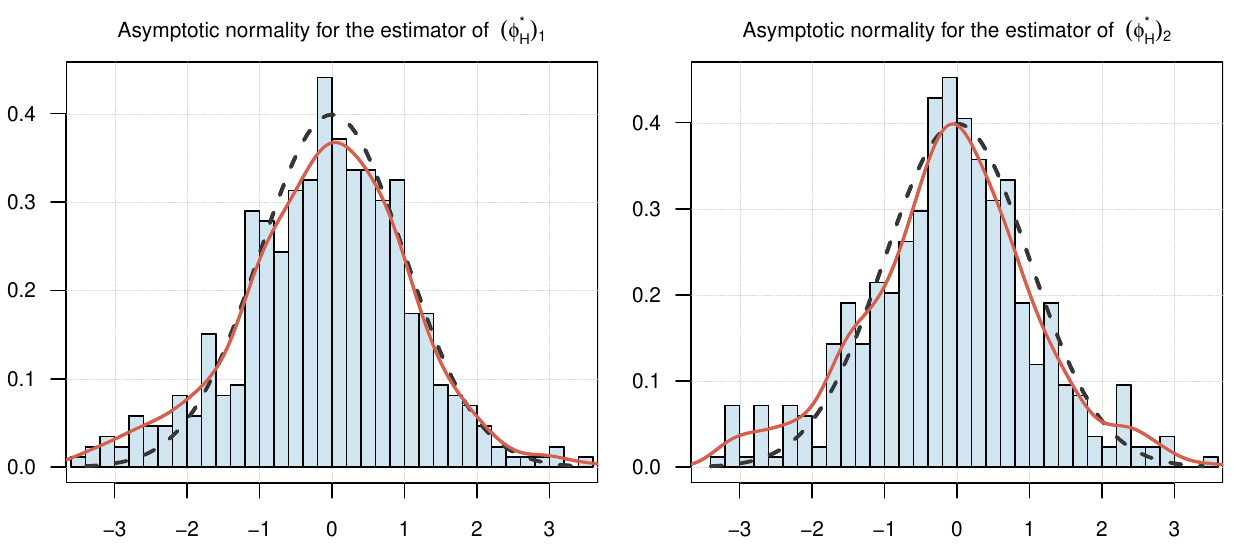}}
\caption{Histogram of $T^{1/2} (\wh\R_{\wh{H}} \wh\S_{\bgamma} \wh\R_{\beta} \wh\bSigma_{\beta} \wh\R_{\beta}^\top \wh\S_{\bgamma}^\top \wh\R_{\wh{H}}^\top)^{-1/2} (\wh\bphi_{\wh{H}} -\bphi_{\wh{H}}^\ast)$ for $(T,d) =(200,50)$, shown for the first coordinate (left panel) and the second coordinate (right panel). The red curves are the empirical density, and the black dotted curves are the density for $\cN(0,1)$.}
\label{fig: simulation_asymp_200_50}
\end{center}
\end{figure}

To better illustrate the asymptotic normality for $\wh\bphi$ in Theorem \ref{thm: ada_LASSO_phi_asymp}, we use the same data generating mechanism as above with $(T,d) =(200,50)$, except that $\bphi^\ast = (0,\, \text{-}0.5,\, 0.5,\, 0,\, 0,\, 0)^\top$, $\X_t$ is exogenous and $\bepsilon_t$ has a diagonal covariance matrix. For ease of presentation, we fix $\wh{H} =\{2,3\}$ which is the index set of true nonzero parameters. The remaining components of the covariance matrix are estimated according to the last part of Section~\ref{sec: practical_implementation}. In particular, Figure~\ref{fig: simulation_asymp_200_50} displays the histogram of $T^{1/2} (\wh\R_{\wh{H}} \wh\S_{\bgamma} \wh\R_{\beta} \wh\bSigma_{\beta} \wh\R_{\beta}^\top \wh\S_{\bgamma}^\top \wh\R_{\wh{H}}^\top)^{-1/2} (\wh\bphi_{\wh{H}} -\bphi_{\wh{H}}^\ast)$. The plots show good normal approximation to the distribution of this quantity, and hence confirm the results in Theorem~\ref{thm: ada_LASSO_phi_asymp} together with the validity of our proposed covariance estimator. Some discrepancies are present on the tails, potentially due to insufficient dimensions $T$ and $d$. We leave potential improvements to future studies.

\subsection{Change point analysis}\label{subsec: sim_change_point}
We demonstrate in this subsection the performance of our dynamic framework with structural changes as described in Section~\ref{subsec: example_change_point}, with simulation results for the threshold model in Section~\ref{subsec: example_threshold} included in the online supplement. We first consider a spatial autoregressive model with a single change:
\begin{equation}
\label{eqn: sim_dgp_change_point}
\y_t = \Bigl\{ \I_d- 0.3\cdot \b{1}\{t \leq 30\} \W_1 - 0.3\cdot \b{1}\{t > 30\} \W_2 \Bigr\}^{-1} \big(\bmu^\ast + \X_t\bbeta^\ast + \bepsilon_t \big) ,
\end{equation}
where $\{\W_1, \W_2, \bmu^\ast, \X_t, \bbeta^\ast\}$ are constructed in the same way as in \eqref{eqn: sim_dgp_general}. We generate $\bepsilon_t$ as \text{i.i.d.} $\cN(0,1)$ and $t_6$, respectively, to demonstrate the robustness of our estimators under heavy-tailed noise. This data generating mechanism \eqref{eqn: sim_dgp_change_point} represents a change on the true spatial weight matrix at time $t=30$ from $0.3\, \W_1$ to $0.3\, \W_2$. We then fit a model
\begin{align}
    & \y_t = \biggl\{ \I_d- \sum_{l=1}^{\lfloor T/\Delta \rfloor -1} z_{1,l,t} \phi_{1,l}^\ast \W_1 - \sum_{l=1}^{\lfloor T/\Delta \rfloor -1} z_{2,l,t} \phi_{2,l}^\ast \W_2 \biggr\}^{-1} \big(\bmu^\ast + \X_t\bbeta^\ast + \bepsilon_t \big) ,
    \;\;\; \text{where}
    \label{eqn: sim_dgp_change_point_saturated} \\
    & z_{1,l,t} = \b{1}\{t\leq t_l\},
    \;\;\; z_{2,l,t} = \b{1}\{t> t_l\},
    \;\;\;
    t_l = \Delta \cdot l. \notag
\end{align}
We consider a grid of candidate change locations, spaced at intervals of $\Delta=5$. From \eqref{eqn: sim_dgp_change_point}, the true value of every $(\phi_{1,l}^\ast, \phi_{2,l}^\ast)$ is  $(0,0)$ except the one corresponding to $t_l=30$. Table \ref{tab: sim_change_point_setting_weak_lag5} presents the results for each $(T,d)$ setting specified therein, together with the results under a stronger change signal, where the true spatial weight matrix switches from $0.5\, \W_1$ to $0.5\, \W_2$.

\begin{table}[ht!]
\setlength{\tabcolsep}{4pt}
\small
\begin{center}
\begin{tabular}{l||cc|cc|cc|cc}
\hline 
\multicolumn{9}{c}{Weak Change Signal ($0.3\W_1$ to $0.3\W_2$)} \\
\hline $\bepsilon_t$ &  \multicolumn{4}{c}{i.i.d. $\cN(0,1)$} & \multicolumn{4}{|c}{i.i.d. $t_6$}  \\
\hline & & & & & & & & \\[-1em] 
{$(T,d)$} & $(50,25)$ & $(50,50)$ & $(100,50)$ & $(100,75)$ & $(50,25)$ & $(50,50)$ & $(100,50)$ & $(100,75)$ \\
\hline & & & & & & & & \\[-1em]
\text{$\wh\bphi$ MSE} & .008 & .004 & .004 & .003 & .010 & .004 & .004 & .003 \\
 & {\small (.007)} & {\small (.003)} & {\small (.002)} & {\small (.002)} & {\small (.009)} & {\small (.003)} & {\small (.003)} & {\small (.002)} \\
 \text{$\wh\bphi$ True-Unique} & .611 & .965 & .668 & .938 & .559 & .944 & .647 & .898  \\
 & {\small (.488)} & {\small (.184)} & {\small (.472)} & {\small (.242)} & {\small (.497)} & {\small (.231)} & {\small (.479)} & {\small (.303)} \\
\hline
\hline
\multicolumn{9}{c}{Strong Change Signal ($0.5\W_1$ to $0.5\W_2$)} \\
\hline $\bepsilon_t$ &  \multicolumn{4}{c}{\text{i.i.d.} $\cN(0,1)$} & \multicolumn{4}{|c}{\text{i.i.d.} $t_6$}  \\
\hline & & & & & & & & \\[-1em] 
{$(T,d)$} & $(50,25)$ & $(50,50)$ & $(100,50)$ & $(100,75)$ & $(50,25)$ & $(50,50)$ & $(100,50)$ & $(100,75)$ \\
\hline & & & & & & & & \\[-1em]
\text{$\wh\bphi$ MSE} & .006 & .001 & .003 & .001 & .007 & .001 & .004 & .001 \\
 & {\small (.011)} & {\small (.001)} & {\small (.006)} & {\small (.001)} & {\small (.012)} & {\small (.003)} & {\small (.007)} & {\small (.002)} \\
 \text{$\wh\bphi$ True-Unique} & .826 & .994 & .846 & .990 & .759 & .984 & .811 & .972  \\
 & {\small (.379)} & {\small (.077)} & {\small (.361)} & {\small (.100)} & {\small (.428)} & {\small (.126)} & {\small (.392)} & {\small (.166)} \\
\hline
\end{tabular}
\end{center}
\caption{Simulation results for the model~\eqref{eqn: sim_dgp_change_point_saturated} with either a weak or strong change signal. \textit{$\wh\bphi$ True-Unique} is defined to be 1 if the only nonzero pair $(\phi_{1,l}^\ast, \phi_{2,l}^\ast)$ corresponds to $t_l=30$. The mean and standard deviation (in brackets) of the corresponding error measures over 500 repetitions are reported.}
\label{tab: sim_change_point_setting_weak_lag5}
\end{table}

Results for both weak and strong change signals display similar patterns. The performance of change detection is slightly affected by heavy-tailed noise and weaker signals, but remains effective overall. Detection accuracy improves with an increasing spatial dimension $d$. However, a larger $T$ appears to undermine the $\wh\bphi$ True-Unique measure, primarily due to the greater number of dynamic variables $z_{1,l,t}$ and $z_{2,l,t}$ included in the setting.

\bigskip

\noindent \textbf{Comparison with \cite{Li2018}}

To further assess the performance of our estimator and compare both the computational cost and accuracy of our method with that in \cite{Li2018}, we consider
\begin{equation}
\label{eqn: sim_dgp_change_point_compare}
\y_t = \Big\{\I_d- \big(-0.3+ \varrho \cdot \b{1}\{t >30\} \big) \W_1 \Big\}^{-1} \big(\bmu^\ast + \X_t\bbeta^\ast + \bepsilon_t \big) ,
\end{equation}
with $T=100$ and other parameters the same as in \eqref{eqn: sim_dgp_change_point}. This data generating process represents a single change on the spatial weight matrix at $t=30$ from $-0.3 \W_1$ to $(-0.3+\varrho)\W_1$. For simplicity, we only examine the change point candidates in $\{10,20,\dots,90\}$, and compare three estimators:
\begin{itemize}
    \item [1.] AL: The adaptive LASSO estimator with dynamic variables $\{z_{t,0,1}, \dots, z_{t,9,1}\}$, where $z_{t,0,1} = 1$ and $z_{t,k,1} = \b{1}\{t > 10 \cdot k\}$ for $k \in [9]$ and $t \in [100]$.
    \item [2.] AL-DC: The adaptive LASSO estimator using the divide-and-conquer scheme in Remark~\ref{remark: sequential_procedure}, with three subsets $\{z_{t,0,1}, z_{t,1,1}, z_{t,2,1}, z_{t,3,1}\}$, $\{z_{t,0,1}, z_{t,4,1}, z_{t,5,1}, z_{t,6,1}\}$, and $\{z_{t,0,1}, z_{t,7,1}, z_{t,8,1}, z_{t,9,1}\}$, where all dynamic variables are the same as in method 1 above. The final estimator is selected from the three results with the smallest estimation error.
    \item [3.] QMLE \citep{Li2018}: The estimation is performed for each candidate in $\{10,20,\dots,90\}$ and the final estimator is the one  with the smallest estimation error within this set.
\end{itemize}

Table~\ref{tab: sim_change_point_compare_Li2018} reports the results for $T=100$, $d = 25, 50$ with $\varrho = 0.5$, where all simulations are repeated 500 times, and Table~\ref{tab: sim_change_point_compare_Li2018_smallchange} presents the results for a smaller change of $\varrho = 0.1$ at $t = 30$. Note that specificity and sensitivity are not reported here, as QMLE leverages prior information of a single change. We use the estimation error $\sum_{t=1}^T\|\wh\bepsilon_t\|^2 /(Td)$ to quantify overall performance.

\begin{table}[ht!]
\setlength{\tabcolsep}{8pt}
\small
\begin{center}
\begin{tabular}{l||ccc|ccc}
\hline $\bepsilon_t$ &  \multicolumn{3}{c}{\text{i.i.d.} $\cN(0,1)$} & \multicolumn{3}{|c}{\text{i.i.d.} $t_6$}  \\
\hline & & & & & & \\[-1em] 
\hline & & & & & & \\[-1em] 
{$d=25$} & AL & AL-DC & QMLE & AL & AL-DC & QMLE \\
\hline & & & & & & \\[-1em]
\text{$\wh\bphi$ MSE} & .027 & \textbf{.016} & .044 & .032 & \textbf{.019} & .042 \\
 & {\small (.027)} & {\small (.018)} & {\small (.008)} & {\small (.031)} & {\small (.023)} & {\small (.005)} \\
\text{Estimation Error} & 1.01 & \textbf{1.00} & 1.11 & 1.51 & \textbf{1.49} & 1.61 \\
 & {\small (.035)} & {\small (.030)} & {\small (.038)} & {\small (.069)} & {\small (.067)} & {\small (.070)} \\
\text{Run Time~(s)} & \textbf{1.43} & 1.82 & 5.43 & \textbf{1.47} & 1.89 & 5.58  \\
 & {\small (.091)} & {\small (.081)} & {\small (.357)} & {\small (.084)} & {\small (.086)} & {\small (.496)} \\
\hline
\hline & & & & & & \\[-1em] 
{$d=50$} & AL & AL-DC & QMLE & AL & AL-DC & QMLE \\
\hline & & & & & & \\[-1em]
\text{$\wh\bphi$ MSE} & .008 & \textbf{.005} & .036 & .010 & \textbf{.007} & .031 \\
 & {\small (.007)} & {\small (.005)} & {\small (.013)} & {\small (.009)} & {\small (.007)} & {\small (.017)} \\
\text{Estimation Error} & .995 & \textbf{.992} & 1.07 & 1.49 & \textbf{1.48} & 1.56  \\
 & {\small (.020)} & {\small (.020)} & {\small (.047)} & {\small (.049)} & {\small (.048)} & {\small (.073)} \\
\text{Run Time~(s)} & \textbf{6.75} & 8.47 & 40.9 & \textbf{6.74} & 8.43 & 41.7  \\
 & {\small (.136)} & {\small (.142)} & {\small (1.25)} & {\small (.114)} & {\small (.090)} & {\small (.648)} \\
\hline
\end{tabular}
\end{center}
\caption{Simulation results for model~\eqref{eqn: sim_dgp_change_point_compare} with $\varrho = 0.5$ using methods: AL, AL-DC, and QMLE. The mean and standard deviation (in brackets) of the measures over 500 repetitions are reported.}
\label{tab: sim_change_point_compare_Li2018}
\end{table}

\begin{table}[ht!]
\setlength{\tabcolsep}{8pt}
\small
\begin{center}
\begin{tabular}{l||ccc|ccc}
\hline $\bepsilon_t$ &  \multicolumn{3}{c}{\text{i.i.d.} $\cN(0,1)$} & \multicolumn{3}{|c}{\text{i.i.d.} $t_6$}  \\
\hline & & & & & & \\[-1em] 
\hline & & & & & & \\[-1em] 
{$d=25$} & AL & AL-DC & QMLE & AL & AL-DC & QMLE \\
\hline & & & & & & \\[-1em]
\text{$\wh\bphi$ MSE} & .033 & \textbf{.005} & .007 & .046 & .007 & \textbf{.003} \\
 & {\small (.033)} & {\small (.007)} & {\small (.017)} & {\small (.044)} & {\small (.010)} & {\small (.004)} \\
\text{Estimation Error} & 1.01 & \textbf{.992} & 1.01 & 1.51 & \textbf{1.48} & 1.50 \\
 & {\small (.034)} & {\small (.028)} & {\small (.049)} & {\small (.071)} & {\small (.066)} & {\small (.070)} \\
\text{Run Time (s)} & \textbf{1.42} & 1.81 & 5.97 & \textbf{1.44} & 1.84 & 6.05  \\
 & {\small (.081)} & {\small (.084)} & {\small (.405)} & {\small (.078)} & {\small (.077)} & {\small (.518)} \\
\hline
\hline & & & & & & \\[-1em] 
{$d=50$} & AL & AL-DC & QMLE & AL & AL-DC & QMLE \\
\hline & & & & & & \\[-1em]
\text{$\wh\bphi$ MSE} & .010 & .003 & \textbf{.002} & .014 & .004 & \textbf{.002} \\
 & {\small (.010)} & {\small (.003)} & {\small (.003)} & {\small (.014)} & {\small (.004)} & {\small (.004)} \\
\text{Estimation Error} & .993 & \textbf{.990} & 1.00 & 1.49 & \textbf{1.48} & 1.49  \\
 & {\small (.020)} & {\small (.019)} & {\small (.034)} & {\small (.050)} & {\small (.050)} & {\small (.053)} \\
\text{Run Time (s)} & \textbf{6.73} & 8.41 & 39.9 & \textbf{6.72} & 8.40 & 41.7  \\
 & {\small (.105)} & {\small (.091)} & {\small (1.08)} & {\small (.098)} & {\small (.078)} & {\small (2.34)} \\
\hline
\end{tabular}
\end{center}
\caption{Simulation results for model~\eqref{eqn: sim_dgp_change_point_compare} with $\varrho = 0.1$ using methods: AL, AL-DC, and QMLE. The mean and standard deviation (in brackets) of the measures over 500 repetitions are reported.}
\label{tab: sim_change_point_compare_Li2018_smallchange}
\end{table}

Our adaptive LASSO estimator offers significant computational advantages over QMLE while achieving comparable or superior estimation accuracy. Among all three methods, the adaptive LASSO estimator without using divide-and-conquer has the shortest run time across all settings; however, its performance deteriorates as the signal weakens. Although the QMLE method is naturally expected to perform well in low signal strength settings due to its likelihood-maximizing nature, it requires an optimization step at each candidate location to maximize the log-likelihood, resulting in high computational costs. We employ the BFGS method for optimization in the simulations and note that the statistical performance of the procedure could possibly be improved, albeit at the expense of even higher computational costs. In contrast, our adaptive LASSO estimator with the divide-and-conquer scheme achieves the best balance between computational efficiency and statistical accuracy.

\subsection{A real data example: stock returns}\label{subsec: real_data_NYSE}
In this subsection, we use our proposed model to analyze the spillover effects among the stock returns of the largest $d=50$ firms traded on the New York Stock Exchange (NYSE) in 2024 and compare it with a selection of benchmark models. The data set comprises $251$ timestamps of daily log-returns, and we use the Fama--French three factors\footnote{For the log-return at time $t$, computed as the logarithm of the ratio of prices at times $t$ and $t-1$, its corresponding factors are taken at time $t-1$.}, namely the size factor (\textit{Rm-Rf}), the small-minus-big factor (\textit{SMB}), and the high-minus-low book-to-market equity factor (\textit{HML}), as our explanatory variables; see more details at: \url{https://mba.tuck.dartmouth.edu/pages/faculty/ken.french/data_library.html}. Strictly speaking, the Fama--French factors are not exogenous to individual NYSE stock returns, but the issue of endogeneity is fortunately mild. Thus, we treat the factors as exogenous for the convenience of this analysis, leaving further scope for the possible specification of instrumental variables in practice.

We construct three candidate spatial weight matrices: the inverse distance between firm headquarters ($\W_1$), firms belonging to the same sectors according to the GICS classifications ($\W_2$), and firms belonging to the same sub-industries ($\W_3$). If the sum of any row exceeds one, each entry in that row is divided by the $L_1$ norm of the row. For a thorough comparison, we examine both in-sample and out-of-sample performance for each model. Specifically, the first $100$ timestamps are used for the in-sample study, while a rolling window of $T=50$ on the remaining data is used for the out-of-sample study. We consider five measures of performance: (1) in-sample error; (2) in-sample run time; (3) out-of-sample one-step ahead prediction error; (4) out-of-sample two-step ahead prediction error; and (5) annualized Sharpe ratio based on one-step ahead prediction\footnote{For the annualized Sharpe ratio: at each timestamp, the portfolio position is set directly as the predicted return based on the one-step prediction; the portfolio profit is computed as the average of each predicted stock return multiplied by its actual return; the annualized Sharpe ratio is then calculated as $\sqrt{255}$ times the ratio of the sample mean to the sample standard deviation of the portfolio profit.}.

Inspired by the key feature of volatility clustering in financial data \citep{Chaietal2020}, we apply our framework in a threshold setting with the threshold variable $z_t$ computed as the average of squared log-returns over $[t-\tau,t-1]$. This can be seen as a sparse threshold autoregressive panel data model. We focus on the case with $\tau=5$, as the model interpretations remain similar for other choices of $\tau$. Moreover, for clearer interpretation, we use only the median of the threshold variable, $0.000197$, as the candidate threshold value. For our in-sample study, the model, denoted as DSAR-1, can be written as
\begin{equation*}
\textit{log-return}_t = \bmu^\ast + \sum_{j=1}^{3} \Big( \phi_{j,0}^\ast + \phi_{j,1}^\ast \b{1}\{ z_t \geq 0.000197 \} \Big) \W_j \, \textit{log-return}_t
+ (\textit{Rm-Rf}_t, \textit{SMB}_t, \textit{HML}_t) \, \bbeta^\ast + \bepsilon_t .
\end{equation*}
We also consider a simple variant, DSAR-2, in which $\W_2$ is the only spatial weight matrix in the model, and  the $0.2$, $0.4$, $0.6$, and $0.8$ empirical quantiles are used to construct the dynamic variables. Specifically, the spatial correlation for $\W_2$ is 
\[
\phi_{2,0}^\ast + \phi_{2,1}^\ast \b{1}\{ z_t \geq 0.000166 \} + \phi_{2,2}^\ast \b{1}\{ z_t \geq 0.000186 \} + \phi_{2,3}^\ast \b{1}\{ z_t \geq 0.000213 \} + \phi_{2,4}^\ast \b{1}\{ z_t \geq 0.000236 \}.
\]

The estimated parameters for our two models DSAR-1 and DSAR-2 are presented in Table~\ref{tab: realdata_NYSE_param}. The nonzero coefficients suggest the varying spillover effects in response to recent overall market fluctuations. For DSAR-1, it is also worth noting that both coefficients $\wh\phi_{1,0}$ and  $\wh\phi_{1,1}$ for $\W_1$ are nonzero. While The co-movement of stock returns based on headquarter distances might seem unconventional, this finding is consistent with \cite{PirinskyWang2006}, who attribute such a phenomenon to the distinct trading behaviors of local residents. The estimated coefficients in DSAR-2 reveal three phases of sector spillover effect, which can be interpreted as an increasing co-movement of equity prices within the same sector as the market becomes more active.

\begin{table}[ht!]
\setlength{\tabcolsep}{4pt}
\small
\begin{center}
\begin{tabular}{l|cccccc|ccc|c}
\hline & & & & & & & & & & \\[-1em] 
 & $\wh\phi_{1,0}$ & $\wh\phi_{1,1}$ & $\wh\phi_{2,0}$ & $\wh\phi_{2,1}$ & $\wh\phi_{3,0}$ & $\wh\phi_{3,1}$ & $\wh\beta_{\textit{Rm-Rf}}$ & $\wh\beta_{\textit{SMB}}$ & $\wh\beta_{\textit{HML}}$ & BIC  \\
\hline & & & & & & & & & & \\[-1em]
\text{DSAR-1} & .015 & -1.12 & -.025 & .000 & .000 & .118 & .001 & -.000 & .001 & -3.77  \\
& {\small (.000)} & {\small (.000)}  & {\small (.000)} &  &  & {\small (.000)} &  &  & &   \\
\hline & & & & & & & & & & \\[-1em] 
\hline & & & & & & & & & & \\[-1em] 
& $\wh\phi_{2,0}$ & $\wh\phi_{2,1}$ & $\wh\phi_{2,2}$ & $\wh\phi_{2,3}$ & $\wh\phi_{2,4}$ &  & $\wh\beta_{\textit{Rm-Rf}}$ & $\wh\beta_{\textit{SMB}}$ & $\wh\beta_{\textit{HML}}$ & BIC  \\
\hline & & & & & & & & & & \\[-1em]
\text{DSAR-2} & .000 & .024 & .000 & .116 & .000 & & .000 & -.000 & .001 & -4.63  \\
& & {\small (.000)} & & {\small (.000)} & & &  & & &   \\
\hline
\end{tabular}
\end{center}
\caption{Estimated coefficients for models DSAR-1 and DSAR-2, with standard errors (in brackets) computed as described in the final part of Section~\ref{sec: practical_implementation}. $\wh\beta_{\textit{Rm-Rf}}$, $\wh\beta_{\textit{SMB}}$ and $\wh\beta_{\textit{HML}}$ denote the estimates of $\bbeta^\ast$ corresponding to $\textit{Rm-Rf}_t$, $\textit{SMB}_t$ and $\textit{HML}_t$, respectively.}
\label{tab: realdata_NYSE_param}
\end{table}

We compare our models, DSAR-1 and DSAR-2, with several existing ones: \cite{LamSouza2020}, which also allows for several spatial weight candidates to be specified but consider only time-invariant weights; \cite{Li2018}, which utilizes QMLE within a single change point framework searching potential change locations among $\{0.2T, 0.4T, 0.6T, 0.8T\}$, with each of the three spatial weight matrices experimented on separately; \cite{LiLin2024}, adapted to panel data, where a threshold spatial autoregressive model is estimated using QMLE, with the aforementioned $z_t$ as the threshold variable and candidates for the threshold value taken from the $\{0.2, 0.4, 0.6, 0.8\}$ empirical quantiles of $\{z_t\}$; and \cite{Liangetal2022} where a time-varying regression coefficient and semiparametric spatial weights are considered, estimated by the method described therein, again experimenting with each of the three spatial weight matrices respectively. We present the performance results of all the aforementioned models in Table~\ref{tab: real_data_NYSE}. 

\begin{table}[ht!]
\setlength{\tabcolsep}{3pt}
\small
\begin{center}
\begin{tabular}{l||ccccc}
\hline
Model & In-sample Error$^*$ & Run Time (s) & One-step Error$^*$ & Two-step Error$^*$ & Sharpe Ratio \\
\hline & & & & & \\[-1em]
\text{DSAR-1} & 4.15 & 3.75 & \textbf{2.47} & \textbf{2.42} & .545 \\
\text{DSAR-2} & 1.87 & \textbf{3.24} & 6.68 & 4.68 & \textbf{.695} \\
\text{Lam \& Souza} & 1.94 & 1,722 & 3.11 & 3.30 & -.654 \\
\text{Li-$\W_1$} & 154 & 20.3 & 603 & 1,630 & -3.25 \\
\text{Li-$\W_3$} & 2.30 & 18.3 & 9,407 & 23,134 & -1.57 \\
\text{Li \& Lin-$\W_1$} & 1,407 & 21.9 & 712 & 1,447 & -1.48 \\
\text{Li \& Lin-$\W_3$} & 401 & 21.4 & 24,561 & 793,957 & -1.60 \\
\text{Liang et al.-$\W_1$} & 1.74 & 2,087 & 2.92 & 2.78 & -1.74 \\
\text{Liang et al.-$\W_2$} & \textbf{1.66} & 1,968 & 2.97 & 2.82 & -.830\\
\text{Liang et al.-$\W_3$} & 1.76 & 1,502 & 3.00 & 2.84 & -.747 \\
\hline
\multicolumn{6}{r}{\scriptsize $^*$All errors are multiplied by $10^{4}$}
\end{tabular}
\end{center}
\caption{Model performance on the stock return data. Lam \& Souza refers to the estimator in \cite{LamSouza2020}, Li-$\W_j$ denotes the estimator in \cite{Li2018} using $\W_j$; Li \& Lin-$\W_j$ denotes the estimator adapted from \cite{LiLin2024} using the corresponding spatial weight matrix $\W_j$; and Liang et al.-$\W_j$ refers to the estimator in \cite{Liangetal2022}.}
\label{tab: real_data_NYSE}
\end{table}

We first remark that we have excluded both \cite{Li2018} and \cite{LiLin2024} with spatial weight matrix $\W_2$, as their in-sample errors explode. This may be due to (near) non-stationarity, as large errors are also observed in Li-$\W_1$, Li-$\W_3$, Li \& Lin-$\W_1$, and Li \& Lin-$\W_3$, suggesting that more sophisticated spatial weight configurations may be necessary for their models to perform adequately.

Overall, both DSAR-1 and DSAR-2 models perform very well across all five performance measures. In particular, DSAR-1, which uses three different spatial weight matrices, achieves the best predictive power, while DSAR-2 attains smaller in-sample errors. Notably, our DSAR-1 and DSAR-2 models achieve the highest Sharpe ratios among all models. We observe that \cite{LamSouza2020} and \cite{Liangetal2022}, which use a sparse correction matrix and semiparametric spatial weights respectively, can achieve comparable or slightly better performance. However, both approaches come with much higher computational complexity, being over 500 times slower than our model fitting. This again highlights the advantage of our proposed approach: achieving strong statistical performance with low computational cost.

\section{Conclusions}\label{sec: conclusion}

In this work, we advance dynamic spatial modeling by introducing a framework that enables data-driven selection of time-varying spatial structures, addressing a critical gap in the spatial econometrics literature. By integrating adaptive penalization with instrumental variables, our framework and method provides practitioners with a principled approach for resolving uncertainty in the specification of spatial weight matrices while detecting structural breaks. Empirical studies demonstrate that our method achieves strong statistical performance at a substantially lower computational cost compared with existing models.

A promising direction for future research is to extend our framework to incorporate time-varying regression coefficients or even spatial autoregressive terms using lagged panels, which would further enhance flexibility. In applied econometrics, our method can provide new insights into evolving spatial systems, with applications ranging from financial contagion to climate risk propagation.

\bigskip

\bibliographystyle{jasa}
\bibliography{ref}
\newpage

\renewcommand{\thesection}{S\arabic{section}}
\renewcommand{\thefigure}{S\arabic{figure}}
\renewcommand{\thetable}{S\arabic{table}}
\renewcommand{\theequation}{S\arabic{equation}}
\setcounter{section}{0}
\setcounter{figure}{0}
\setcounter{table}{0}
\setcounter{equation}{0}

\section*{\centering Supplement to ``Inference on Dynamic Spatial Autoregressive Models with Change Point Detection''}

In this supplementary material, we provide additional mathematical details for Section~\ref{sec: full_matrix_notations} in the main text, extra simulation results, and additional explanations for the real data analysis. Proofs of the main results stated in the main text are also included, together with auxiliary results and their proofs.

    \section{Additional details for the model}
    
	\subsection{Additional details for Section \ref{sec: full_matrix_notations}}

    Together with (\ref{eqn: spatiallag_rewrite1_augmented}), the least squares problem in (\ref{eqn: phi_ls}) can be described as
    \begin{align}
        \wt{\bphi} &= \argmin_{\bphi}\; \frac{1}{2T}
        \Big\| \B^\top\y - \B^\top\V \bphi - \B^\top\X_{\bbeta(\bphi)} \Vec{\I_d} \Big\|^2 .
        \label{eqn: phi_ls_matrix}
    \end{align}
    With the least squares estimator $\wt{\bphi}$, the problem in (\ref{eqn: phi_ada_lasso}) in matrix notation is
    \begin{align}
        \wh{\bphi} &= \argmin_{\bphi}\; \frac{1}{2T}
        \Big\| \B^\top\y - \B^\top\V \bphi - \B^\top\X_{\bbeta(\bphi)} \Vec{\I_d} \Big\|^2
        + \lambda \u^\top |\bphi| ,
        \label{eqn: phi_ada_lasso_matrix} \\
        & \;\;\;\;\; \text{subj. to} \;\;\;
        \|\bLambda_t \bPhi\|_{\infty} < 1, \;\;\; \text{with} \;\;\;
        |\z_t^\top \bphi| < 1. \notag
    \end{align}
    Note the squared errors in (\ref{eqn: phi_ls_matrix}) and (\ref{eqn: phi_ada_lasso_matrix}) are still implicit in $\bphi$ due to $\bbeta(\bphi)$.
	
	Note from (\ref{eqn: profiled_beta_matrix}) and the definition of $\B$ in (\ref{eqn: def_B}), we have
	\begin{align*}
		&\hspace{12pt}
		\B^\top\Big(\I_d \otimes \Big\{ \Big(\I_T \otimes \Big\{(\y^\nu)^\top\B^\nu (\B^\nu)^\top \X \big[\X^\top \B^\nu (\B^\nu)^\top \X \big]^{-1}\Big\} \Big) (\X_1, \dots, \X_T)^\top \Big\} \Big)\Vec{\I_d} \\
		&=
		T^{-1/2}d^{-a/2} \Big(\I_d \otimes \Big\{ (\B_1 - \bar{\B}, \dots, \B_T - \bar{\B}) (\I_T\otimes \bgamma)\\
		&\hspace{12pt}
		\Big(\I_T \otimes \Big\{ (\y^\nu)^\top\B^\nu (\B^\nu)^\top \X \big[\X^\top \B^\nu (\B^\nu)^\top \X \big]^{-1}\Big\} \Big) (\X_1, \dots, \X_T)^\top \Big\} \Big)\Vec{\I_d} \\
		&=
		T^{-1/2}d^{-a/2} \Big(\I_d \otimes \Big\{ \sum_{t=1}^T (\B_t - \bar{\B}) \bgamma (\y^\nu)^\top\B^\nu (\B^\nu)^\top \X \big[\X^\top \B^\nu (\B^\nu)^\top \X \big]^{-1} \X_t^\top  \Big\} \Big)\Vec{\I_d} \\
		&=
		T^{-1/2}d^{-a/2} \Vec{\sum_{t=1}^T (\B_t - \bar{\B}) \bgamma (\y^\nu)^\top\B^\nu (\B^\nu)^\top \X \big[\X^\top \B^\nu (\B^\nu)^\top \X \big]^{-1} \X_t^\top} \\
		&=
		T^{-1/2}d^{-a/2} \Big(\sum_{t=1}^T \X_t \otimes (\B_t - \bar{\B}) \bgamma \Big) \Vec{(\y^\nu)^\top\B^\nu (\B^\nu)^\top \X \big[\X^\top \B^\nu (\B^\nu)^\top \X\big]^{-1}} \\
		&=
		T^{-1/2}d^{-a/2} \Big(\sum_{t=1}^T \X_t \otimes (\B_t - \bar{\B}) \bgamma \Big) \big[\X^\top \B^\nu (\B^\nu)^\top \X \big]^{-1} \X^\top\B^\nu (\B^\nu)^\top \y^\nu .
	\end{align*}
	Similarly, by the definition of $\Y_W$,
	\begin{align*}
		&\hspace{12pt}
		\B^\top\Big(\I_d \otimes \Big\{ \Big(\I_T \otimes \Big\{ \Big(\sum_{j=1}^p \sum_{k=0}^{l_j} \phi_{j,k} (\y_{j,k}^\nu)^\top (\W_j^\otimes)^\top \Big) \\
		&\hspace{36pt}
		\B^\nu (\B^\nu)^\top \X \big[\X^\top \B^\nu (\B^\nu)^\top \X \big]^{-1} \Big\} \Big) (\X_1, \dots, \X_T)^\top \Big\} \Big)\Vec{\I_d} \\
		&=
		T^{-1/2}d^{-a/2}\Big(\I_d \otimes \Big\{ (\B_1 - \bar{\B}, \dots, \B_T - \bar{\B}) (\I_T \otimes \bgamma ) \\
		&\hspace{12pt}
		\cdot \Big(\I_T \otimes \Big\{ \Big(\sum_{j=1}^p \sum_{k=0}^{l_j} \phi_{j,k} (\y_{j,k}^\nu)^\top (\W_j^\otimes)^\top \Big) \\
        &\hspace{12pt}
        \cdot \B^\nu (\B^\nu)^\top \X \big[\X^\top \B^\nu (\B^\nu)^\top \X \big]^{-1} \Big\} \Big) (\X_1, \dots, \X_T)^\top \Big\} \Big)\Vec{\I_d} \\
		&=
		T^{-1/2}d^{-a/2}\Big(\I_d \otimes \Big\{ \sum_{t=1}^T  (\B_t - \bar{\B}) \bgamma \Big(\sum_{j=1}^p \sum_{k=0}^{l_j} \phi_{j,k} (\y_{j,k}^\nu)^\top (\W_j^\otimes)^\top \Big) \\
		&\hspace{12pt}
		\B^\nu (\B^\nu)^\top \X \big[\X^\top \B^\nu (\B^\nu)^\top \X \big]^{-1} \X_t^\top \Big\} \Big) \Vec{\I_d} \\
		&=
		T^{-1/2}d^{-a/2} \Vec{ \sum_{t=1}^T (\B_t - \bar{\B}) \bgamma (\bphi^\top \Y_W^\top) \B^\nu (\B^\nu)^\top \X \big[\X^\top \B^\nu (\B^\nu)^\top \X \big]^{-1} \X_t^\top } \\
		&=
		T^{-1/2}d^{-a/2} \Big(\sum_{t=1}^T \X_t \otimes (\B_t - \bar{\B}) \bgamma \Big) \big[\X^\top \B^\nu (\B^\nu)^\top \X \big]^{-1} \X^\top \B^\nu (\B^\nu)^\top \Y_W \bphi .
	\end{align*}
	With (\ref{eqn: profiled_beta_matrix}), the term inside the squared Euclidean norm of (\ref{eqn: phi_ls_matrix}) can hence be further written as
	\begin{equation}
		\label{eqn: simplify_LASSO}
		\begin{split}
			&\hspace{12pt}
			\B^\top\y - \B^\top\V \bphi - \B^\top\X_{\bbeta(\bphi)} \Vec{\I_d} \\
			&=
			\B^\top\y - \B^\top\V \bphi - \B^\top\Big(\I_d \otimes \Big\{ \Big(\I_T \otimes \Big\{(\y^\nu)^\top\B^\nu (\B^\nu)^\top \X \big[\X^\top \B^\nu (\B^\nu)^\top \X \big]^{-1}\Big\} \Big) \\
            &\hspace{12pt}
            \cdot \big(\X_1, \dots, \X_T \big)^\top \Big\} \Big)\Vec{\I_d} \\
			&\hspace{12pt}
			+ \B^\top\Big(\I_d \otimes \Big\{ \Big(\I_T \otimes \Big\{ \Big(\sum_{j=1}^p \sum_{k=0}^{l_j} \phi_{j,k} (\y_{j,k}^\nu)^\top (\W_j^\otimes)^\top \Big) \\
			&\hspace{25pt}
			\B^\nu (\B^\nu)^\top \X \big[\X^\top \B^\nu (\B^\nu)^\top \X \big]^{-1} \Big\} \Big) (\X_1, \dots, \X_T)^\top \Big\} \Big)\Vec{\I_d} \\
			&=
			\B^\top\y - T^{-1/2}d^{-a/2} \Big(\sum_{t=1}^T \X_t \otimes (\B_t - \bar{\B}) \bgamma \Big) \big[\X^\top \B^\nu (\B^\nu)^\top \X \big]^{-1} \X^\top\B^\nu (\B^\nu)^\top \y^\nu \\
			&-
			\Big\{ \B^\top\V -
			T^{-1/2}d^{-a/2} \Big(\sum_{t=1}^T \X_t \otimes (\B_t - \bar{\B}) \bgamma \Big) \big[\X^\top \B^\nu (\B^\nu)^\top \X \big]^{-1} \X^\top \B^\nu (\B^\nu)^\top \Y_W \Big\} \bphi \\
			&=
			\B^\top\y - \bXi\y^\nu -(\B^\top\V -\bXi\Y_W ) \bphi.
		\end{split}
	\end{equation}

\subsection{Identification of the model (\ref{eqn: spatiallag_rewrite1_augmented})}
We show that the coefficients $\bphi^\ast$ and $\bbeta^\ast$ in model \eqref{eqn: spatiallag_rewrite1_augmented} are identified under Assumption (I1). Assume that the two sets of parameters $(\check{\bphi}, \check{\bbeta})$ and $(\Acute{\bphi}, \Acute{\bbeta})$ both satisfy model (\ref{eqn: spatiallag_rewrite1_augmented}), then we have
\[
\B^\top\V \check{\bphi} + \B^\top\X_{\check{\bbeta}} \Vec{\I_d} = \B^\top\V \Acute{\bphi} + \B^\top\X_{\Acute{\bbeta}} \Vec{\I_d} .
\]
By noticing that $\B^\top\X_{\bbeta} \Vec{\I_d} = \B^\top\wt{\X} \bbeta$, we may rearrange the above and arrive at
\[
\0 = \B^\top\V (\check{\bphi} -\Acute{\bphi}) +
\B^\top\wt{\X} (\check{\bbeta} - \Acute{\bbeta}) =
\begin{pmatrix}
\B^\top\V & \B^\top\wt{\X}
\end{pmatrix}
\begin{pmatrix}
\check{\bphi} -\Acute{\bphi} \\
\check{\bbeta} - \Acute{\bbeta}
\end{pmatrix}.
\]
Taking expectation and left-multiplying by $(\Q^\top\Q)^{-1}\Q^\top$ on both sides, we obtain $\check{\bphi} = \Acute{\bphi}$ and $\check{\bbeta} = \Acute{\bbeta}$. This implies $\bphi^\ast$ and $\bbeta^\ast$ are identified in model (\ref{eqn: spatiallag}), and hence $\bmu^\ast$ is uniquely identified.

It is worth to remind that in practice, we should be careful in constructing dynamic variables to due to the issue of multicollinearity, as reflected in $\B^\top \V$ above.

\subsection{Additional details to Section~\ref{subsec: example_threshold}}
In this subsection, we discuss the model framework with regime changes based on multiple threshold variables with multiple threshold values. For illustrations, consider the following spatial autoregressive model with $(k+1)$ regimes, where $k$ can be unknown:
\begin{equation}
\label{eqn: threshold_spatial_coef_multiple}
\y_t =
\bmu^\ast + \Big(\phi_1^\ast \b{1}\{q_t \leq \gamma_1^\ast\} + \phi_2^\ast \b{1}\{\gamma_1^\ast < q_t \leq \gamma_2^\ast\} + \dots + \phi_k^\ast \b{1}\{q_t > \gamma_k^\ast\} \Big) \W_1 \y_t +\X_t\bbeta^\ast + \bepsilon_t .
\end{equation}
Model \eqref{eqn: threshold_spatial_coef_multiple} can be written in the form of \eqref{eqn: threshold_spatial_coef_rewrite_saturated}, with the consistency of $\{\phi_{1,l}^\ast \}_{l\in[L]}$ guaranteed by Corollary~\ref{corollary: threshold_consistency_multiple}. This also implies that the estimation on $k$ is consistent.
\begin{corollary}
\label{corollary: threshold_consistency_multiple}
(Consistency on the number of threshold regimes and multiple threshold values estimation)
Let all assumptions in Theorem \ref{thm: ada_LASSO_phi_asymp} hold. For model \eqref{eqn: threshold_spatial_coef_multiple}, let $\{ \wh\phi_{1,l} \}_{l\in[L]}$ denote the adaptive LASSO solution for $\{ \phi_{1,l}^\ast \}_{l\in[L]}$ in \eqref{eqn: threshold_spatial_coef_rewrite_saturated}. Then, $\wh{k} := |\{\gamma_l : \wh\phi_{1,l} \neq 0\}|$ estimates $k$ consistently. Moreover, for every $i\in[\wh{k}]$, the $i$-th smallest element in $\{\gamma_l : \wh\phi_{1,l} \neq 0\}$ estimates $\gamma_i^\ast$ consistently.
\end{corollary}

As we often have limited prior knowledge on the parameter space in practice, we recommend using our framework in an exploratory way. This should help researchers discover more reasonable threshold structures in the data.

\section{Additional simulations}

\subsection{Different methods to select the tuning parameter}

In this subsection, we conduct Monte Carlo experiments to study the sensitivity of the estimator's performance with respect to different methods for selecting the tuning parameter $\lambda$ in the adaptive LASSO problem. Specifically, we compare our BIC criterion in \eqref{eqn: BIC_lambda} with two alternative choices: (i) three-fold cross validation minimizing the least squares error, which effectively uses the first term in \eqref{eqn: BIC_lambda} as the criterion (five-fold and ten-fold cross validation are also experimented, yielding similar results and thus omitted here); (ii) the BIC criterion in \eqref{eqn: BIC_lambda_null}, with variables constructed under the model framework \eqref{eqn: spatiallag}, which is a reasonable choice when $L$ is small.

We use a similar data generating process to~\eqref{eqn: sim_dgp_general}:
\begin{equation}
\label{eqn: supp_spec_sens_data}
\y_t = \Big\{\I_d- \big(0 +0.2\, z_{1,1,t} +0\, z_{1,2,t}\big) \W_1 - \big(0 +0.2\, z_{2,1,t} +0\, z_{2,2,t} \big) \W_2 \Big\}^{-1} \big(\bmu^\ast + \X_t\bbeta^\ast + \bepsilon_t \big),
\end{equation}
where $\{z_{1,1,t}\}$, $\{z_{1,2,t}\}$, $\{z_{2,1,t}\}$, $\{z_{2,2,t}\}$, $\bmu^\ast$, $\X_t$, $\bbeta^\ast$, $\W_1$, $\W_2$, and $\bepsilon_t$ are generated in the same way as described after~\eqref{eqn: sim_dgp_general} in the main text, and $z_{1,0,t}=z_{2,0,t}=1$ for all $t\in[T]$. We first fit the model using these dynamic variables, i.e., $L=6$. For a more thorough comparison between the two criteria in \eqref{eqn: BIC_lambda} and \eqref{eqn: BIC_lambda_null}, we also perform estimation using additional dynamic variables $\{z_{1,3,t}\}, \dots, \{z_{1,9,t}\}, \{z_{2,3,t}\}, \dots, \{z_{2,9,t}\}$, all generated independently from $\cN(0,1)$. In this case, we have that $L=20$ and that the true parameters corresponding to these extra  dynamic variables are all zero. For both scenarios, we experiment with $T=50$ and $d=25,50$, respectively, and each setting is repeated 500 times. Table~\ref{tab: sim_general_setting_tuning_param_L6} reports the error measures of the parameter estimators for all settings under $L=6$, and Table~\ref{tab: sim_general_setting_tuning_param_L20} for $L=20$.

\begin{table}[htp!]
\setlength{\tabcolsep}{8pt}
\begin{center}
\begin{tabular}{l||ccc|ccc}
\hline &  & $d=25$ &  &  & $d=50$ &  \\ \cline{2-7}
& CV & BIC & BIC2 & CV & BIC & BIC2 \\
\hline & & & & & & \\[-1em]
$\wh\bphi$ MSE & .001 & .000 & .000 & .001 & .000 & .000 \\
 & {\small (.001)} & {\small (.001)} & {\small (.000)} & {\small (.001)} & {\small (.000)} & {\small (.000)} \\
 $\wh\bbeta$ MSE & .001 & .001 & .001 & .001 & .001 & .001  \\
 & {\small (.000)} & {\small (.000)} & {\small (.000)} & {\small (.000)} & {\small (.000)} & {\small (.000)}  \\
 $\wh\bmu$ MSE & .021 & .021 & .021 & .022 & .021 & .021  \\
 & {\small (.005)} & {\small (.004)} & {\small (.004)} & {\small (.005)} & {\small (.004)} & {\small (.004)}  \\
 $\wh\bphi$ Specificity & .463 & .743 & .743 & .467 & .745 & .745 \\
 & {\small (.326)} & {\small(.326)} & {\small (.326)} & {\small (.328)} & {\small (.325)} & {\small (.325)}  \\
 $\wh\bphi$ Sensitivity & 1.000 & .999 & 1.000 & 1.000 & 1.000 & 1.000  \\
 & {\small (.000)} & {\small (.022)} & {\small (.000)} & {\small (.000)} & {\small (.000)} & {\small (.000)}  \\
\hline
\end{tabular}
\end{center}
\caption{Simulation results for different methods of selecting the tuning parameter. The data generating process follows~\eqref{eqn: supp_spec_sens_data} with $T=50$, and estimation is performed using $L=6$ dynamic variables. The columns labeled ``CV'', ``BIC'', and ``BIC2'' correspond to selection based on least squares residuals, \eqref{eqn: BIC_lambda_null}, and \eqref{eqn: BIC_lambda}, respectively. The mean and standard deviation (in brackets) of the corresponding error measures over 500 repetitions are reported.}
\label{tab: sim_general_setting_tuning_param_L6}
\end{table}

\begin{table}[htp!]
\setlength{\tabcolsep}{8pt}
\begin{center}
\begin{tabular}{l||ccc|ccc}
\hline &  & $d=25$ &  &  & $d=50$ &  \\ \cline{2-7}
& CV & BIC & BIC2 & CV & BIC & BIC2 \\
\hline & & & & & & \\[-1em]
$\wh\bphi$ MSE & .002 & .001 & .001 & .006 & .001 & .001 \\
 & {\small (.001)} & {\small (.001)} & {\small (.001)} & {\small (.004)} & {\small (.002)} & {\small (.002)} \\
 $\wh\bbeta$ MSE & .001 & .001 & .001 & .001 & .001 & .001  \\
 & {\small (.001)} & {\small (.001)} & {\small (.001)} & {\small (.000)} & {\small (.000)} & {\small (.000)}  \\
 $\wh\bmu$ MSE & .023 & .021 & .021 & .021 & .020 & .020  \\
 & {\small (.007)} & {\small (.006)} & {\small (.006)} & {\small (.005)} & {\small (.004)} & {\small (.004)}  \\
 $\wh\bphi$ Specificity & .489 & .886 & .886 & .495 & .906 & .906  \\
 & {\small (.196)} & {\small(.142)} & {\small (.142)} & {\small (.196)} & {\small (.114)} & {\small (.114)}  \\
 $\wh\bphi$ Sensitivity & .996 & .959 & .960 & 1.000 & 1.000 & 1.000  \\
 & {\small (.045)} & {\small (.137)} & {\small (.136)} & {\small (.000)} & {\small (.000)} & {\small (.000)}  \\
\hline
\end{tabular}
\end{center}
\caption{Simulation results for the same data generating process as in Table~\ref{tab: sim_general_setting_tuning_param_L6}, with model fitting performed using $L=20$ dynamic variables.}
\label{tab: sim_general_setting_tuning_param_L20}
\end{table}

Both tables show that selecting $\lambda$ using either of the two BIC criteria greatly outperforms selection based on the least squares criterion (``CV'') in terms of specificity, while all three methods achieve high sensitivity. This is reasonable, as the BIC criteria are specifically designed to address model overfitting and are thus less likely to select truly zero parameters. We also observe that the results under the two BIC criteria are very similar, which is expected since even $L = 20$ in Table~\ref{tab: sim_general_setting_tuning_param_L20} is still considered small. The improvement of our proposed criterion ``BIC2'' over ``BIC'' is mild, but both tables demonstrate the strength of our approach and suggest that ``BIC'' may also be used effectively when $L$ is not large.

\subsection{Sensitivity of parameter estimators to endogeneity}

\begin{figure}[htp!]
  \centering
  \includegraphics[width=1\textwidth]{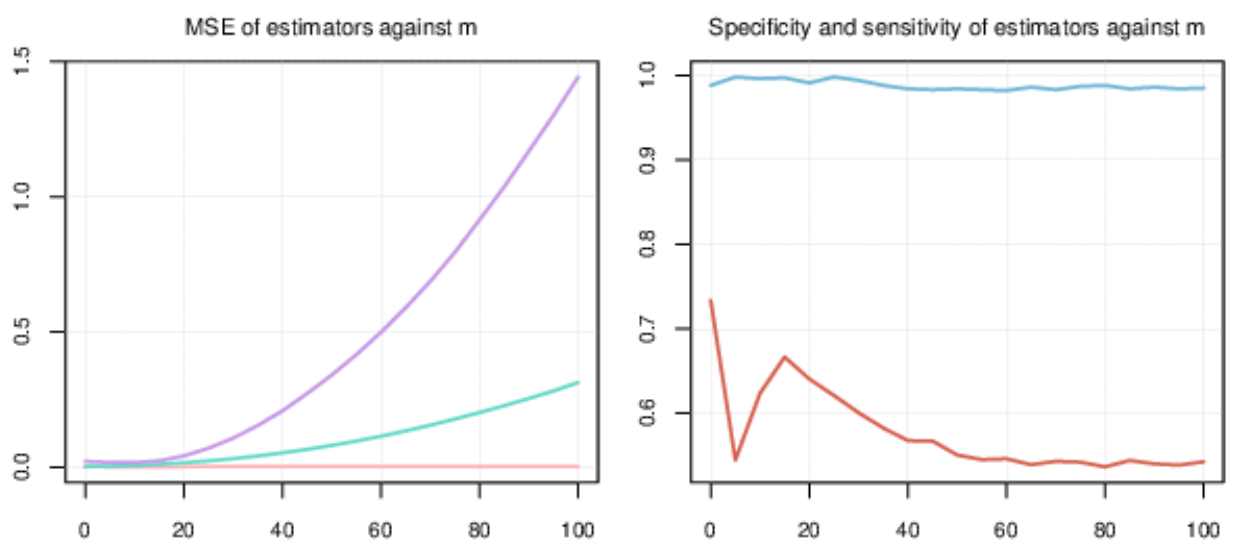}
  \caption{Estimator performance for model~\eqref{eqn: sim_dgp_endo} with $m$ varying from $0$ to $100$, under the setting $(T,d)=(50,25)$. The left panel shows the MSE for $\wh\bphi$ (pink), $\wh\bbeta$ (teal), and $\wh\bmu$ (purple); the right panel shows the specificity (red) and sensitivity (blue) of $\wh\bphi$. Each value is averaged over 500 repetitions.}
  \label{fig: simulation_endo_50_25}

  \vspace*{\floatsep}

  \includegraphics[width=1\textwidth]{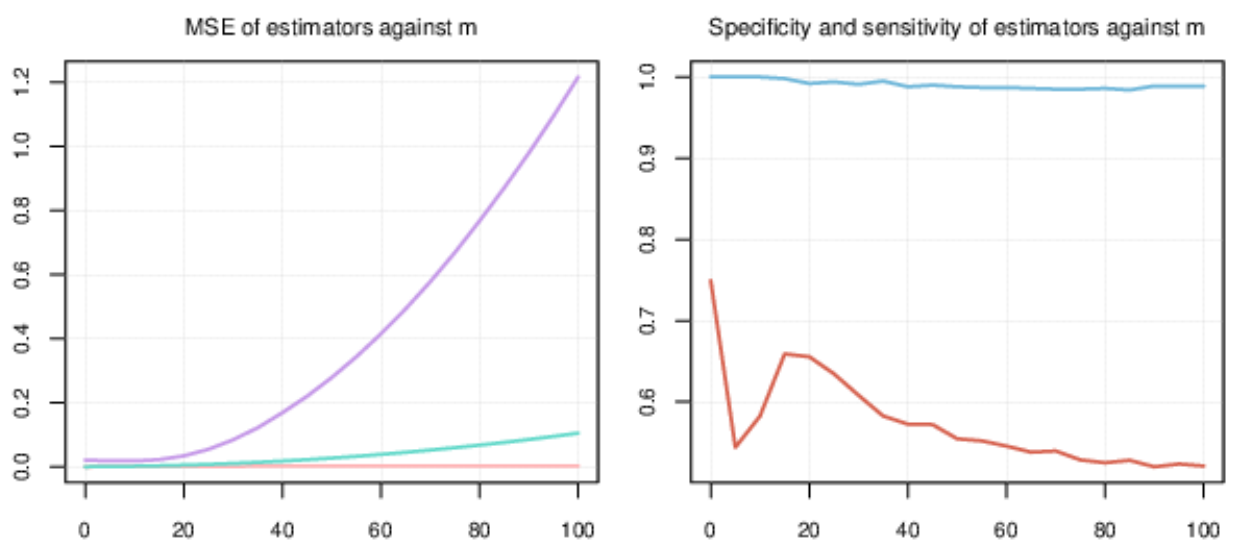}
  \caption{Estimator performance for model \eqref{eqn: sim_dgp_endo} with $m$ varying from $0$ to $100$.}
  \label{fig: simulation_endo_50_50}
\end{figure}

In the main text, we demonstrated the effectiveness of our parameter estimators under mild endogeneity. To further investigate the impact of endogeneity, we consider the model
\begin{equation}
\label{eqn: sim_dgp_endo}
\y_t = \Big\{\I_d- \big(0 +0.2\, z_{1,1,t} +0\, z_{1,2,t}\big) \W_1 - \big(0 +0.2\, z_{2,1,t} +0\, z_{2,2,t} \big) \W_2 \Big\}^{-1} \big(\bmu^\ast + \X_t\bbeta^\ast + \bepsilon_t \big) ,
\end{equation}
where $\{z_{1,1,t}\}$, $\{z_{1,2,t}\}$, $\{z_{2,1,t}\}$, $\{z_{2,2,t}\}$, $\bmu^\ast$, $\bbeta^\ast$, $\W_1$, $\W_2$, and $\bepsilon_t$ are again generated in the same way as described after~\eqref{eqn: sim_dgp_general} in the main text. The covariate $\X_t$ also has three features, with each entry independently drawn from the standard normal distribution, except that the third feature is made endogenous by adding $m\; \bepsilon_t$. Thus, the parameter $m$ controls the level of endogeneity, and we experiment with values of $m$ ranging from $0$ to $100$. The results for the dimension settings $(T,d) = (50,25)$ and $(50,50)$ are summarized in Figure~\ref{fig: simulation_endo_50_25} and Figure~\ref{fig: simulation_endo_50_50}, respectively.

The performance of all estimators improves slightly as the spatial dimension increases, although the general patterns remain consistent across both settings. From the left panels in both figures, we observe that all MSE's worsen as the covariates become more endogenous. Notably, the performance of the estimator $\wh\bmu$ is most affected by endogeneity, as the estimation errors in $\wh\bphi$ and $\wh\bbeta$ propagate into $\wh\bmu$. The right panels of both figures show that while the sensitivity of $\wh\bphi$ remains consistently high, its specificity gradually declines as endogeneity increases. Interestingly, the specificity curve exhibits a significant dip when $m$ is relatively small, a phenomenon that warrants further investigation in future work.

\subsection{Experiments on the threshold spatial autoregressive models}
We present numerical results for the threshold spatial autoregressive model introduced in Section~\ref{subsec: example_threshold}. Specifically, we consider the data generating process
\begin{equation}
\label{eqn: sim_dgp_threshold}
\y_t = \big(\I_d- 0.3\, z_{1,t} \W_1 - 0.8\, z_{2,t} \W_2 \big)^{-1} \big(\bmu^\ast + \X_t\bbeta^\ast + \bepsilon_t \big) ,
\end{equation}
where $\W_1$, $\W_2$, $\bmu^\ast$, $\X_t$, $\bbeta^\ast$ and $\bepsilon_t$ are generated in the same way as in \eqref{eqn: sim_dgp_general}. The indicators $z_{1,t}= \b{1}\{q_t\leq \gamma^\ast\}$ and $z_{2,t}= \b{1}\{q_t> \gamma^\ast\}$ are constructed using a threshold variable $q_t$ and threshold value $\gamma^\ast$. Thus, \eqref{eqn: sim_dgp_threshold} defines a threshold spatial autoregressive model that has changes in both the coefficients and the spatial weight matrices. To simultaneously estimate the parameters and the threshold value, we fit a model of the form
\begin{align}
    & \y_t = \Big(\I_d- \sum_{l=1}^{19} z_{1,l,t} \phi_{1,l}^\ast \W_1 - \sum_{l=1}^{19} z_{2,l,t} \phi_{2,l}^\ast \W_2 \Big)^{-1} \big(\bmu^\ast + \X_t\bbeta^\ast + \bepsilon_t \big) ,
    \;\;\; \text{where}
    \label{eqn: sim_dgp_threshold_saturated} \\
    & z_{1,l,t} = \b{1}\{q_t\leq \wh{\gamma}_l\},
    \;\;\; z_{2,l,t} = \b{1}\{q_t> \wh{\gamma}_l\},
    \;\;\;
    \wh{\gamma}_l = \text{$(5\%\cdot l)$-th empirical quantile of $\{q_t\}_{t\in[T]}$.} \notag
\end{align}
As discussed in Section \ref{subsec: example_threshold}, we expect $\phi_{1,1}^\ast, \dots, \phi_{1,19}^\ast$ to be all zero, except for the one corresponding to $z_{1,l,t}$ where $\wh{\gamma}_l$ is closest to the true threshold $\gamma^\ast$. Let $l^\ast$ denote the index of this $\wh{\gamma}_l$. Similarly, among $\phi_{2,1}^\ast, \dots, \phi_{2,19}^\ast$, all entries should be zero except for the one corresponding to $z_{2,l^\ast,t}$. Moreover, we expect $(\phi_{1, l^\ast}^\ast, \phi_{2, l^\ast}^\ast)\approx (0.3, 0.8)$.

We consider two types of threshold variables in our experiments:
\begin{enumerate}
    \item \textit{(AR(5))} $q_t$ follows an AR(5) process with i.i.d.\ $\cN(0,1)$ innovations, with $\gamma^\ast = 0.3$;
    \item \textit{(Self-exciting on mean)} $q_t = \1^\top \y_{t-1} / d$, i.e., the regime switches are driven in a self-exciting manner based on the mean of the previous observation, with $\gamma^\ast = 1.5$.
\end{enumerate}
Results for $d=50, 75$ and $T=100, 150$ are reported in Table \ref{tab: sim_threshold_setting}, where the estimators $\wh\bphi$ and $\wh{\gamma}_l$ are obtained from fitting the model \eqref{eqn: sim_dgp_threshold_saturated}. The evaluation metrics are defined as follows:
\begin{align*}
    & \text{$\wh\bphi$ MSE} := \text{Mean squared error of $\wh\bphi$ with $\bphi$ all zero except for $(\phi_{1, l^\ast}^\ast, \phi_{2, l^\ast}^\ast)=(0.3, 0.8)$} , \\
    & \text{$\wh\bphi$ True-Unique} := \b{1}\{\text{$\wh\phi_{1, l^\ast}$ and $\wh\phi_{2, l^\ast}$ are uniquely identified as the only nonzero in $\wh\bphi$}\} ,\\
    & \text{$\wh{\gamma}_l$ MSE} := \text{Mean squared error of $\wh{\gamma}_l$ relative to the true threshold value $\gamma^\ast$}.
\end{align*}
The metric $\wh\bphi$ True-Unique is crucial for demonstrating the validity of our algorithm, as it reflects both specificity and sensitivity. In addition, it assesses whether the estimated threshold value is unique. When computing the MSE of $\wh{\gamma}_l$ in cases where multiple indices $l$ are identified with different $\wh{\gamma}_l$ values, we select the index $l$ corresponding to the largest $\wh\phi_{1, l^\ast}$. 

\begin{table}[ht!]
\setlength{\tabcolsep}{4pt}
\footnotesize
\begin{center}
\begin{tabular}{l||cc|cc|cc|cc}
\hline $q_t$ setting &  \multicolumn{4}{c}{AR(5)} & \multicolumn{4}{|c}{Self-exciting on mean}  \\
\hline & & & & & & & & \\[-1em] 
{$(T,d)$} & $(100,50)$ & $(100,75)$ & $(150,50)$ & $(150,75)$ & $(100,50)$ & $(100,75)$ & $(150,50)$ & $(150,75)$ \\
\hline & & & & & & & & \\[-1em]
\text{$\wh\bphi$ MSE} & .002 & .003 & .017 & .020 & .009 & .007 & .003 & .001 \\
 & {\small (.002)} & {\small (.003)} & {\small (.013)} & {\small (.011)} & {\small (.011)} & {\small (.010)} & {\small (.006)} & {\small (.003)} \\
 \text{$\wh\bphi$ True-Unique} & .562 & .439 & .706 & .844  & .537 & .621 & .797 & .938  \\
 & {\small (.497)} & {\small (.500)} & {\small (.456)} & {\small (.363)} & {\small (.499)} & {\small (.485)} & {\small (.403)} & {\small (.242)} \\
 \text{$\wh{\gamma}_l$ MSE} & .343 & .066 & .207 & .025 & .128 & .031 & .080 & .005  \\
 & {\small (.870)} & {\small (.309)} & {\small (.707)} & {\small (.193)} & {\small (.881)} & {\small (.126)} & {\small (.802)} & {\small (.023)} \\
\hline
\end{tabular}
\end{center}
\caption{Simulation results for fitting the threshold model~\eqref{eqn: sim_dgp_threshold_saturated}. The mean and standard deviation (in brackets) of the corresponding error measures over 500 repetitions are reported.}
\label{tab: sim_threshold_setting}
\end{table}

Table \ref{tab: sim_threshold_setting} confirms that our procedure can efficiently estimate both the threshold value simultaneously in one go. Although the estimator of $\gamma^\ast$ is coarse, accurate only up to the $5\%$ empirical quantile of the threshold variable, the results show that increasing the data dimensions improves the performance of $\wh{\gamma}_l$. In practice, re-estimation using a finer grid based on such initial threshold estimate could  further enhance accuracy.

\subsubsection*{Comparison with \cite{LiLin2024}}

We also compare our method with the quasi-maximum likelihood estimator (QMLE) for the threshold spatial autoregressive model (TSAR) in panel data, which is a variant of the TSAR model in \cite{LiLin2024} adapted to the time series setup. We consider the model
\begin{equation}
\label{eqn: sim_dgp_threshold_LiLin2024}
\y_t = \left\{
    \begin{array}{ll}
	\bmu^\ast + (\rho^\ast +\varrho^\ast) \W_1 \y_t + \X_t\bbeta^\ast + \bepsilon_t, & \hbox{$q_t \leq \gamma^\ast$;} \\
	\bmu^\ast + \rho^\ast \W_1 \y_t +\X_t\bbeta^\ast + \bepsilon_t, & \hbox{$q_t > \gamma^\ast$.}
    \end{array}
    \right. 
\end{equation}
This model represents a change in regime for the spatial correlation coefficients, from $(\rho^\ast +\varrho^\ast)$ to $\rho^\ast$, controlled by the threshold variable $\{q_t\}$. This setup mirrors the one for change point models in Section~\ref{subsec: sim_change_point} of the main paper. Specifically, we generate $\bmu^\ast$, $\W_1$, $\bmu^\ast$, $\X_t$, $\bbeta^\ast$ and $\bepsilon_t$ as described previously in \eqref{eqn: sim_dgp_threshold}, and also generate $q_t$ using both an AR(5) process and in a self-exciting manner, respectively. We fix $T=100$, $\gamma^\ast=0.05$, and $(\rho^\ast, \varrho^\ast) = (-0.3, 0.3)$, and experiment with three different methods, as described below:
\begin{itemize}
    \item [1.] AL: The adaptive LASSO estimator with dynamic variables $\{z_{t,0,1}, \dots, z_{t,9,1}\}$, where $z_{t,0,1}=1$, $z_{t,k,1}=\b{1}\{q_t\leq \wh\gamma_k\}$ for $k\in[9]$, $t\in[100]$, and $\wh\gamma_k$ is the $(10\% \cdot k)$-th empirical quantile of $\{q_t\}_{t \in [100]}$.
    \item [2.] AL-DC: The adaptive LASSO estimator using the divide-and-conquer scheme in Remark~\ref{remark: sequential_procedure}, with three subsets $\{z_{t,0,1}, z_{t,1,1}, z_{t,2,1}, z_{t,3,1}\}$, $\{z_{t,0,1}, z_{t,4,1}, z_{t,5,1}, z_{t,6,1}\}$, and $\{z_{t,0,1}, z_{t,7,1}, z_{t,8,1}, z_{t,9,1}\}$, where all dynamic variables are the same as in method 1 above. The final estimator is selected from the three results with the smallest estimation error.
    \item [3.] QMLE \citep{LiLin2024}: The estimation is performed for each $\{\wh\gamma_k\}_{k\in[9]}$, where $\wh\gamma_k$ is the same as in AL. The optimization step is carried out using the Nelder–Mead algorithm, and the final estimator is selected as the one with the smallest estimation error among them.
\end{itemize}

Table~\ref{tab: sim_change_point_compare_LiLin2024} reports the results for $T=100$ and $d=25,50$, where all simulations are repeated 500 times.

\begin{table}[ht!]
\setlength{\tabcolsep}{8pt}
\small
\begin{center}
\begin{tabular}{l||ccc|ccc}
\hline $\bepsilon_t$ &  \multicolumn{3}{c}{AR(5)} & \multicolumn{3}{|c}{Self-exciting on mean}  \\
\hline & & & & & & \\[-1em] 
\hline & & & & & & \\[-1em] 
{$d=25$} & AL & AL-DC & QMLE & AL & AL-DC & QMLE \\
\hline & & & & & & \\[-1em]
\text{$\wh\bphi$ MSE} & .036 & \textbf{.011} & .027 & .037 & \textbf{.014} & .030 \\
 & {\small (.041)} & {\small (.015)} & {\small (.011)} & {\small (.038)} & {\small (.016)} & {\small (.089)} \\
\text{Estimation Error} & 1.00 & \textbf{.994} & 1.05 & 1.00 & \textbf{.994} & 1.21 \\
 & {\small (.031)} & {\small (.028)} & {\small (.040)} & {\small (.031)} & {\small (.028)} & {\small (.492)} \\
\text{Run Time (s)} & \textbf{1.57} & 2.00 & 15.1 & \textbf{1.40} & 1.78 & 12.2  \\
 & {\small (.271)} & {\small (.330)} & {\small (3.85)} & {\small (.088)} & {\small (.080)} & {\small (.240)} \\
\hline
\hline & & & & & & \\[-1em] 
{$d=50$} & AL & AL-DC & QMLE & AL & AL-DC & QMLE \\
\hline & & & & & & \\[-1em]
\text{$\wh\bphi$ MSE} & .012 & \textbf{.004} & .030 & .013 & \textbf{.006} & .321 \\
 & {\small (.011)} & {\small (.006)} & {\small (.011)} & {\small (.011)} & {\small (.008)} & {\small (.279)} \\
\text{Estimation Error} & .988 & \textbf{.986} & 1.04 & .992 & \textbf{.991} & 2.91 \\
 & {\small (.020)} & {\small (.020)} & {\small (.031)} & {\small (.019)} & {\small (.019)} & {\small (1.58)} \\
\text{Run Time (s)} & \textbf{6.60} & 8.25 & 44.3 & \textbf{6.52} & 8.16 & 44.1  \\
 & {\small (.109)} & {\small (.090)} & {\small (.681)} & {\small (.103)} & {\small (.095)} & {\small (.861)} \\
\hline
\end{tabular}
\end{center}
\caption{Simulation results for the model~\eqref{eqn: sim_dgp_threshold_LiLin2024} with $T=100$ using three methods: AL, AL-DC, and QMLE. The mean and standard deviation (in brackets) of the measures over 500 repetitions are reported.}
\label{tab: sim_change_point_compare_LiLin2024}
\end{table}

From this table, we observe that our adaptive LASSO estimator with the divide-and-conquer scheme achieves the best estimation accuracy while having high computational efficiency. We thus reach the same conclusions as in Section~\ref{subsec: sim_change_point} for Tables~\ref{tab: sim_change_point_compare_Li2018} and \ref{tab: sim_change_point_compare_Li2018_smallchange}.

\subsection{Experiments on the divide-and-conquer scheme in Remark~\ref{remark: sequential_procedure}}
\begin{figure}[ht!]
\begin{center}
\centerline{\includegraphics[width=0.9\textwidth,scale=0.5]{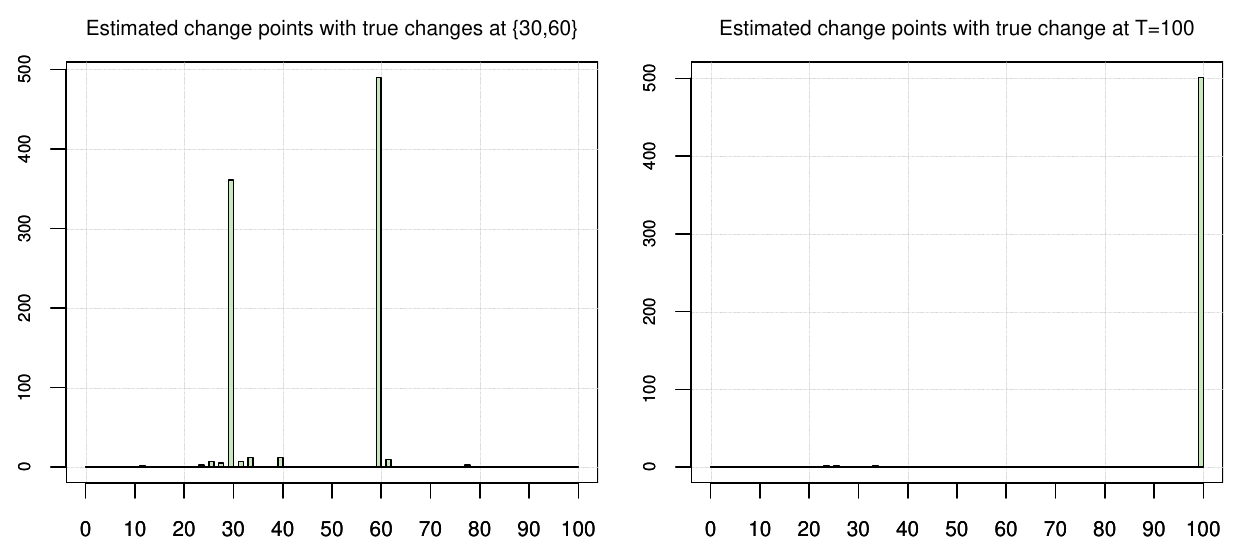}}
\caption{Histograms of estimated change locations under true models \eqref{eqn: sim_dgp_remark1} (left panel) and \eqref{eqn: sim_dgp_remark1_nochange} (right panel). Both experiments are repeated 500 times.}
\label{fig: simulation_Remark1_100_75}
\end{center}
\end{figure}

We demonstrate here the numerical performance of the divide-and-conquer scheme depicted in Remark~\ref{remark: sequential_procedure}. Under $(T,d) =(100,75)$, consider an extension of \eqref{eqn: sim_dgp_change_point} with two change points:
\begin{equation}
\label{eqn: sim_dgp_remark1}
\y_t = \big(\I_d- 0.8\, \b{1}_{\{t \leq 30\}} \W_1 + 0.9\, \b{1}_{\{t \leq 60\}} \W_1 + 0.9\, \b{1}_{\{t > 60\}} \W_2 \big)^{-1} \big(\bmu^\ast + \X_t\bbeta^\ast + \bepsilon_t \big) .
\end{equation}
That is, the spatial weight matrix changes from $-0.1 \W_1$ to $-0.9 \W_1$ at $t=30$, followed by a change from $-0.9 \W_1$ to $-0.9 \W_2$ at $t=60$. Suppose it is only known a priori that on $\c{T} =\{2, 4, \dots, 98, 100\}$\footnote{Change point identified at the last observed time point $T=100$ represents no change in the structure.}, the spatial weight matrix might change from $\W_1$ to $\W_2$ and the spatial correlation coefficients might change as well. We wish to estimate the number of changes and the change locations. Following Remark~\ref{remark: sequential_procedure}, we construct the ordered sets $\c{T}_1 = \{2,4,\dots, 20\}$, $\c{T}_2 = \{20,22, \dots, 40\}$, $\c{T}_3 = \{40, 42, \dots, 60\}$, $\c{T}_4 = \{60, 62, \dots, 80\}$ and $\c{T}_5 = \{80, 82, \dots, 100\}$. Note that we add ``overlaps'' between adjacent sets to circumvent falsely identifying change candidates at the margin. Now, the size of each set is $11$, which is reasonable according to the numerical results in Tables \ref{tab: sim_change_point_setting_weak_lag5}. For each $j\in[5]$, consider\footnote{Note that the last dynamic variable in $\c{T}_5$ for $\W_2$, $z_{5,2,11,t}$, is 0 for all $t$, so we directly specify $\phi_{5,2,11}^*$ as 0.}
\begin{align}
    & \y_t = \Big(\I_d- \sum_{l=1}^{|\c{T}_j|} z_{j,1,l,t} \phi_{j,1,l}^\ast \W_1 - \sum_{l=1}^{|\c{T}_j|} z_{j,2,l,t} \phi_{j,2,l}^\ast \W_2 \Big)^{-1} \big(\bmu^\ast + \X_t\bbeta^\ast + \bepsilon_t \big) ,
    \;\;\;  \text{where}
    \label{eqn: sim_dgp_remark1_saturated} \\
    & z_{j,1,l,t} := \b{1}_{\{t \leq (\c{T}_j)_l\}},
    \;\;\; z_{j,2,l,t} := \b{1}_{\{t > (\c{T}_j)_l\}} .\notag
\end{align}
As in Remark~\ref{remark: sequential_procedure}, all time points corresponding to nonzero estimates of $\phi_{j,1,l}^\ast$ or $\phi_{j,2,l}^\ast$ for $j\in[5], l\in[|\c{T}_j|]$ are identified and collected to form a refined candidate set $\wt{\c{T}}$, where marginal time points (20, 40, 60,80) are discarded if they are not identified in all $\c{T}_j$ containing them. Finally, consider
\begin{align}
    & \y_t = \Big(\I_d- \sum_{l=1}^{|\wt{\c{T}}|} z_{1,l,t} \phi_{1,l}^\ast \W_1 - \sum_{l=1}^{|\wt{\c{T}}|} z_{2,l,t} \phi_{2,l}^\ast \W_2 \Big)^{-1} \big(\bmu^\ast + \X_t\bbeta^\ast + \bepsilon_t \big) ,
    \;\;\; \text{where}
    \label{eqn: sim_dgp_remark1_saturated_refine} \\
    & z_{1,l,t} = \b{1}_{\{t \leq \wt{\c{T}}_l\}},
    \;\;\; z_{2,l,t} = \b{1}_{\{t > \wt{\c{T}}_l\}} .\notag
\end{align}
Then, change points are estimated as the timestamps corresponding to nonzero estimates for $\phi_{1,l}^\ast$ or $\phi_{2,l}^\ast$. The histogram for the estimated change locations over 500 repetitions is shown in the left panel of Figure~\ref{fig: simulation_Remark1_100_75} and is encouraging. To further quantify the performance of our scheme, we use the Adjusted Rand index (ARI) of the estimated time segmentation against the truth\footnote{The estimated time segmentation assigns the same labels to time points between the estimated change points, with different labels assigned after each change point. For true time partitioning with changes at $\{30,60\}$, the intervals $\{1,2, \dots, 30\}$, $\{31,32, \dots, 60\}$, and $\{61,62, \dots, 100\}$ are labelled as $1$, $2$, and $3$, respectively.}\citep{Rand1971,HubertArabie1985}, a measure frequently used by change point researchers \citep{WangSamworth2018}. The average ARI across all runs is 0.901, again suggesting that our scheme is performing very well.

We also consider \eqref{eqn: sim_dgp_remark1} under no change or, equivalently, one change at $t=100$:
\begin{equation}
\label{eqn: sim_dgp_remark1_nochange}
\y_t = \big(\I_d+ 0.9 \cdot \b{1}_{\{t \leq T\}} \W_1\big)^{-1} \big(\bmu^\ast + \X_t\bbeta^\ast + \bepsilon_t \big) .
\end{equation}
We follow the same exact procedure to estimate \eqref{eqn: sim_dgp_remark1}, and the histogram for the estimated change points over 500 runs is shown in the right panel of Figure~\ref{fig: simulation_Remark1_100_75}. In $98\%$ of the experiments, exactly $T=100$ is identified, meaning no change is detected, which corresponds to a $2\%$ false change discovery rate. Furthermore, the average ARI\footnote{The true time partition, when there are no changes, labels all time points as 1.} is $0.980$.

\section{Additional real data example: enterprise profits}\label{subsec: real_data_NBS}
In this case study, we use our proposed model to analyze the total profits of enterprises for a selection of provincial regions in China. Our panel data covers $T=86$ monthly periods from March 2016 to August 2024 and 25 provinces and 4 direct-administered municipalities (i.e., $d=29$).

\begin{figure}[!thp]
\includegraphics[width=\textwidth]{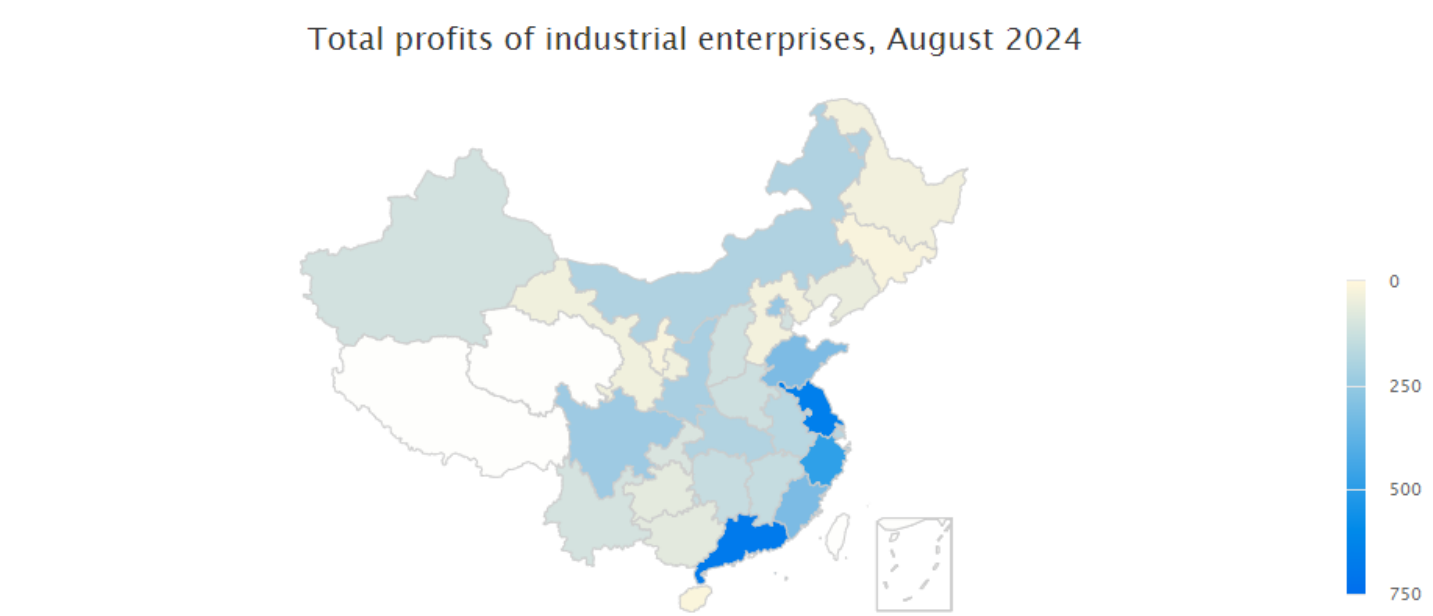}
\caption{Illustration of the total profits of industrial enterprises within considered Chinese provinces and direct-administered municipalities, in 100 million yuans.}
\label{Fig: realdata_NBS_map}
\end{figure}

Due to missingness, we exclude January and February data. The 25 provinces included in the data are Hebei, Shanxi, Inner Mongolia, Liaoning, Jilin, Heilongjiang, Jiangsu, Zhejiang, Anhui, Fujian, Jiangxi, Shandong, Henan, Hubei, Hunan, Guangdong, Guangxi, Hainan, Sichuan, Guizhou, Yunnan, Shaanxi, Gansu, Ningxia, Xinjiang, and the four direct-administered municipalities are Beijing, Tianjin, Shanghai and Chongqing. A snippet of the total profits for August 2024 is shown in Figure \ref{Fig: realdata_NBS_map}, where the map is produced using the R package \texttt{hchinamap}. From the estimation of our null model, it is revealed from $\wh\bmu$ that Guangdong, Beijing, Jiangsu and Shanghai have significantly larger spatial fixed effects than other provinces or municipalities. 

The set of covariates (all standardized) consists of Consumer Price Index (\textit{CPI}), Purchasing Price Index for industrial producers (\textit{PPI}) and output of electricity (\textit{elec}). The data is available at the National Bureau of Statistics of China: \url{https://data.stats.gov.cn/english/}.

We consider three spatial weight matrix candidates, with each row standardized by its $L_1$ norm if the row sum exceeds one: inverse distance matrix using inverse of geographical distances between locations computed by the Geodesic WGS-84 System ($\W_1$), contiguity matrix ($\W_2$), municipality matrix such that all direct-administered municipalities are neighbors ($\W_3$).

We treat the covariates as exogenous for two reasons: \textit{CPI} and \textit{PPI} are largely independent of the internal economic activities specific to enterprises within each province or municipality, and electricity supply as a public utility is often price inelastic. Using the aforementioned covariates and spatial weight matrices, we first specify a time-invariant spatial autoregressive model as our null model:
\begin{equation}
\label{eqn: realdata_NBS_null}
\textit{profit}_t = \bmu^\ast + (\phi_{1,0}^\ast \W_1 + \phi_{2,0}^\ast \W_2 + \phi_{3,0}^\ast \W_3) \, \textit{profit}_t + (\textit{CPI}_t, \textit{PPI}_t, \textit{elec}_t) \, \bbeta^\ast + \bepsilon_t.
\end{equation}
We estimate the parameters in \eqref{eqn: realdata_NBS_null} using our adaptive LASSO estimators. The estimated coefficients $\{\wh\phi_{1,0}, \wh\phi_{2,0}, \wh\phi_{3,0}, \wh\bbeta\}$ are presented in Table \ref{tab: realdata_NBS_null}, together with the BIC computed according to \eqref{eqn: BIC_lambda_null}. The table also shows the standard errors of $\wh\phi_{1,0}$ and $\wh\phi_{3,0}$ based on Theorem \ref{thm: ada_LASSO_phi_asymp}. The respective p-values for testing $\wh\phi_{1,0} = 0$ and $\wh\phi_{3,0} = 0$ are both less than $0.0001$, revealing some spillovers among the neighbors of provinces and municipalities. Interestingly, $\wh\phi_{3,0}$ suggests a negative spillover effect among the four direct-administered municipalities, which could be explained by that the enterprises within municipalities are main competitors in the market.

\begin{table}[ht!]
\setlength{\tabcolsep}{4pt}
\small
\begin{center}
\begin{tabular}{l|ccc|ccc|c}
\hline & & & & & & &  \\[-1em] 
 & $\wh\phi_{1,0}$ & $\wh\phi_{2,0}$ & $\wh\phi_{3,0}$ & $\wh\beta_{\textit{CPI}}$ & $\wh\beta_{\textit{PPI}}$ & $\wh\beta_{\textit{elec}}$ & BIC  \\
\hline & & & & & & &  \\[-1em]
\text{Null Model} & 15.184 & .000 & -.285 & .021 & .053 & .394 & 2.790  \\
& {\small (3.725)} &  & {\small (.066)} &  &  &  &   \\
\hline
\end{tabular}
\end{center}
\caption{Estimated coefficients for model \eqref{eqn: realdata_NBS_null}, with standard errors (in brackets) computed according to the last part of Section \ref{sec: practical_implementation}. $\wh\beta_{\textit{CPI}}$, $\wh\beta_{\textit{PPI}}$ and $\wh\beta_{\textit{elec}}$ denote the estimates of $\bbeta^\ast$ corresponding to $\textit{CPI}_t$, $\textit{PPI}_t$ and $\textit{elec}_t$, respectively.}
\label{tab: realdata_NBS_null}
\end{table}

Hereafter, we refer to \eqref{eqn: realdata_NBS_null} as the null model. The rest of the analysis is performed in an exploratory fashion such that we consider spatial autoregressive models of the form \eqref{eqn: spatiallag} with some $l_j$ and dynamic variables $\{ z_{j,k,t}\}$. We consider the following models:
{\small
\begin{equation}
\label{eqn: realdata_NBS_alternative}
\begin{split}
    \text{Model 1} &: \textit{profit}_t = \bmu^\ast + \sum_{j=1}^{3} \Big( \phi_{j,0}^\ast + \sum_{k=1}^{15} \phi_{j,k}^\ast \b{1}\{ t\leq 5+5k \} \Big) \W_j \, \textit{profit}_t + (\textit{CPI}_t, \textit{PPI}_t, \textit{elec}_t) \, \bbeta^\ast + \bepsilon_t ;\\
    \text{Model 2} &: \textit{profit}_t = \bmu^\ast + \sum_{j=1}^{3} \Big( \phi_{j,0}^\ast + \sum_{k=1}^{9} \phi_{j,k}^\ast \b{1}\{ \text{sd}(\textit{profit}_{t-5}) \leq \gamma_k \} \Big) \W_j \, \textit{profit}_t
    + (\textit{CPI}_t, \textit{PPI}_t, \textit{elec}_t) \, \bbeta^\ast + \bepsilon_t , \\
    &\hspace{6pt}
    \text{where } (\gamma_1, \dots, \gamma_9) =(.177, .193, .202, .207, .217, .219, .231, .250, .302) ;\\
    \text{Model 3} &: \textit{profit}_t = \bmu^\ast + \sum_{j=1}^{3} \Big( \phi_{j,0}^\ast + \sum_{k=1}^{5} \phi_{j,k}^\ast \b{1}\{\text{$t$ divides $2k$}\} \Big) \W_j \, \textit{profit}_t + (\textit{CPI}_t, \textit{PPI}_t, \textit{elec}_t) \, \bbeta^\ast + \bepsilon_t ; \\
    \text{Model 4} &: \textit{profit}_t = \bmu^\ast + \Big( \phi_{1,0}^\ast \alpha_{3,1}(\W_1) + \phi_{2,0}^\ast \alpha_{3,2}(\W_1) + \phi_{3,0}^\ast \alpha_{3,3}(\W_1) \Big)\, \textit{profit}_t + (\textit{CPI}_t, \textit{PPI}_t, \textit{elec}_t) \, \bbeta^\ast + \bepsilon_t .
\end{split}
\end{equation}
}

Model 1 represents a spatial autoregressive model with the spatial weight matrix potentially changing at $(10, 15, \dots, 75, 80)$. Model 2 is a self-exciting threshold spatial autoregressive model with the standard deviation of $\textit{profit}_{t-5}$ as the threshold variable. The sequence of threshold value is in fact the empirical quantile, from $10\%$ to $90\%$, of $\text{sd}(\textit{profit}_{t-5})$. Model 3 is similar to the null model but accounts for monthly spillovers for lags of two, four, six, eight and ten months. Model 4 adapts our framework to time-invariant and nonlinear spatial weight matrices, where $\alpha_{3,1}(\W_1)$, $\alpha_{3,2}(\W_1)$ and $\alpha_{3,3}(\W_1)$ denote the matrices formed by series expansion using the order-3 normalized Laguerre functions (inspired by \cite{Sun2016}) based on $\{ (\W_1)_{ij}^{-1} \}_{i,j\in[29]}$.

Table \ref{tab: realdata_NBS_alternative} reports the estimated parameters and BIC for each model. The nonzero $\wh\phi_{1,9}$ for Model 1 corresponds to a change in the spillovers featured by $\W_1$ in March 2020, potentially suggesting inactive economic activities due to COVID-19 starting at the beginning of 2020.

\begin{table}[ht!]
	\setlength{\tabcolsep}{4pt}
    \small
	\begin{center}
		\begin{tabular}{l|cccc|ccc|c}
			\hline & & & & & & &  \\[-1em] 
			& \multicolumn{4}{c|}{ nonzero $\wh\phi_{j,k}$ } & $\wh\beta_{\textit{CPI}}$ & $\wh\beta_{\textit{PPI}}$ & $\wh\beta_{\textit{elec}}$ & BIC  \\
			\hline & & & & & & &  \\[-1em]
			\text{Model 1} & $\wh\phi_{1,9}= 18.030$ & & & & .044 & .038 & .309 & 2.978  \\
			& \hspace{25pt} {\small $(13.601)$} & & & & &  &  &   \\
			\text{Model 2} & $\wh\phi_{1,3}= 9.721$ & $\wh\phi_{1,9}= 18.997$ & $\wh\phi_{3,7}= -.522$ & & .030 & .038 & .368 & 2.744  \\
			& \hspace{25pt} {\small $(3.487)$} & \hspace{25pt} {\small $(.717)$} & \hspace{25pt} {\small $(.001)$} & &  &  &  &   \\
			\text{Model 3} & $\wh\phi_{1,0}= 15.424$ & $\wh\phi_{2,1}= .335$ & $\wh\phi_{2,3}= -.331$ & $\wh\phi_{3,5}= -.401$ & .023 & .052 & .398 & 2.747  \\
			& \hspace{25pt} {\small (.000)} & \hspace{25pt} {\small (.000)}  & \hspace{25pt} {\small (.000)} & \hspace{25pt} {\small (.000)} &  &  &  &   \\
			\text{Model 4} & $\wh\phi_{1,0}= .048$ & $\wh\phi_{2,0}= -.034$ & $\wh\phi_{3,0}= .114$ &  & .043 & .056 & .512 & 2.848  \\
			& \hspace{25pt} {\small (.001)} & \hspace{25pt} {\small (.003)}  & \hspace{25pt} {\small (.001)} &  &  &  &  &   \\
			\hline
		\end{tabular}
	\end{center}
	\caption{Estimated coefficients for different models specified in \eqref{eqn: realdata_NBS_alternative}, with standard errors (in brackets) computed according to the last part of Section \ref{sec: practical_implementation}. Refer to Table \ref{tab: realdata_NBS_null} for the definitions of $\wh\beta_{\textit{CPI}}$, $\wh\beta_{\textit{PPI}}$ and $\wh\beta_{\textit{elec}}$.}
	\label{tab: realdata_NBS_alternative}
\end{table}

\begin{figure}[!htbp]
	\includegraphics[width=\textwidth]{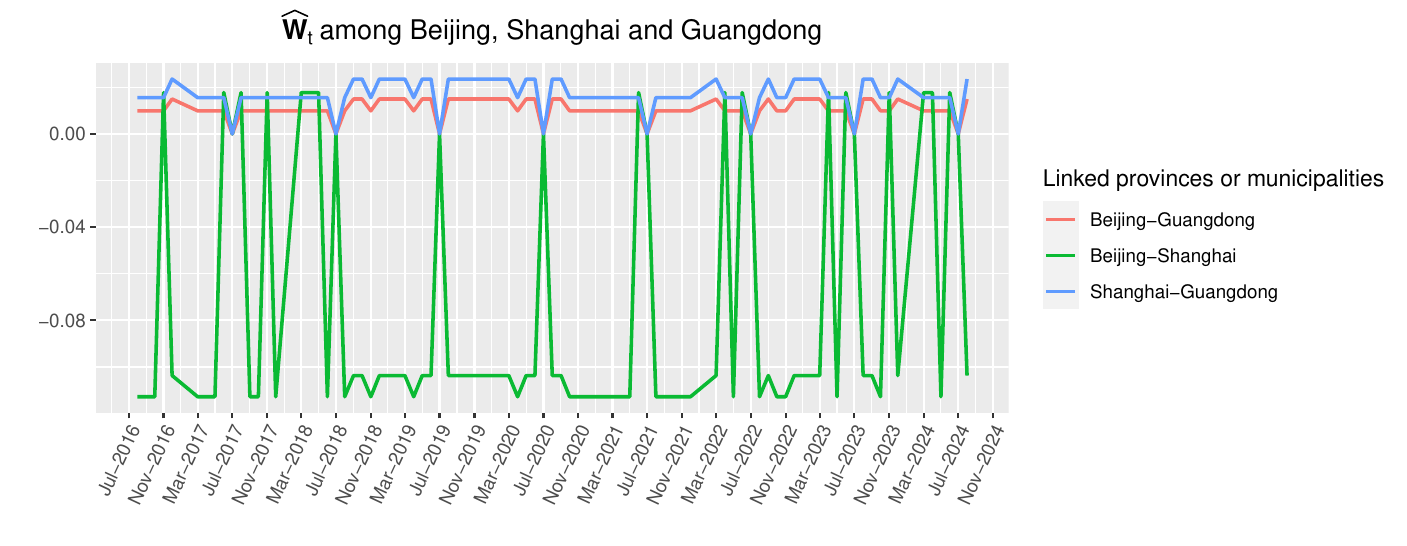}
	\caption{Illustration of $\wh\W_t$ of Model 2 in \eqref{eqn: realdata_NBS_alternative} among Beijing, Shanghai and Guangdong, from August 2016 to August 2024.}
	\label{Fig: realdata_NBS_thres}
\end{figure}

Model 2 has the best BIC among all models shown here, including the null model. From Table~\ref{tab: realdata_NBS_alternative}, four threshold regions are identified as
\[
\wh\W_t = \begin{cases}
    28.718\, \W_1 - 0.522\, \W_3, & \text{sd}(\textit{profit}_{t-5}) \leq 0.202; \\
    18.997\, \W_1 - 0.522\, \W_3, & 0.202 < \text{sd}(\textit{profit}_{t-5}) \leq 0.231; \\
    18.997\, \W_1, & 0.231 < \text{sd}(\textit{profit}_{t-5}) \leq 0.302; \\
    0, & \text{sd}(\textit{profit}_{t-5}) > 0.302.
\end{cases}
\]
For better illustration, the series of $\wh\W_t$ among Beijing, Shanghai and Guangdong are plotted in Figure \ref{Fig: realdata_NBS_thres}. We see that the spillovers between Beijing and Shanghai is more significant than their respective spillovers with Guangdong.

\begin{figure}[!htbp]
	\includegraphics[width=\textwidth]{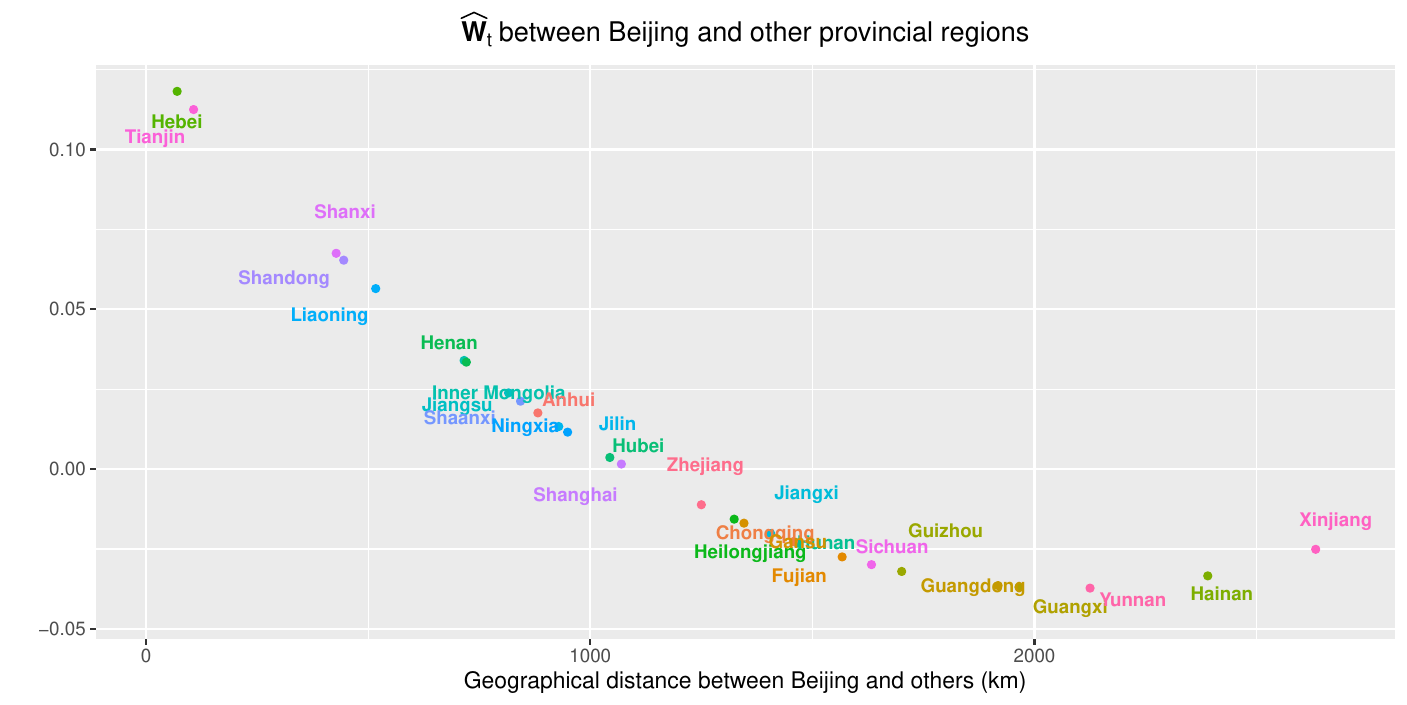}
	\caption{Illustration of (time-invariant) $\wh\W_t$ of Model 4 in \eqref{eqn: realdata_NBS_alternative} between Beijing and other provincial regions, against their geographical distances.}
	\label{Fig: realdata_NBS_laguerre_Beijing}
\end{figure}

On Model 3, Table \ref{tab: realdata_NBS_alternative} suggests that the effect of $\W_1$ (representing domestic spillovers) remains constant, that of $\W_2$ (representing more local spillovers) persists every two months but roughly cancels out every half year, and the ``municipality spillover'' by $\W_3$ occurs every December. The various spillover patterns featured by the expert spatial weight matrices are intriguing and warrant further investigation.

Model 4 considers a time-invariant spillover effect. An example of the estimated spatial weight matrix is displayed in Figure \ref{Fig: realdata_NBS_laguerre_Beijing}. It depicts how the spillovers diminish with the geographical distance. Lastly, the analysis on the total profits data serves to demonstrate our dynamic spatial autoregressive framework. More comprehensive investigations are required to further understand the spatial relations among industrial enterprises in Chinese provinces and direct-administered municipalities.

	\section{Proofs of theorems and auxiliary results}
	
	\subsection{Lemmas and their proofs}
	
	To prove our main theorems, recall first $B_{t,ij}$ and $X_{t,ij}$ represent the $(i,j)$ entry of $\B_t$ and $\X_t$ respectively. Define $\cM = \bigcap_{i=1}^{13} \cA_i$, where
	\begin{equation}
    \label{eqn: set_A_in_M}
	\begin{split}
			\cA_1 &= \Bigg\{\max_{i,q \in[d]} \max_{j,l \in[r]} \bigg| \frac{1}{T}\sum_{t=1}^T [B_{t,ij}X_{t,ql} - \b{E}(B_{t,ij}X_{t,ql})]\bigg| < c_T \Bigg\} ,\\
			\cA_2 &= \Bigg\{\max_{i,q \in[d]} \max_{j\in[r]} \bigg| \frac{1}{T}\sum_{t=1}^T B_{t,ij} \epsilon_{t,q}\bigg| < c_T \Bigg\} ,\\
			\cA_3 &= \Bigg\{ \max_{j\in[r]} \bigg| \frac{1}{T}\sum_{t=1}^T \sum_{q=1}^d B_{t,qj}\epsilon_{t,q} \bigg| < c_T d^{\frac{1}{2} + \frac{1}{2w}} \Bigg\} ,\\
			\cA_4 &= \Bigg\{\max_{i\in[d]} \max_{j\in[r]} \big|  \bar{B}_{\cdot,ij} - \b{E}[B_{t,ij}] \big| < c_T \Bigg\} ,\\
			\cA_5 &= \Bigg\{\max_{q \in[d]} \big| \bar{\epsilon}_{\cdot, q} \big| < c_T \Bigg\} ,\\
			\cA_6 &= \Bigg\{\max_{i\in[d]} \max_{j\in[r]} \big|  \bar{X}_{\cdot,ij} \big| < c_T \Bigg\} ,\\
			\cA_7 &= \Bigg\{ \max_{j\in[r]} \bigg| \sum_{q=1}^d \bar{B}_{\cdot, qj} \bar{\epsilon}_{\cdot,q} \bigg| < 2^{1/2} c_T d^{1/2} \log^{1/2}(T \vee d) S_\epsilon (\mu_{b,\text{max}} + c_T) \Bigg\} ,\\
			\cA_8 &= \Bigg\{\max_{m\in[p]} \max_{n\in[l_m] \cup\{0\}}\max_{i,q \in[d]} \max_{j \in[v]} \max_{l \in[r]} \bigg| \frac{1}{T}\sum_{t=1}^T [z_{m,n,t} B_{t,ij}X_{t,ql} - \b{E}(z_{m,n,t} B_{t,ij}X_{t,ql})]\bigg| < c_T \Bigg\} ,\\
			\cA_9 &= \Bigg\{\max_{m\in[p]} \max_{n\in[l_m] \cup\{0\}} \max_{i\in[d]} \max_{j\in[r]} \bigg| \frac{1}{T}\sum_{t=1}^T z_{m,n,t} X_{t,ij} \bigg| < c_T \Bigg\} ,\\
            \cA_{10} &= \Bigg\{\max_{m\in[p]} \max_{n\in[l_m] \cup\{0\}} \max_{i,q \in[d]} \max_{j\in[v]} \bigg| \frac{1}{T}\sum_{t=1}^T z_{m,n,t} B_{t,ij} \epsilon_{t,q}\bigg| < c_T \Bigg\} ,\\
            \cA_{11} &= \Bigg\{\max_{m\in[p]} \max_{n\in[l_m] \cup\{0\}} \max_{q\in[d]} \bigg| \frac{1}{T}\sum_{t=1}^T z_{m,n,t} \epsilon_{t,q} \bigg| < c_T \Bigg\} ,\\
			\cA_{12} &= \Bigg\{\max_{m\in[p]} \max_{n\in[l_m] \cup\{0\}} \max_{i\in[d]} \max_{j\in[v]} \bigg| \frac{1}{T}\sum_{t=1}^T [z_{m,n,t} B_{t,ij} -\b{E}(z_{m,n,t} B_{t,ij})] \bigg| < c_T \Bigg\} ,\\
			\cA_{13} &= \Bigg\{\max_{m\in[p]} \max_{n\in[l_m] \cup\{0\}} \bigg| \frac{1}{T}\sum_{t=1}^T z_{m,n,t} \bigg| < c_T \vee z_\text{max} \Bigg\} ,
		\end{split}
	\end{equation}
	where $\bar{B}_{\cdot,ij} := T^{-1}\sum_{t=1}^T B_{t,ij}$, $\bar{X}_{\cdot,ij} := T^{-1}\sum_{t=1}^T X_{t,ij}$, $\bar{\epsilon}_{\cdot,q} := T^{-1}\sum_{t=1}^T \epsilon_{t,q}$, $\hspace{2pt}$ $\mu_{b,\text{max}} := \max_{i,j} \big| \b{E}[B_{t,ij}] \big|$ being a constant implied by Assumption (M1), and $z_\text{max}$ is the upper bound for $\{z_{j,k,t}\}$ implied in (M2) with $z_\text{max}=1$ for $\{z_{j,0,t}\}$ by default. Our main theoretical results depict the properties of estimators on the set $\cM$ which holds with probability approaching 1 as $T,d\to \infty$ by Assumption (R10), as shown in Lemma \ref{lemma: LamSouza_theorem_S1} which is similar to Theorem S.1 of \cite{LamSouza2020}.
	
	To prove Lemma \ref{lemma: LamSouza_theorem_S1}, we first quote a Nagaev-type inequality for functional dependent data from Theorem 2(ii), 2(iii) and Section 4 of \cite{Liuetal2013}, presented as the following lemma.

	\begin{lemma}\label{lemma: nagaev_inequality}
		For a zero mean time series process $\x_t = \f(\cF_t)$ defined in (\ref{eqn: functional_x_b_epsilon}) with dependence measure $\theta_{t,q,i}^x$ defined in (\ref{eqn: def_functional_dependence}), assume $\Theta_{m,2w}^x \leq Cm^{-\alpha}$ as in Assumption (M1). Then there exists constants $C_1$, $C_2$ and $C_3$ independent of $n,\, T$ and the index $i$ such that
		\[
		\b{P}\bigg(\Big| \frac{1}{T}\sum_{t=1}^T x_{t,i} \Big| > n\bigg) \leq \frac{C_1 T^{w(1/2 - \wt{\alpha})}}{(Tn)^w} + C_2 \exp(-C_3 T^{\wt{\beta}} n^2) ,
		\]
		where $\wt{\alpha} = \alpha \wedge (1/2 - 1/w)$ and $\wt{\beta} = (3+ 2\wt{\alpha}w)/(1+w)$.
		
		Furthermore, assume another zero mean time series process $\{\e_t\}$ (can be the same process $\{\x_t\}$) with $\Theta_{m,2w}^e$ as in (M1). Then provided $\max_i \|x_{t,i}\|_{2w}, \, \max_j \|e_{t,j}\|_{2w} \leq c_0 < \infty$ where $c_0$ is a constant, the above Nagaev-type inequality holds for the product process $\{x_{t,i} e_{t,j} - \b{E}(x_{t,i} e_{t,j})\}$.
		
		The above results also hold for any zero mean non-stationary process $\x_t = \f_t(\cF_t)$ provided that $\max_i \|x_{t,i}\|_{w}< \infty$ and $\Theta_{m,2w}^{x,\ast} \leq Cm^{-\alpha}$, where $\Theta_{m,q}^{x,\ast}$ is uniform tail sum defined as
		\[
		\Theta_{m,q}^{x,\ast} := \sum_{t=m}^\infty \max_{i} \theta_{t,q,i}^{x,\ast} := \sum_{t=m}^\infty \max_{i} \sup_t \big\|x_{t,i} - x_{t,i}'\big\|_q .
		\]
	\end{lemma}

	We present Lemma \ref{lemma: LamSouza_theorem_S1} below. Note that we assume $\alpha > 1/2 - 1/w$ which can be relaxed at the cost of more complicated rates and longer proofs presented here, and it simplifies the form of Lemma \ref{lemma: nagaev_inequality} as $w(1/2 - \wt{\alpha}) = \wt{\beta} = 1$.
	
	\begin{lemma}\label{lemma: LamSouza_theorem_S1}
		Let Assumptions (M1), (M2) (or (M2')), and (R2) hold and $\alpha > 1/2 - 1/w$ in Assumption (M1). Suppose for the application of the Nagaev-type inequality in Lemma \ref{lemma: nagaev_inequality} for the processes in $\cM = \bigcap_{i=1}^{13} \cA_i$ where $\cA_i$ is defined in (\ref{eqn: set_A_in_M}), the constants $C_1$, $C_2$ and $C_3$ are the same. Then with $g \geq \sqrt{3/C_3}$ where $g$ is the constant defined in $c_T = gT^{-1/2} \log^{1/2} (T \vee d)$, we have
		\[
		\b{P}(\cM) \geq 1- 14 C_1 r^2 vL \Big(\frac{C_3}{3}\Big)^{w/2} \frac{d^2}{T^{w/2 - 1} \log^{w/2}(T \vee d)} - \frac{14 C_2 r^2 vL d^2}{T^3 \vee d^3} - \frac{2r}{T \vee d}.
		\]
	\end{lemma}
	\noindent\textbf{\textit{Proof of Lemma \ref{lemma: LamSouza_theorem_S1}.}}
	As shown in Theorem S.1 of \cite{LamSouza2020}, the tail condition in Assumption (M1) implies $\|\cdot\|_{2w}$ is bounded for the processes $\{B_{t,ij} X_{t,ql} - \b{E}(B_{t,ij} X_{t,ql})\}$, $\{B_{t,ij} \epsilon_{t,q}\}$, $\{X_{t,ij}\}$ and $\{\epsilon_{t,q}\}$, and further with Assumption (R2) we have
	\begin{align*}
		\b{P}(\cA_1^c) &\leq C_1 r^2 \Big(\frac{C_3}{3}\Big)^{w/2} \frac{d^2}{T^{w/2 - 1} \log^{w/2}(T\vee d)} + \frac{C_2 r^2 d^2}{T^3\vee d^3} ,\\
		\b{P}(\cA_2^c) &\leq C_1 r \Big(\frac{C_3}{3}\Big)^{w/2} \frac{d^2}{T^{w/2 - 1} \log^{w/2}(T\vee d)} + \frac{C_2 r d^2}{T^3\vee d^3} ,\\
		\b{P}(\cA_3^c) &\leq C_1 \Big(\frac{C_3}{3}\Big)^{w/2} \frac{r}{T^{w/2 - 1} \log^{w/2}(T\vee d)} + \frac{C_2 r}{T^3\vee d^3} ,\\
		\b{P}(\cA_4^c) &\leq C_1 r \Big(\frac{C_3}{3}\Big)^{w/2} \frac{d}{T^{w/2 - 1} \log^{w/2}(T\vee d)} + \frac{C_2 r d}{T^3\vee d^3} ,\\
		\b{P}(\cA_5^c) &\leq C_1 \Big(\frac{C_3}{3}\Big)^{w/2} \frac{d}{T^{w/2 - 1} \log^{w/2}(T\vee d)} + \frac{C_2 d}{T^3\vee d^3} ,\\
		\b{P}(\cA_6^c) &\leq C_1 r \Big(\frac{C_3}{3}\Big)^{w/2} \frac{d}{T^{w/2 - 1} \log^{w/2}(T\vee d)} + \frac{C_2 r d}{T^3\vee d^3} ,\\
		\b{P}(\cA_7^c) &\leq \frac{2r}{T\vee d} + \b{P}(\cA_4^c) +\b{P}(\cA_5^c) .
	\end{align*}

	Consider the remaining sets. First let Assumption (M2) hold and it is easy to see $\|\cdot\|_{2w}$ is bounded for the processes $\{z_{m,n,t} B_{t,ij} X_{t,ql} - \b{E}(z_{m,n,t} B_{t,ij} X_{t,ql})\}$, $\{z_{m,n,t} B_{t,ij} \epsilon_{t,q}\}$, $\{z_{m,n,t} X_{t,ij}\}$ and $\{z_{m,n,t} \epsilon_{t,q}\}$, and their uniform tail sums satisfy the condition in Lemma \ref{lemma: nagaev_inequality}. Thus, apply Lemma \ref{lemma: nagaev_inequality} first on $\cA_8^c$ and we have by the union bound,
	\begin{align*}
		\b{P}(\cA_8^c) &\leq \sum_{m\in[p]} \sum_{n\in[l_m]}\sum_{i,q \in[d]} \sum_{j \in[v]} \sum_{l \in[r]} \b{P}\Bigg(\bigg| \frac{1}{T}\sum_{t=1}^T [z_{m,n,t} B_{t,ij}X_{t,ql} - \b{E}(z_{m,n,t} B_{t,ij}X_{t,ql})]\bigg| \geq c_T \Bigg) \\
		&\leq
		d^2 rv L \Big( \frac{C_1 T}{(T c_T)^w} + C_2 \exp(-C_3 T c_T^2) \Big) \\
            &\leq
		C_1 rv L\Big(\frac{C_3}{3}\Big)^{w/2}\frac{d^2}{T^{w/2 - 1} \log^{w/2}(T\vee d)} + \frac{C_2 rv L d^2}{T^3\vee d^3} .
	\end{align*}
	Similarly using Lemma \ref{lemma: nagaev_inequality}, we have
	\begin{align*}
		\b{P}(\cA_9^c) &\leq
		C_1 r L\Big(\frac{C_3}{3}\Big)^{w/2}\frac{d}{T^{w/2 - 1} \log^{w/2}(T\vee d)} + \frac{C_2 r L d}{T^3\vee d^3} , \\
		\b{P}(\cA_{10}^c) &\leq
		C_1 v L\Big(\frac{C_3}{3}\Big)^{w/2}\frac{d^2}{T^{w/2 - 1} \log^{w/2}(T\vee d)} + \frac{C_2 v L d^2}{T^3\vee d^3} ,\\
		\b{P}(\cA_{11}^c) &\leq
		C_1 L\Big(\frac{C_3}{3}\Big)^{w/2}\frac{d}{T^{w/2 - 1} \log^{w/2}(T\vee d)} + \frac{C_2 L d}{T^3\vee d^3} ,\\
		\b{P}(\cA_{12}^c) &\leq
		C_1 v L\Big(\frac{C_3}{3}\Big)^{w/2}\frac{d}{T^{w/2 - 1} \log^{w/2}(T\vee d)} + \frac{C_2 v L d}{T^3\vee d^3} ,
	\end{align*}
	while $\b{P}(\cA_{13}^c) =0$ by Assumption (M2). On the other hand, if we have Assumption (M2'), the above results remain valid, except that
	\[
	\b{P}(\cA_{13}^c) \leq
	C_1 L\Big(\frac{C_3}{3}\Big)^{w/2}\frac{1}{T^{w/2 - 1} \log^{w/2}(T\vee d)} + \frac{C_2 L}{T^3\vee d^3} ,
	\]
	by applying Lemma \ref{lemma: nagaev_inequality} given the bounded support and tail sum assumption in (M2'). For any $\{z_{j,0,t}\}$, we may treat it as a non-stochastic basis as in (M2) and the result follows. Lastly, by $\b{P}(\cM) \geq 1- \sum_{i=1}^{13}\b{P}(\cA_i^c)$ we complete the proof of Lemma \ref{lemma: LamSouza_theorem_S1}.
	$\square$

	We present the following lemma with a short proof as well, and we will frequently utilize the defined notation $\V_{\H,K}$ (which is the same definition in Theorem \ref{thm: ada_LASSO_phi_asymp}) in the proof of main theorems.
	\begin{lemma}\label{lemma: V_H_kronecker}
		For any $n\times d$ matrix $\H = (\h_1, \dots, \h_n)^\top$ and any $d \times K$ matrix $\M$, define
		\[
		\V_{\H, K} =
		\begin{pmatrix}
			\I_K \otimes \h_1 \\
			\vdots \\
			\I_K \otimes \h_n
		\end{pmatrix} .
		\]
		We then have $\H\M = \Big(\I_n \otimes \textnormal{vec}(\M)^\top \Big) \V_{\H, K}$.
	\end{lemma}
	\noindent\textbf{\textit{Proof of Lemma \ref{lemma: V_H_kronecker}.}}
	Notice that
	\[
	\Big(\I_n \otimes \textnormal{vec}(\M)^\top \Big) \V_{\H, K} =
	\begin{pmatrix}
		\vec{\M}^\top (\I_K \otimes \h_1) \\
		\vdots \\
		\vec{\M}^\top (\I_K \otimes \h_n)
	\end{pmatrix} .
	\]
	The $j$-th row of this matrix (as a column vector) is hence 
    \[(\I_K \otimes \h_j^\top)\vec{\M} = \vec{\h_j^\top \M} = \vec{\M^\top \h_j} = \M^\top \h_j,\]
    which is the $j$-th row of $\H\M$ indeed. $\square$

	\begin{remark}\label{remark: V_H_kronecker}
		With the notation of $\V_{\H,K}$, we may write any $d\times K$ matrix $\M$ as
		\[
		\M =\I_d \M = \Big(\I_d \otimes \textnormal{vec}(\M)^\top \Big) \V_{\I_d, K},
		\]
		which will be useful if we are interested in the interaction only between $\A, \,\M$ in $\A\B\M$, with $\A\in \b{R}^{l\times r}$ and $\B\in \b{R}^{r\times d}$, since we have
		\[
		\A\B\M = (\B^\top \A^\top)^\top \M =\V_{\B^\top, l}^\top \Big(\I_d \otimes \textnormal{vec}(\A^\top) \Big) \M
		= \V_{\B^\top, l}^\top \Big(\I_d \otimes \textnormal{vec}(\A^\top) \textnormal{vec}(\M)^\top \Big) \V_{\I_d, K}.
		\]
		Moreover, notice that in Lemma \ref{lemma: V_H_kronecker}, if $K=1$, i.e. $\M$ is a vector, then $\V_{\H,1} =\textnormal{vec}(\H^\top)$ and Lemma \ref{lemma: V_H_kronecker} simply coincides with the fact that
		\[
		\B\M = \textnormal{vec}(\M^\top \B^\top) =\textnormal{vec}(\M^\top \B^\top \I_r) = \big(\I_r \otimes \M^\top \big) \textnormal{vec}(\B^\top) =\Big(\I_r \otimes \textnormal{vec}(\M)^\top \Big) \V_{\B,1}.
		\]
		Thus, we can interpret $\V_{\H,K}$ to be a ``$K$-block vectorization'' of $\H$, as a generalization of vectorization.
	\end{remark}

	\begin{lemma}\label{lemma: asymp_I2}
		Let the assumptions in Theorem \ref{thm: ada_LASSO_phi_asymp} hold. Let $\R_{\beta}$ and $\bSigma_{\beta}$ be defined in Theorem \ref{thm: ada_LASSO_phi_asymp}. For $I_2 =\big[\b{E}(\X_t^\top \B_t) \b{E}(\B_t^\top \X_t)\big]^{-1} T^{-2} \X^\top \B^\nu (\B^\nu)^\top \bepsilon^\nu$, we have $I_2$ asymptotically normal with rate $T^{-1/2} d^{-(1-b)/2}$ such that
		\[
		T^{1/2} (\R_{\beta} \bSigma_{\beta} \R_{\beta}^\top)^{-1/2} I_2 \xrightarrow{D} \cN(\0, \I_r) .
		\]
	\end{lemma}
	\noindent\textbf{\textit{Proof of Lemma \ref{lemma: asymp_I2}.}}
	Given any non-zero $\balpha\in \b{R}^r$ with $\|\balpha\|_1 \leq c<\infty$, we construct below the asymptotic normality of $\balpha^\top I_2$ which is $T^{1/2} d^{(1-b)/2}$-convergent. First, we can further decompose
	\begin{align*}
		\balpha^\top I_2 &= \big[\b{E}(\X_t^\top \B_t) \b{E}(\B_t^\top \X_t)\big]^{-1} \big(T^{-1} \X^\top \B^\nu -\b{E}(\X_t^\top \B_t) \big) T^{-1} (\B^\nu)^\top \bepsilon^\nu \\
		&\hspace{12pt}
		+ \big[\b{E}(\X_t^\top \B_t) \b{E}(\B_t^\top \X_t)\big]^{-1} \b{E}(\X_t^\top \B_t) T^{-1} (\B^\nu)^\top \bepsilon^\nu ,
	\end{align*}
	with the second term dominating the first by (\ref{eqn: rate_diff_U_U0}). Recall that
    $$\R_{\beta} = \big[\b{E}(\X_t^\top \B_t) \b{E}(\B_t^\top \X_t)\big]^{-1} \b{E}(\X_t^\top \B_t),$$
    then
	\[
	\balpha^\top I_2 = \frac{1}{T} \sum_{t=1}^T \balpha^\top \R_{\beta} (\B_t- \b{E}(\B_t))^\top \bepsilon_t (1+o_P(1)).
	\]
	
	To construct the asymptotic normality of $\balpha^\top I_2$, we want to show the following as in (\ref{eqn: cond_Wu2011_F6}) that
	\begin{equation}
		\label{eqn: cond_Wu2011_I2}
		\sum_{t\geq 0}\Big\| P_0( \balpha^\top \R_{\beta} (\B_t- \b{E}(\B_t))^\top \bepsilon_t ) \Big\|_2 <\infty,
	\end{equation}
	so that Theorem 3 (ii) of \cite{Wu2011} can then be applied. With the definition $s_2:= \balpha^\top \R_{\beta} \bSigma_{\beta} \R_{\beta}^\top \balpha$, we have $T^{1/2} s_2^{-1/2} \balpha^\top I_2 \xrightarrow{D} \cN(0, 1)$ and hence
	\begin{equation}
		\label{eqn: asymp_I2}
		T^{1/2} (\R_{\beta} \bSigma_{\beta} \R_{\beta}^\top)^{-1/2} I_2 \xrightarrow{D} \cN(\0, \I_r).
	\end{equation}
	
	Similar to the proof of (\ref{eqn: cond_Wu2011_F6}), we have (\ref{eqn: cond_Wu2011_I2}) hold by Assumption (R7) and
	\begin{align*}
		&\hspace{12pt}
		\Big\| P_0( \balpha^\top \R_{\beta} (\B_t- \b{E}(\B_t))^\top \bepsilon_t )\Big\|_2 \\
		&=
		\Big\|\balpha^\top \R_{\beta} \big\{P_0((\B_t- \b{E}(\B_t))^\top) \b{E}_0(\bepsilon_t) \big\} + \balpha^\top \R_{\beta} \big\{ \b{E}_{-1}((\B_t- \b{E}(\B_t))^\top) P_0(\bepsilon_t )\big\} \Big\|_2 \\
		&\leq
		\Big\{2 \balpha^\top \R_{\beta} \b{E}\Big\{ P_0((\B_t- \b{E}(\B_t))^\top)\, \b{E}_0(\bepsilon_t)\, \b{E}_0(\bepsilon_t^\top)\, P_0(\B_t- \b{E}(\B_t)) \Big\} \R_{\beta}^\top \balpha \Big\}^{1/2} \\
		&\hspace{12pt}
		+ \Big\{2 \balpha^\top \R_{\beta} \b{E}\Big\{ \b{E}_{-1}((\B_t- \b{E}(\B_t))^\top) \, P_0(\bepsilon_t )\, P_0(\bepsilon_t^\top) \, \b{E}_{-1}(\B_t- \b{E}(\B_t)) \Big\} \R_{\beta}^\top \balpha \Big\}^{1/2} \\
		&=
		O\Big(\big\|\balpha \big\|_1 \big\|\R_{\beta} \big\|_\infty \Big) \Big( d\cdot \max_{j\in[d]} \b{E}^{1/2}( \b{E}_0^2(\epsilon_{t,j})) \,
        \max_{j\in[d]} \max_{k\in[v]} \big\| P_0(B_{t,jk}) \big\|_2 + d\cdot \sigma_\text{max} \max_{j\in[d]} \big\|P_0 (\epsilon_{t,j})\big\|_2 \Big) \\
		&= O\Big( \max_{j\in[d]} \big\|P_0^\epsilon (\epsilon_{t,j})\big\|_2 + \max_{j\in[d]} \max_{k\in[v]} \big\| P_0^b (B_{t,jk}) \big\|_2\Big),
	\end{align*}
	where the second last equality used Assumption (R2), and the last used $\big\|\R_{\beta} \big\|_\infty = O(d^{-2} \cdot d)= O(d^{-1})$ by (\ref{eqn: rate_inverse_expectation_XBBX}) and Assumption (R3).
	
	It remains to show $\balpha^\top I_2$ is indeed of order $T^{-1/2} d^{-(1-b)/2}$. To this end, we only need to show $s_2$ is of order $d^{-(1-b)}$. First, we have $\R_{\beta}\R_{\beta}^\top =\big[\b{E}(\X_t^\top \B_t) \b{E}(\B_t^\top \X_t)\big]^{-1}$ which has all eigenvalues of order $d^{-2}$ from (\ref{eqn: rate_inverse_expectation_XBBX}) and Assumption (R3). Consider any $j$-th diagonal element of $\bSigma_{\beta}$, we have
	\begin{align*}
		(\bSigma_{\beta})_{jj} &= \sum_{\tau} \b{E}\big\{(\B_t- \b{E}(\B_t))_{\cdot j}^\top \bepsilon_t \bepsilon_{t+\tau}^\top (\B_{t+\tau}- \b{E}(\B_t))_{\cdot j}\big\} \\
		&=
		\sum_{\tau} \tr\big\{\b{E}((\B_{t+\tau}- \b{E}(\B_t))_{\cdot j} (\B_{t+\tau}- \b{E}(\B_t))_{\cdot j}^\top) \, \b{E}(\bepsilon_t \bepsilon_{t+\tau}^\top) \big\} ,
	\end{align*}
	which is finite and has order exactly $d^{1+b}$ by Assumption (R8). Notice $\bSigma_{\beta}$ is $r\times r$, the order of eigenvalues of $\bSigma_{\beta}$ is hence exactly $d^{1+b}$. The order of $s_2$ is $d^{b-1}$ by
	\[
	\|\balpha \|_1^2 \lambda_\text{min}(\R_{\beta} \R_{\beta}^\top) \lambda_\text{min}( \bSigma_{\beta}) \leq s_2 \leq \|\balpha \|_1^2 \lambda_\text{max}(\R_{\beta} \R_{\beta}^\top) \lambda_\text{max}( \bSigma_{\beta}).
	\]
	This completes the proof of Lemma \ref{lemma: asymp_I2}.
	$\square$

	\subsection{Proofs of corollaries}
	
	\noindent\textbf{\textit{Proof of Corollary \ref{corollary: threshold_consistency}.}}
	This is direct from Theorem \ref{thm: ada_LASSO_phi_asymp}. $\square$

    \noindent\textbf{\textit{Proof of Corollary \ref{corollary: threshold_consistency_multiple}.}}
	This is direct from Theorem \ref{thm: ada_LASSO_phi_asymp}. $\square$
	
	\noindent\textbf{\textit{Proof of Corollary \ref{corollary: change_consistency}.}}
	This is direct from Theorem \ref{thm: ada_LASSO_phi_asymp}. $\square$

    \noindent\textbf{\textit{Proof of Corollary \ref{corollary: change_consistency_multiple}.}}
	This is direct from Theorem \ref{thm: ada_LASSO_phi_asymp}. $\square$

	\subsection{Proofs of theorems}
	
	\noindent\textbf{\textit{Proof of Theorem \ref{thm: LASSO_rate}.}}
	From (\ref{eqn: spatiallag_rewrite2}) and (\ref{eqn: profiled_beta_matrix}), we have
	\begin{align*}
		\bbeta(\bphi^\ast) &= \Big(\X^\top \B^\nu (\B^\nu)^\top \X \Big)^{-1} \X^\top \B^\nu (\B^\nu)^\top \Big(\y^\nu - \sum_{j=1}^p \sum_{k=0}^{l_j} \phi_{j,k}^\ast \W_j^\otimes \y_{j,k}^\nu \Big) \\
		&=
		\Big(\X^\top \B^\nu (\B^\nu)^\top \X \Big)^{-1} \X^\top \B^\nu (\B^\nu)^\top \Big(\1_T \otimes \bmu^\ast  + \X \bbeta^\ast + \bepsilon^\nu \Big) \\
		&=
		\bbeta^\ast + \Big(\X^\top \B^\nu (\B^\nu)^\top \X \Big)^{-1} \X^\top \B^\nu (\B^\nu)^\top \bepsilon^\nu .
	\end{align*}
	We now define a diagonal matrix $\D_{z_{j,k}} := \diag(z_{j,k,1}\I_d, \dots, z_{j,k,T}\I_d) \in \b{R}^{dT\times dT}$, with diagonal blocks $z_{j,k,1}\I_d, \dots, z_{j,k,T}\I_d$, and $\bPi^{\ast \otimes} := \big(\I_{Td} - \sum_{j=1}^p \sum_{k=0}^{l_j} \phi_{j,k}^\ast \W_j^\otimes \D_{z_{j,k}} \big)^{-1}$. We then have $\y^\nu = \bPi^{\ast \otimes}\big(\1_T \otimes \bmu^\ast + \X \bbeta^\ast + \bepsilon^\nu\big)$. Thus,
	\begin{align*}
		&\hspace{5mm}
		\bbeta(\wt{\bphi}) = \Big(\X^\top \B^\nu (\B^\nu)^\top \X \Big)^{-1} \X^\top \B^\nu (\B^\nu)^\top \Big(\y^\nu - \sum_{j=1}^p \sum_{k=0}^{l_j} \wt{\phi}_{j,k} \W_j^\otimes \y_{j,k}^\nu \Big) \\
		&=
		\bbeta(\bphi^\ast) + \Big(\X^\top \B^\nu (\B^\nu)^\top \X \Big)^{-1} \X^\top \B^\nu (\B^\nu)^\top \Big( \sum_{j=1}^p \sum_{k=0}^{l_j} (\phi_{j,k}^\ast - \wt{\phi}_{j,k}) \W_j^\otimes \y_{j,k}^\nu \Big) \\
		&=
		\bbeta(\bphi^\ast) + \Big(\X^\top \B^\nu (\B^\nu)^\top \X \Big)^{-1} \X^\top \B^\nu (\B^\nu)^\top \\
        &\hspace{60pt}
        \cdot \Big( \sum_{j=1}^p \sum_{k=0}^{l_j} (\phi_{j,k}^\ast - \wt{\phi}_{j,k}) \W_j^\otimes \D_{z_{j,k}} \bPi^{\ast \otimes}\big(\1_T \otimes \bmu^\ast +\X \bbeta^\ast + \bepsilon^\nu\big)\Big).
	\end{align*}
	We can hence decompose $\bbeta(\wt{\bphi}) - \bbeta^\ast = \sum_{j=1}^5 I_j$ where
	\begin{align*}
		& I_1 := \big[\b{E}(\X_t^\top \B_t) \b{E}(\B_t^\top \X_t)\big]^{-1} \big[\b{E}(\X_t^\top \B_t) \b{E}(\B_t^\top \X_t) - T^{-2} \X^\top \B^\nu (\B^\nu)^\top \X\big] (\bbeta(\wt{\bphi}) - \bbeta^\ast) ,\\
		& I_2 := \big[\b{E}(\X_t^\top \B_t) \b{E}(\B_t^\top \X_t)\big]^{-1} T^{-2} \X^\top \B^\nu (\B^\nu)^\top \bepsilon^\nu ,\\
		& I_3 := \big[\b{E}(\X_t^\top \B_t) \b{E}(\B_t^\top \X_t)\big]^{-1} T^{-2} \X^\top \B^\nu (\B^\nu)^\top \sum_{j=1}^p \sum_{k=0}^{l_j} (\phi_{j,k}^\ast - \wt{\phi}_{j,k}) \W_j^\otimes \D_{z_{j,k}} \bPi^{\ast \otimes} \X \bbeta^\ast ,\\
		& I_4 := \big[\b{E}(\X_t^\top \B_t) \b{E}(\B_t^\top \X_t)\big]^{-1} T^{-2} \X^\top \B^\nu (\B^\nu)^\top \sum_{j=1}^p \sum_{k=0}^{l_j} (\phi_{j,k}^\ast - \wt{\phi}_{j,k}) \W_j^\otimes \D_{z_{j,k}} \bPi^{\ast \otimes} \bepsilon^\nu , \\
		& I_5 := \big[\b{E}(\X_t^\top \B_t) \b{E}(\B_t^\top \X_t)\big]^{-1} T^{-2} \X^\top \B^\nu (\B^\nu)^\top \sum_{j=1}^p \sum_{k=0}^{l_j} (\phi_{j,k}^\ast - \wt{\phi}_{j,k}) \W_j^\otimes \D_{z_{j,k}} \bPi^{\ast \otimes} (\1_T \otimes \bmu^\ast) .
	\end{align*}
	Notice we can take any $t\in[T]$ for $\b{E}(\X_t^\top \B_t)$ and $\b{E}(\B_t^\top \X_t)$ due to Assumption (M1). To bound $I_1$ to $I_4$, we first have
	\begin{equation}
		\label{eqn: rate_inverse_expectation_XBBX}
		\big\| \big[\b{E}(\X_t^\top \B_t) \b{E}(\B_t^\top \X_t)\big]^{-1} \big\|_1 \leq \frac{r^{1/2}}{\lambda_{r} \big[\b{E}(\X_t^\top \B_t) \b{E}(\B_t^\top \X_t)\big]} \leq \frac{r^{1/2}}{d^2 u^2} ,
	\end{equation}
	where $\lambda_{r} \big[\b{E}(\X_t^\top \B_t) \b{E}(\B_t^\top \X_t)\big] = \sigma_r^2(\b{E}(\X_t^\top \B_t)) \geq d^2 u^2$ with $u>0$ being a constant by Assumption (R3), and $\lambda_{j}(\cdot)$ and $\sigma_j(\cdot)$ denoting the $j$-th largest eigenvalue and singular value of a matrix respectively.
	
	Next, define
	\begin{equation*}
		\U = \I_d \otimes T^{-1}\sum_{t=1}^T \vec{\B_t - \bar{\B}} \x_t^\top,
		\;\;\;\;\;
		\U_0 = \I_d \otimes \b{E}(\bb_t\x_t^\top).
	\end{equation*}
	By Lemma \ref{lemma: V_H_kronecker}, we then have
	\begin{align*}
		T^{-1}\X^\top \B^v &= T^{-1} \sum_{t=1}^T \X_t^\top (\B_t - \bar{\B}) \\
		&= T^{-1} \sum_{t=1}^T \Big\{\big( \I_d \otimes \x_t^\top \big) \V_{\I_d, r}\Big\}^\top \Big\{\big( \I_d \otimes \vec{\B_t - \bar{\B}}^\top \big) \V_{\I_d, v}\Big\} \\
		&= T^{-1} \sum_{t=1}^T \V_{\I_d, r}^\top \big( \I_d \otimes \x_t \big)  \big( \I_d \otimes \vec{\B_t - \bar{\B}}^\top \big) \V_{\I_d, v}
		= \V_{\I_d, r}^\top \U^\top \V_{\I_d, v} .
	\end{align*}
	Similarly we have $\b{E}(\X_t^\top \B_t) = \V_{\I_d, r}^\top \U_0^\top \V_{\I_d, v}$. Thus on the set $\cM$ in Lemma \ref{lemma: LamSouza_theorem_S1}, with (\ref{eqn: rate_inverse_expectation_XBBX}) it holds that
	\begin{equation}
		\label{eqn: rate_I1}
		\begin{split}
            &\hspace{12pt}
			\big\|I_1 \big\|_1 \leq \frac{r^{1/2}}{d^2 u^2}
			\cdot \big\| \V_{\I_d, r}^\top \U_0^\top \V_{\I_d, v} \V_{\I_d, v}^\top \U_0
			- \V_{\I_d, r}^\top \U^\top \V_{\I_d, v} \V_{\I_d, v}^\top \U \big\|_1 \cdot \big\|\V_{\I_d, r}(\bbeta(\wt{\bphi}) - \bbeta^\ast) \big\|_1 \\
			&=
			O\Big(\frac{1}{d}\Big) \Big(\big\| \V_{\I_d, r}^\top (\U_0 - \U)^\top \V_{\I_d, v} \V_{\I_d, v}^\top \U_0 \big\|_1 + \big\|\V_{\I_d, r}^\top \U^\top \V_{\I_d, v} \V_{\I_d, v}^\top (\U_0 - \U) \big\|_1 \Big)\, \big\|\bbeta(\wt{\bphi}) - \bbeta^\ast \big\|_1 \\
			&=
			O\Big(\frac{1}{d}\Big) \Big\{ d \cdot \big\|\U_0 - \U \big\|_\text{max}\cdot \big\|\U_0 \big\|_\text{max} \\
			&\hspace{12pt}
			+ d \cdot \big\|\U_0 - \U \big\|_\text{max}\cdot \Big(\big\|\U_0 - \U\big\|_\text{max} + \big\|\U_0 \big\|_\text{max} \Big) \Big\}\cdot \big\|\bbeta(\wt{\bphi}) - \bbeta^\ast \big\|_1
			=
			O\Big(c_T \big\|\bbeta(\wt{\bphi}) - \bbeta^\ast \big\|_1 \Big) ,
		\end{split}
	\end{equation}
	where the last equality used Assumption (R3) which implies $\big\|\U_0 \big\|_\text{max}$ is bounded by some constant, and Lemma \ref{lemma: LamSouza_theorem_S1} (using $\cA_1, \, \cA_4, \, \cA_6$) that
	\begin{equation}
		\label{eqn: rate_diff_U_U0}
		\begin{split}
			&\hspace{12pt}
			\big\|\U - \U_0\big\|_\text{max} =
			\bigg\|\frac{1}{T} \sum_{t=1}^T \big[ \vec{\B_t - \bar{\B}} \x_t^\top \big] - \b{E}(\bb_t\x_t^\top)
			\bigg\|_\text{max} \\
			&=
			\bigg\|\frac{1}{T} \sum_{t=1}^T \bb_t \x_t^\top - \b{E}(\bb_t\x_t^\top) - \vec{\bar{\B}}\frac{1}{T} \sum_{t=1}^T \x_t^\top \bigg\|_\text{max}
			\leq
			c_T + \bigg\|\vec{\bar{\B}}\frac{1}{T} \sum_{t=1}^T \x_t^\top \bigg\|_\text{max} \\
			&\leq
			c_T + \Big(\big\|\vec{\bar{\B}} -  \b{E}[\B_t]\big\|_\text{max} + \big\| \b{E}[\B_t] \big\|_\text{max}\Big)\bigg\| \frac{1}{T} \sum_{t=1}^T \x_t^\top \bigg\|_\text{max}
			\leq c_T + c_T(c_T + \mu_{b,\text{max}}) .
		\end{split}
	\end{equation}

	Similarly for $I_2$, we have on the set $\cM$ that
	\begin{equation}
		\label{eqn: rate_I2}
		\begin{split}
			&\hspace{12pt}
			\big\|I_2 \big\|_1 \leq \frac{r^{1/2}}{d^2 u^2}
			\cdot \big\| \V_{\I_d, r}^\top \U^\top \V_{\I_d, v} \big\|_1 \cdot \big\|T^{-1} (\B^\nu)^\top \bepsilon^\nu \big\|_1 \\
			&=
			O\Big(\frac{1}{d}\Big) \big\|T^{-1} (\B^\nu)^\top \bepsilon^\nu \big\|_1
			= O\Big(\frac{1}{d}\Big) \bigg\| \frac{1}{T} \sum_{t=1}^T (\B_t^\top - \bar{\B}^\top)\bepsilon_t \bigg\|_1 \\
			&= O_p\Big(\frac{1}{d}\Big) \Big(c_T d^{\frac{1}{2} + \frac{1}{2w}} + c_T d^{1/2} \log^{1/2}(T \vee d) S_\epsilon (\mu_{b,\text{max}} + c_T) \Big)
			= O_p\big(c_T d^{-\frac{1}{2} + \frac{1}{2w}}\big),
		\end{split}
	\end{equation}
	where the first equality used $\cA_1, \, \cA_4, \, \cA_6$ in $\cM$, the third used $\cA_3, \, \cA_7$ in $\cM$, and the last used Assumption (R10).

	For $I_3$, first recall that
	\begin{equation}
		\label{eqn: def_bPi_ast}
		\bPi_t^\ast = (\I_d - \W_t^\ast)^{-1} = \big(\I_{d} - \sum_{j=1}^p \sum_{k=0}^{l_j} \phi_{j,k}^\ast z_{j,k,t} \W_j \big)^{-1} ,
	\end{equation}
	and hence from Assumption (M2) (resp. (M2')) we have $\|\bPi_t^\ast\|_\infty \leq 1/(1-\eta) = O(1)$ (resp. $\|\bPi_t^\ast\|_\infty  = O_P(1)$) using that $(\I_d - \W_t^\ast)$ is strictly diagonally dominant. Then by (\ref{eqn: rate_inverse_expectation_XBBX}), we have on the set $\cM$ that
	\begin{equation}
		\label{eqn: rate_I3}
		\begin{split}
			&\hspace{12pt}
			\big\|I_3 \big\|_1 \leq \frac{r^{1/2}}{d^2 u^2}
			\cdot \big\| \V_{\I_d, r}^\top \U^\top \V_{\I_d, v} \big\|_1 \cdot \bigg\|T^{-1} (\B^\nu)^\top \sum_{j=1}^p \sum_{k=0}^{l_j} (\phi_{j,k}^\ast - \wt{\phi}_{j,k}) \W_j^\otimes \D_{z_{j,k}} \bPi^{\ast \otimes} \X \bigg\|_1 \cdot \big\|\bbeta^\ast \big\|_1 \\
			&=
			O\Big(\frac{1}{d}\Big) \bigg\| \frac{1}{T} \sum_{j=1}^p \sum_{k=0}^{l_j} (\phi_{j,k}^\ast - \wt{\phi}_{j,k}) (\B^\nu)^\top \W_j^\otimes \D_{z_{j,k}} \bPi^{\ast \otimes} \X \bigg\|_1 \\
			&=
			O\Big(\frac{1}{d}\Big) \bigg\|\frac{1}{T} \sum_{j=1}^p \sum_{k=0}^{l_j} (\phi_{j,k}^\ast - \wt{\phi}_{j,k}) \sum_{t=1}^T z_{j,k,t} (\B_t^\top - \bar{\B}^\top) \W_j \bPi_t^\ast \X_t \bigg\|_1 \\
			&=
			O\Big(\frac{1}{d}\Big) \bigg\{\max_{q\in[r]} \max_{s\in[v]} \Big| \frac{1}{T} \sum_{j=1}^p \sum_{k=0}^{l_j} (\phi_{j,k}^\ast - \wt{\phi}_{j,k}) \sum_{t=1}^T \sum_{i=1}^d z_{j,k,t} \W_{j,\cdot i}^\top (\B_{t,\cdot s} - \bar{\B}_{\cdot s}) \X_{t,\cdot q}^\top \bPi_{t,i\cdot}^\ast \Big| \bigg\} \\
			&=
			O\Big(\frac{1}{d}\Big) \Big(\big\| \bphi^\ast - \wt\bphi \big\|_1 \\
            &\hspace{40pt}
            \cdot \max_{q\in[r], s\in[v], j\in[p]} \max_{k\in[l_j] \cup\{0\}} \max_{m,n\in[d]} \Big|\frac{1}{T} \sum_{t=1}^T z_{j,k,t} (B_{t,ms} - \bar{B}_{ms}) X_{t,nq} \Big| \sum_{i=1}^d \|\W_{j,\cdot i}\|_1  \|\bPi_{t,i\cdot}^\ast\|_1 \Big) \\
			&=
			O_p\Big(\big\|\bphi^\ast - \wt\bphi \big\|_1 \big[c_T + 1 + c_T(c_T + \mu_{b,\text{max}})\big] \Big) =
			O_p\Big(\big\|\bphi^\ast - \wt\bphi \big\|_1 \Big),
		\end{split}
	\end{equation}
	where the first equality used $\cA_1, \, \cA_4, \, \cA_6$ in $\cM$, the second last used $\cA_4, \, \cA_8, \, \cA_9$ in $\cM$ and Assumptions (R1) and (R3). Similarly, for $I_4$ on the set $\cM$,
	\begin{equation}
		\label{eqn: rate_I4}
		\begin{split}
			&\hspace{12pt}
			\big\|I_4 \big\|_1 \leq \frac{r^{1/2}}{d^2 u^2}
			\cdot \big\| \V_{\I_d, r}^\top \U^\top \V_{\I_d, v} \big\|_1 \cdot \big\|T^{-1} (\B^\nu)^\top \sum_{j=1}^p \sum_{k=0}^{l_j} (\phi_{j,k}^\ast - \wt{\phi}_{j,k}) \W_j^\otimes \D_{z_{j,k}} \bPi^{\ast \otimes} \bepsilon^\nu \big\|_1 \\
			&=
			O\Big(\frac{1}{d}\Big) \bigg\{\max_{s\in[v]} \Big| \frac{1}{T} \sum_{j=1}^p \sum_{k=0}^{l_j} (\phi_{j,k}^\ast - \wt{\phi}_{j,k}) \sum_{t=1}^T \sum_{i=1}^d z_{j,k,t} \W_{j,\cdot i}^\top (\B_{t,\cdot s} - \bar{\B}_{\cdot s}) \bepsilon_t^\top \bPi_{t,i\cdot}^\ast \Big| \bigg\} \\
			&=
			O\Big(\frac{1}{d}\Big) \bigg\{\big\|\bphi^\ast - \wt\bphi \big\|_1 \\
            &\hspace{50pt}
            \cdot \max_{s\in[v]} \max_{j\in[p]} \max_{k\in[l_j] \cup\{0\}} \max_{m,q\in[d]} \Big|\frac{1}{T} \sum_{t=1}^T z_{j,k,t} (B_{t,ms} - \bar{B}_{ms}) \epsilon_{t,q} \Big| \, \sum_{i=1}^d \|\W_{j,\cdot i}\|_1  \|\bPi_{t,i\cdot}^\ast\|_1 \bigg\} \\
			&=
			O_p\Big(\big\|\bphi^\ast - \wt\bphi \big\|_1 \big[c_T + c_T(c_T + \mu_{b,\text{max}})\big] \Big) =
			O_p\Big( c_T \big\|\bphi^\ast - \wt\bphi \big\|_1 \Big),
		\end{split}
	\end{equation}
	where the first equality used $\cA_1, \, \cA_4, \, \cA_6$ in $\cM$, the second last used $\cA_4, \, \cA_{10}, \, \cA_{11}$ in $\cM$ and Assumptions (R1) and (R3). For $I_5$ we also have on the set $\cM$,
	\begin{equation}
		\label{eqn: rate_I5}
		\begin{split}
			&\hspace{12pt}
			\big\|I_5 \big\|_1 \leq \frac{r^{1/2}}{d^2 u^2}
			\cdot \big\| \V_{\I_d, r}^\top \U^\top \V_{\I_d, v} \big\|_1 \cdot \big\|T^{-1} (\B^\nu)^\top \sum_{j=1}^p \sum_{k=0}^{l_j} (\phi_{j,k}^\ast - \wt{\phi}_{j,k}) \W_j^\otimes \D_{z_{j,k}} \bPi^{\ast \otimes} (\1_T \otimes \bmu^\ast) \big\|_1 \\
			&=
			O\Big(\frac{1}{d}\Big) \bigg\{\max_{s\in[v]} \Big| \frac{1}{T} \sum_{j=1}^p \sum_{k=0}^{l_j} (\phi_{j,k}^\ast - \wt{\phi}_{j,k}) \sum_{t=1}^T \sum_{i=1}^d z_{j,k,t} \W_{j,\cdot i}^\top (\B_{t,\cdot s} - \bar{\B}_{\cdot s}) \bmu^{\ast\top} \bPi_{t,i\cdot}^\ast \Big| \bigg\} \\
			&=
			O\Big(\frac{1}{d}\Big) \bigg\{\big\|\bphi^\ast - \wt\bphi \big\|_1 \\
            &\hspace{40pt}
            \cdot \max_{s\in[v]} \max_{j\in[p]} \max_{k\in[l_j] \cup\{0\}} \max_{m\in[d]} \Big|\frac{1}{T} \sum_{t=1}^T z_{j,k,t} (B_{t,ms} - \bar{B}_{ms}) \Big| \, \|\bmu^\ast\|_\text{max} \, \sum_{i=1}^d \|\W_{j,\cdot i}\|_1  \|\bPi_{t,i\cdot}^\ast\|_1 \bigg\} \\
			&=
			O_p\Big(\big\|\bphi^\ast - \wt\bphi \big\|_1 \big[c_T + (z_\text{max} c_T \vee c_T) \big]\Big) =
			O_p\Big( c_T \big\|\bphi^\ast - \wt\bphi \big\|_1 \Big),
		\end{split}
	\end{equation}
	where the first equality used $\cA_1, \, \cA_4, \, \cA_6$ in $\cM$, the second last used $\cA_4, \, \cA_{12}, \, \cA_{13}$ in $\cM$ and Assumptions (R1),
	(R3), and $\b{E}(z_{j,k,t} \B_t) =\0$ from (M2') if (M2) is not satisfied.

	From (\ref{eqn: rate_I1}), (\ref{eqn: rate_I2}), (\ref{eqn: rate_I3}) and (\ref{eqn: rate_I4}), combining with Lemma \ref{lemma: LamSouza_theorem_S1}, we have
	\begin{equation}
		\label{eqn: beta_consistency}
		\big\|\bbeta(\wt{\bphi}) - \bbeta^\ast\big\|_1 \leq \sum_{j=1}^5 \big\|I_j \big\|_1 =
		O_p\Big( \big\|\wt{\bphi} - \bphi^\ast \big\|_1 + c_T d^{-\frac{1}{2} + \frac{1}{2w}} \Big).
	\end{equation}

	For the remaining of the proof of \ref{thm: LASSO_rate}, we work on the rate for $\big\|\wt\bphi - \bphi^\ast \big\|_1$. From (\ref{eqn: simplify_phi_ls_matrix}), we have
	\begin{align*}
		\wt{\bphi} &= \big[( \B^\top\V - \bXi \Y_W )^\top ( \B^\top\V - \bXi \Y_W ) \big]^{-1} ( \B^\top\V - \bXi \Y_W )^\top (\B^\top\y - \bXi \y^\nu) \\
		&=
		\big[( \B^\top\V - \bXi \Y_W )^\top ( \B^\top\V - \bXi \Y_W ) \big]^{-1} ( \B^\top\V - \bXi \Y_W )^\top (\B^\top\X_{\bbeta^\ast} \Vec{\I_d} + \B^\top\bepsilon - \bXi \y^\nu) \\
		&\hspace{12pt}
		+ \big[( \B^\top\V - \bXi \Y_W )^\top ( \B^\top\V - \bXi \Y_W ) \big]^{-1} ( \B^\top\V - \bXi \Y_W )^\top (\bXi \Y_W) \bphi^\ast + \bphi^\ast \\
		&=
		\bphi^\ast + \big[( \B^\top\V - \bXi \Y_W )^\top ( \B^\top\V - \bXi \Y_W ) \big]^{-1} ( \B^\top\V - \bXi \Y_W )^\top \B^\top\bepsilon \\
		&\hspace{12pt}
		- T^{-1/2}d^{-a/2} \cdot \big[( \B^\top\V - \bXi \Y_W )^\top ( \B^\top\V - \bXi \Y_W ) \big]^{-1} ( \B^\top\V - \bXi \Y_W )^\top \\
		&\hspace{12pt}
		\cdot \text{vec}\bigg\{\sum_{t=1}^T (\B_t - \bar{\B}) \bgamma (\bepsilon^\nu)^\top \B^\nu (\B^\nu)^\top \X  (\X^\top \B^\nu (\B^\nu)^\top \X)^{-1} \X_t^\top \bigg\} ,
	\end{align*}
	where the second equality used (\ref{eqn: spatiallag_rewrite1_augmented}), and the third used the fact that $\B^\top\X_{\bbeta(\bphi)} \Vec{\I_d} = \bXi\y^\nu - \bXi\Y_W\bphi$ from (\ref{eqn: simplify_LASSO}) and $\bbeta(\bphi^\ast) = \bbeta^\ast + \big(\X^\top \B^\nu (\B^\nu)^\top \X \big)^{-1} \X^\top \B^\nu (\B^\nu)^\top \bepsilon^\nu$. Thus, we may decompose
	\begin{align*}
		\wt\bphi - \bphi^\ast &= D_1 - D_2,
		\,\,\, \text{where}\\
		D_1 &= \big[( \B^\top\V - \bXi \Y_W )^\top ( \B^\top\V - \bXi \Y_W ) \big]^{-1} ( \B^\top\V - \bXi \Y_W )^\top \B^\top\bepsilon , \\
		D_2	&= T^{-1/2}d^{-a/2} \cdot \big[( \B^\top\V - \bXi \Y_W )^\top ( \B^\top\V - \bXi \Y_W ) \big]^{-1} ( \B^\top\V - \bXi \Y_W )^\top \\
		&\hspace{12pt} \cdot \bigg\{\sum_{t=1}^T \Big(\I_d \otimes (\B_t - \bar{\B}) \bgamma \Big) \X_t \bigg\} (\X^\top \B^\nu (\B^\nu)^\top \X)^{-1} \X^\top \B^\nu (\B^\nu)^\top \bepsilon^\nu .
	\end{align*}

	To bound the above, recall first the following definitions (from the statement of Theorem \ref{thm: ada_LASSO_phi_asymp}) for $j\in[p], k\in[l_j]\cup\{0\}$,
	\begin{align*}
		&\U_{\x, j,k} := \frac{1}{T}\sum_{t=1}^T z_{j,k,t} \vec{\B_t -\bar{\B}} \x_t^\top , \,\,\,
		\U_{\bmu, j,k} := \frac{1}{T}\sum_{t=1}^T z_{j,k,t} \vec{\B_t -\bar{\B}} \bmu^{\ast \top} , \\
		&\U_{\bepsilon, j,k} := \frac{1}{T}\sum_{t=1}^T z_{j,k,t} \vec{\B_t -\bar{\B}} \bepsilon_t^\top .
	\end{align*}

	Consider $\bXi\Y_W$, from its definition we have,
	\begin{align*}
		\bXi\Y_W &=
		T^{-1/2}d^{-a/2} \Big(\sum_{t=1}^T \X_t \otimes (\B_t - \bar{\B}) \bgamma \Big) \big[\X^\top \B^\nu (\B^\nu)^\top \X \big]^{-1} \X^\top \B^\nu (\B^\nu)^\top \\
		&\hspace{12pt}
		\cdot (\W_1^\otimes \y_{1,0}^\nu, \dots, \W_1^\otimes \y_{1,l_1}^\nu, \dots, \W_p^\otimes \y_{p,0}^\nu, \dots, \W_p^\otimes \y_{p,l_p}^\nu) \\
		&=
		T^{-1/2}d^{-a/2} \Big(\sum_{t=1}^T \X_t \otimes (\B_t - \bar{\B}) \bgamma \Big) \big[\X^\top \B^\nu (\B^\nu)^\top \X \big]^{-1} \X^\top \B^\nu \\
		&\hspace{12pt}
		\cdot \Bigg\{\sum_{t=1}^T z_{1,0,t}(\B_t - \bar{\B})^\top \W_1 \y_t, \cdots, \sum_{t=1}^T z_{p,l_p,t}(\B_t - \bar{\B})^\top \W_p \y_t \Bigg\}
	\end{align*}
	From (\ref{eqn: def_bPi_ast}), we have $\y_t = \bPi_t^\ast \bmu^\ast + \bPi_t^\ast \X_t\bbeta^\ast +\bPi_t^\ast \bepsilon_t$. It hence holds for any $j\in[p], k\in[l_j] \cup\{0\}$ by Lemma \ref{lemma: V_H_kronecker} that
	\begin{equation}
		\label{eqn: rewrite_XiYW}
		\begin{split}
			&\hspace{12pt}
			\frac{1}{T}\sum_{t=1}^T z_{j,k,t}(\B_t - \bar{\B})^\top \W_j \y_t \\
            &= \frac{1}{T}\sum_{t=1}^T z_{j,k,t} \V_{\W_j^\top, v}^\top \big(\I_d \otimes \vec{\B_t -\bar{\B}} \big) (\bPi_t^\ast \bmu^\ast + \bPi_t^\ast \X_t\bbeta^\ast +\bPi_t^\ast \bepsilon_t) \\
			&=
			\V_{\W_j^\top, v}^\top (\I_d\otimes \U_{\x, j,k}) \V_{\bPi_t^\ast, r} \bbeta^\ast +
			\V_{\W_j^\top, v}^\top (\I_d\otimes \U_{\bmu, j,k}) \vec{\bPi_t^{\ast \top}} +
			\V_{\W_j^\top, v}^\top (\I_d\otimes \U_{\bepsilon, j,k}) \vec{\bPi_t^{\ast \top}} .
		\end{split}
	\end{equation}

	Consider also $\B^\top \V$, we have
	\begin{align*}
		&\hspace{12pt}
		\B^\top \V =
		T^{-1/2}d^{-a/2} \Big(\I_d \otimes \big\{(\B_1 - \bar{\B}, \dots, \B_T - \bar{\B}) (\I_T \otimes \bgamma )\big\} \Big) \\
		&\hspace{12pt}
		\cdot \Bigg\{ \Big[\I_d \otimes (z_{1,0,1}\y_1, \dots, z_{1,0,T}\y_T)^\top \Big] \vec{\W_1^\top}, \dots, \Big[\I_d \otimes (z_{1,l_1,1}\y_1, \dots, z_{1,l_1,T}\y_T)^\top \Big] \Vec{\W_1^\top}, \\
		&\hspace{12pt}
		\dots, \Big[\I_d \otimes (z_{p,0,1}\y_1, \dots, z_{p,0,T}\y_T)^\top \Big] \vec{\W_p^\top}, \dots, \Big[\I_d \otimes (z_{p,l_p,1}\y_1, \dots, z_{p,l_p,T}\y_T)^\top \Big] \vec{\W_p^\top} \Bigg\} \\
		&=
		T^{-1/2}d^{-a/2}
		\Bigg\{\Big[\I_d \otimes (\B_1 - \bar{\B}, \dots, \B_T - \bar{\B}) (\I_T \otimes \bgamma ) (z_{1,0,1}\y_1, \dots, z_{1,0,T}\y_T)^\top \Big]\vec{\W_1^\top}, \\
		&\hspace{12pt}
		\dots, \Big[\I_d \otimes (\B_1 - \bar{\B}, \dots, \B_T - \bar{\B}) (\I_T \otimes \bgamma ) (z_{p,l_p,1}\y_1, \dots, z_{p,l_p,T}\y_T)^\top \Big]\vec{\W_p^\top} \Bigg\} \\
		&=
		T^{-1/2}d^{-a/2}
		\Bigg\{\Big[\I_d \otimes \sum_{t=1}^T z_{1,0,t} (\B_t-\bar{\B}) \bgamma \y_t^\top \Big]\Vec{\W_1^\top}, \\
        &\hspace{80pt}
        \dots, \Big[\I_d \otimes \sum_{t=1}^T z_{p,l_p,t} (\B_t-\bar{\B}) \bgamma \y_t^\top \Big]\vec{\W_p^\top} \Bigg\} .
	\end{align*}
	Similar to $\bXi \Y_W$, for $j\in[p], k\in[l_p] \cup\{0\}$ we have
	\begin{equation}
		\label{eqn: rewrite_BV}
		\begin{split}
			&\hspace{12pt}
			\frac{1}{T} \sum_{t=1}^T z_{j,k,t} (\B_t-\bar{\B}) \bgamma \y_t^\top = \frac{1}{T} \sum_{t=1}^T z_{j,k,t} (\B_t-\bar{\B}) \bgamma (\bmu^{\ast \top}\bPi_t^{\ast \top} + \bbeta^{\ast \top} \X_t^\top \bPi_t^{\ast \top} +\bepsilon_t^\top \bPi_t^{\ast \top}) \\
			&=
			\frac{1}{T} \sum_{t=1}^T z_{j,k,t}(\bgamma^\top \otimes \I_d) \, \vec{\B_t-\bar{\B}} \bmu^{\ast \top}\bPi_t^{\ast \top} + \frac{1}{T} \sum_{t=1}^T z_{j,k,t}(\bgamma^\top \otimes \I_d) \, \vec{\B_t-\bar{\B}} \x_t^\top (\bbeta^\ast \otimes \I_d) \bPi_t^{\ast \top} \\
			&\hspace{12pt}
			+ \frac{1}{T} \sum_{t=1}^T z_{j,k,t}(\bgamma^\top \otimes \I_d) \, \vec{\B_t-\bar{\B}} \bepsilon_t^\top \bPi_t^{\ast \top} \\
			&=
			(\bgamma^\top \otimes \I_d)\U_{\bmu, j,k} \bPi_t^{\ast \top} + (\bgamma^\top \otimes \I_d)\U_{\x, j,k} (\bbeta^\ast \otimes \I_d) \bPi_t^{\ast \top} + (\bgamma^\top \otimes \I_d) \U_{\bepsilon, j,k} \bPi_t^{\ast \top} .
		\end{split}
	\end{equation}

	With (\ref{eqn: rewrite_XiYW}) and (\ref{eqn: rewrite_BV}), recall the following in the statement of Theorem \ref{thm: ada_LASSO_phi_asymp},
	\begin{align*}
		\H_{10} &= \Big\{\Big(\I_d \otimes  (\bgamma^\top \otimes \I_d) \b{E}\big(\U_{\x, 1,0} (\bbeta^\ast \otimes \I_d) \bPi_t^{\ast \top} \big)\Big) \textnormal{vec}(\W_1^\top) \\
        &\hspace{24pt}
		+ \Big(\I_d \otimes (\bgamma^\top \otimes \I_d) \b{E}\big( \U_{\bmu, 1,0} \bPi_t^{\ast \top} \big) \Big) \textnormal{vec}(\W_1^\top), \\
		&\hspace{12pt}
		\dots, \Big(\I_d \otimes (\bgamma^\top \otimes \I_d) \b{E}\big( \U_{\x, p,l_p} (\bbeta^\ast \otimes \I_d) \bPi_t^{\ast \top} \big)\Big) \textnormal{vec}(\W_p^\top) \\
		&\hspace{24pt}
        + \Big(\I_d \otimes (\bgamma^\top \otimes \I_d) \b{E}\big( \U_{\bmu, p,l_p} \bPi_t^{\ast \top} \big) \Big) \textnormal{vec}(\W_p^\top) \Big\} , \\
		\H_{20} &= \b{E}(\X_t \otimes \B_t\bgamma) \big[\b{E}(\X_t^\top \B_t) \b{E}(\B_t^\top \X_t)\big]^{-1} \b{E}(\X_t^\top \B_t) \\
		&\hspace{12pt}
		\cdot \Big\{\V_{\W_1^\top, v}^\top \b{E}\big[( \I_d \otimes \U_{\x, 1,0}) \V_{\bPi_t^\ast, r} \big] \bbeta^\ast + \V_{\W_1^\top, v}^\top \b{E}\big[ (\I_d \otimes \U_{\bmu, 1,0}) \textnormal{vec}(\bPi_t^{\ast \top}) \big], \\
		&\hspace{12pt}
		\dots, \V_{\W_p^\top, v}^\top  \b{E}\big[ (\I_d \otimes \U_{\x, p,l_p}) \V_{\bPi_t^\ast, r} \big] \bbeta^\ast + \V_{\W_p^\top, v}^\top \b{E}\big[ (\I_d \otimes \U_{\bmu, p,l_p}) \textnormal{vec}(\bPi_t^{\ast \top}) \big] \Big\} ,
	\end{align*}
	which are essentially $T^{-1/2}d^{a/2} \B^\top\V$ and $T^{-1/2}d^{a/2} \bXi\Y_W$ at the population level, respectively.
	
	For the rest of the proof of Theorem \ref{thm: LASSO_rate}, we find the rate of $D_2$, followed by constructing the asymptotic normality of the dominating term in the expansion of $D_1$. For $D_2$, we further decompose $D_2 = F_1 +F_2 -F_3$ where
	\begin{align*}
		F_1 &= [(\H_{20} -\H_{10})^\top (\H_{20} -\H_{10})]^{-1} \Big\{(\H_{20} -\H_{10})^\top (\H_{20} -\H_{10}) \\
		&\hspace{12pt}
		- T^{-1}d^a (\B^\top\V -\bXi \Y_W )^\top (\B^\top\V - \bXi\Y_W) \Big\}D_2, \\
		F_2 &= [(\H_{20} -\H_{10})^\top (\H_{20} -\H_{10})]^{-1} \Big[(T^{-1/2}d^{a/2} \B^\top\V -\H_{10}) -(T^{-1/2}d^{a/2} \bXi \Y_W -\H_{20}) \Big]^\top \\
		&\hspace{12pt}
		\cdot \bigg\{\frac{1}{T} \sum_{t=1}^T \Big(\I_d \otimes (\B_t - \bar{\B}) \bgamma \Big) \X_t \bigg\} (\X^\top \B^\nu (\B^\nu)^\top \X)^{-1} \X^\top \B^\nu (\B^\nu)^\top \bepsilon^\nu ,\\
		F_3 &= [(\H_{20} -\H_{10})^\top (\H_{20} -\H_{10})]^{-1} (\H_{20} -\H_{10})^\top \\
		&\hspace{12pt}
		\cdot \bigg\{\frac{1}{T} \sum_{t=1}^T \Big(\I_d \otimes (\B_t - \bar{\B}) \bgamma \Big) \X_t \bigg\} (\X^\top \B^\nu (\B^\nu)^\top \X)^{-1} \X^\top \B^\nu (\B^\nu)^\top \bepsilon^\nu .
	\end{align*}
	To bound the L1 norm of $F_1$ to $F_3$, first observe that by Assumptions (R1) and (R4) we have
	\begin{align*}
		\sigma^2_L(\H_{10}) &\geq \sigma^2_L\Bigg(
		\Big\{\Big(\I_d \otimes  (\bgamma^\top \otimes \I_d) \, \b{E}\big( \U_{\x, 1,0} (\bbeta^\ast \otimes \I_d) \bPi_t^{\ast \top} \big) \Big) \Vec{\W_1^\top}, \\
		&\hspace{12pt}
		\dots, \Big(\I_d \otimes (\bgamma^\top \otimes \I_d) \, \b{E}\big( \U_{\x, p,l_p} (\bbeta^\ast \otimes \I_d) \bPi_t^{\ast \top} \big)\Big) \Vec{\W_p^\top} \Big\}
		\Bigg) \\
		&\geq
		\sigma_L^2(\D_W) \cdot \sigma^2_{d^2} \Bigg(
		\Big\{\Big(\I_d \otimes  (\bgamma^\top \otimes \I_d) \, \b{E}\big( \U_{\x, 1,0} (\bbeta^\ast \otimes \I_d) \bPi_t^{\ast \top} \big)\Big), \\
		&\hspace{12pt}
		\dots, \Big(\I_d \otimes (\bgamma^\top \otimes \I_d) \, \b{E}\big( \U_{\x, p,l_p} (\bbeta^\ast \otimes \I_d) \bPi_t^{\ast \top} \big)\Big) \Big\}
		\Bigg)
		\geq C d \cdot d^a = C d^{1+a},
	\end{align*}
	where $C>0$ is a generic constant. Similarly, by Assumptions (R1), (R3), (R5) and (R6),
	\begin{align*}
		\sigma_L(\H_{20}) &\geq \sigma_r\Big( \b{E}(\X_t \otimes \B_t\bgamma) \Big) \,\sigma_r\Big( \big[\b{E}(\X_t^\top \B_t) \b{E}(\B_t^\top \X_t)\big]^{-1} \Big) \, \sigma_r\Big( \b{E}(\X_t^\top \B_t)\Big) \\
		&\hspace{12pt}
		\cdot \sigma_\text{min}\Big( \Big\{\V_{\W_1^\top, r}^\top \b{E}\big[ (\I_d \otimes \U_{\x, 1,0}) \V_{\bPi_t^\ast, r} \big] \bbeta^\ast, \dots, \V_{\W_p^\top, r}^\top \b{E}\big[ (\I_d \otimes \U_{\x, p,l_p}) \V_{\bPi_t^\ast, r}\big] \bbeta^\ast \Big\} \Big) \\
		&\geq
		\frac{C d^{1+a} \cdot d}{\lambda_\text{max}[\b{E}(\X_t^\top \B_t) \b{E}(\B_t^\top \X_t)]} \cdot \sigma_v[\b{E}(\ddot\G)] \cdot \sigma_L(\D_W) \\
        &\geq
        \frac{C d^{1+a} \cdot d\cdot d^{1/2}}{\lambda_\text{max}[\b{E}(\X_t^\top \B_t) \b{E}(\B_t^\top \X_t)]}
		\geq C d^{1/2+a},
	\end{align*}
	with some arbitrary constant $C>0$. Notice $\H_{20}$ has the smallest singular value of order larger then that for $\H_{10}$, so $\sigma_L^2(\H_{20} - \H_{10}) \geq C d^{1+a}$ for some $C>0$. Thus,
	\begin{equation}
		\label{eqn: bound_inv_H2010}
		\Big\| [(\H_{20} -\H_{10})^\top (\H_{20} -\H_{10})]^{-1} \Big\|_1 \leq \frac{L^{1/2}}{\lambda_\text{min}[(\H_{20} -\H_{10})^\top (\H_{20} -\H_{10})]} \leq \frac{L^{1/2}}{C d^{1+a}} .
	\end{equation}

	Consider $F_1$ first and we hence have on $\cM$,
	\begin{equation}
		\label{eqn: rate_F1}
		\begin{split}
			\|F_1\|_1 &\leq \frac{L^{1/2} \cdot L}{C d^{1+a}}\Bigg\{ \Big\|(\H_{20} - T^{-1/2}d^{a/2} \bXi\Y_W) + (T^{-1/2}d^{a/2} \B^\top\V -\H_{10})\Big\|_\text{max} \cdot \big\|\H_{20} -\H_{10}\big\|_1 \\
			&\hspace{6pt}
			+\Big\|T^{-1/2}d^{a/2} (\B^\top\V -\bXi \Y_W )\Big\|_\text{max} \\
            &\hspace{12pt}
            \cdot \Big\|(\H_{20} - T^{-1/2}d^{a/2} \bXi\Y_W) + (T^{-1/2}d^{a/2} \B^\top\V -\H_{10})\Big\|_1 \Bigg\} \|D_2\|_1 \\
			&= O\Big\{\frac{L^{3/2}}{d^{1+a}} [c_T\cdot d^2 + 1 \cdot (c_T d^2 + c_T d^2)] \Big\} \|D_2\|_1
			= O\Big(c_T L^{3/2} d^{1-a} \,\|D_2\|_1 \Big) ,
		\end{split}
	\end{equation}
	where the last line used the following rates to be shown later,
	\begin{align}
		\big\| \H_{20}- T^{-1/2}d^{a/2} \bXi\Y_W \big\|_\text{max} &= O_P(c_T),
		\label{eqn: rate_H20_error} \\
		\big\| \H_{10}- T^{-1/2}d^{a/2} \B^\top\V \big\|_\text{max} &= O_P(c_T) .
		\label{eqn: rate_H10_error}
	\end{align}
	
	For neat presentation, we define the following terms whose norms will be bounded and involved in (\ref{eqn: rate_H20_error}) and later,
	\begin{align*}
		& \A_1 := \frac{1}{T}\sum_{t=1}^T \X_t \otimes (\B_t - \bar{\B}) \bgamma, \hspace{32pt}
		\A_1^0 := \b{E}(\X_t \otimes \B_t\bgamma) ,\\
		&\A_2 := \Big(\frac{1}{T}\X^\top \B^\nu \frac{1}{T}(\B^\nu)^\top \X \Big)^{-1}, \hspace{14pt}
		\A_2^0 := \big[\b{E}(\X_t^\top \B_t) \b{E}(\B_t^\top \X_t)\big]^{-1} ,\\
		& \A_3 := \frac{1}{T}\X^\top \B^\nu, \hspace{89pt}
		\A_3^0 := \b{E}(\X_t^\top \B_t) ,\\
		& \A_{4,j,k} := \frac{1}{T}\sum_{t=1}^T z_{j,k,t}(\B_t - \bar{\B})^\top \W_j (\bPi_t^\ast \bmu^\ast + \bPi_t^\ast \X_t\bbeta^\ast), \\
		& \A_{4,j,k}^0 := \b{E}(\A_{4,j,k}), \hspace{74pt}
		\A_{5,j,k} := \frac{1}{T}\sum_{t=1}^T z_{j,k,t}(\B_t - \bar{\B})^\top \W_j \bPi_t^\ast \bepsilon_t .
	\end{align*}
	On $\cM$, we immediately have from Lemma \ref{lemma: LamSouza_theorem_S1} (using $\cA_1, \cA_4, \cA_6$),
	\begin{equation}
		\label{eqn: rate_max_norm_A1_diff}
		\big\|\A_1 - \A_1^0\big\|_\text{max} = O\Big(c_T + c_T(c_T + \mu_{b,\text{max}})\Big) = O(c_T),
	\end{equation}
	which also gives $\big\|\A_1 \big\|_\text{max} \leq \big\|\A_1^0 \big\|_\text{max}+ \big\|\A_1 - \A_1^0\big\|_\text{max} = O(1+ c_T) =O(1)$. Hence with Assumptions (R5) and (R10), we also have on $\cM$ that
	\begin{equation}
		\label{eqn: rate_1_norm_A1}
		\big\|\A_1 \big\|_1 \leq \big\|\A_1^0 \big\|_1+ \big\|\A_1 - \A_1^0\big\|_1 = O(d^{1+a} +c_T d^2) =O(d^{1+a}) .
	\end{equation}
	
	Similarly, with Lemma \ref{lemma: LamSouza_theorem_S1} (using $\cA_1$) and Assumption (R3) we have on $\cM$,
	\begin{equation}
		\label{eqn: rate_1_norm_A3_all}
		\big\|\A_3 -\A_3^0 \big\|_1 = O(c_T d), \,\,\,
		\big\|\A_3^0 \big\|_1 = O(d), \,\,\,
		\big\|\A_3 \big\|_1 \leq \big\|\A_3 -\A_3^0 \big\|_1 +\big\|\A_3^0 \big\|_1 = O(d).
	\end{equation}
	
	Rewrite $\A_2^0 = (\A_3^0 \A_3^{0 \top})^{-1}$ and $\A_2 = (\A_3 \A_3^{\top})^{-1}$, by Assumption (R3),
	\begin{equation}
		\label{eqn: rate_1_norm_A2_0}
		\big\|\A_2^0 \big\|_1 \leq \frac{r^{1/2}}{\lambda_\text{min}(\A_3^0 \A_3^{0 \top})} = O(d^{-2}).
	\end{equation}
	Moreover, from (\ref{eqn: rate_1_norm_A3_all}) we have on $\cM$
	\begin{align*}
		\big\|(\A_2^0)^{-1} - \A_2^{-1}\big\|_1 &\leq \big\|\A_3^0 \A_3^{0 \top} - \A_3 \A_3^{\top} \big\|_1 \leq \big\|\A_3^0 -\A_3\big\|_1 \big\|\A_3^{0 \top}\big\|_1 + \big\|\A_3 \big\|_1 \big\|\A_3^{0 \top} -\A_3^\top \big\|_1 \\
		&= O(c_T d\cdot d + d\cdot c_T d) = O(c_T d^2).
	\end{align*}
	Thus, rewriting $\A_2 -\A_2^0 = (\A_2 -\A_2^0)[(\A_2^0)^{-1} - \A_2^{-1}]\A_2^0 +\A_2^0 [(\A_2^0)^{-1} - \A_2^{-1}]\A_2^0$ and we have on $\cM$,
	\begin{equation}
		\label{eqn: rate_1_norm_A2_diff}
		\big\|\A_2 -\A_2^0 \big\|_1 = o\Big(\big\|\A_2 -\A_2^0 \big\|_1 \Big) + O(c_Td^2\cdot d^{-4}) = O(c_T d^{-2}).
	\end{equation}
	
	Consider now $\A_{4,j,k}^0$ for any $j\in[p], k\in[l_j] \cup\{0\}$. First, we have on $\cM$,
	\begin{align*}
		&\hspace{12pt}
		\Big\| \b{E}\Big( \frac{1}{T} \sum_{t=1}^T z_{j,k,t}(\B_t - \bar{\B})^\top \W_j \bPi_t^\ast \bmu^\ast \Big)\Big\|_1 \leq
		\Big\| \b{E}\Big( \frac{1}{T} \sum_{t=1}^T z_{j,k,t}(\B_t - \bar{\B})^\top\Big) \Big\|_\text{max} \big\| \W_j\big\|_1 \big\|\bPi_t^\ast \bmu^\ast \big\|_1 \\
		&=
		O\Big(d\cdot \big\| \bPi_t^\ast\big\|_\infty \big\| \bmu^\ast \big\|_\text{max} \Big) = O(d),
	\end{align*}
	where the last equality used Assumption (R1). Thus on $\cM$,
	\begin{equation}
		\label{eqn: rate_1_norm_A4_0}
		\begin{split}
            &\hspace{12pt}
			\big\|\A_{4,j,k}^0 \big\|_1 \leq \big\|\V_{\W_j^\top, v}^\top \big[\I_d \otimes \b{E}(\U_{\x, j,k})\big] \V_{\bPi_t^\ast, r} \bbeta^\ast \big\|_1 + \Big\| \b{E}\Big( \frac{1}{T} \sum_{t=1}^T z_{j,k,t}(\B_t - \bar{\B})^\top \W_j \bPi_t^\ast \bmu^\ast \Big)\Big\|_1 \\
			&\leq
			\big\|\V_{\W_j^\top, v}^\top \big[\I_d \otimes \b{E}(\U_{\x, j,k})\big] \big\|_\text{max} \big\|\V_{\bPi_t^\ast, r} \big\|_1 \big\| \bbeta^\ast \big\|_1 +O(d) \\
			&\leq
			\max_{i,n\in[d]} \max_{m\in[v]} \max_{q\in[r]}  \W_{j,\cdot i}^\top \b{E}\Big\{ \frac{1}{T} \sum_{t=1}^T z_{j,k,t} (\B_t-\bar{\B})_{\cdot m} X_{t,nq} \Big\} \big\|\V_{\bPi_t^\ast, r} \big\|_1 \big\| \bbeta^\ast \big\|_1 +O(d) = O(d),
		\end{split}
	\end{equation}
	where the last equality used Assumption (R1). Furthermore, we have on $\cM$ that
	\begin{equation}
		\label{eqn: rate_1_norm_A4_diff}
		\begin{split}
			&\hspace{12pt}
			\big\|\A_{4,j,k} -\A_{4,j,k}^0 \big\|_1 \\
			&\leq
			\Big\|\frac{1}{T}\sum_{t=1}^T z_{j,k,t}(\B_t - \bar{\B})^\top -\b{E}\Big\{ \frac{1}{T}\sum_{t=1}^T z_{j,k,t}(\B_t - \bar{\B})^\top \Big\} \Big\|_\text{max} \big\|\W_j \big\|_1 \big\|\bPi_t^\ast \bmu^\ast \big\|_1 \\
			&\hspace{12pt}
			+ \big\|\V_{\W_j^\top, v}^\top \big\{\I_d \otimes [\U_{\x, j,k} -\b{E}(\U_{\x, j,k})]\big\} \V_{\bPi_t^\ast, r} \bbeta^\ast \big\|_1 \\
			&=
			O\Big([c_T +1\cdot c_T + 1\cdot c_T + 1\cdot (c_T \vee 0) ] \,d \Big) + O\Big(\big\|\V_{\bPi_t^\ast, r} \big\|_1 \big\| \bbeta^\ast \big\|_1 \Big)\\
			&\hspace{12pt}
			\cdot \big\|\W_j^\top \big\|_1 \max_{n,m\in[d]} \max_{s\in[v]} \max_{q\in[r]} \Big|\frac{1}{T} \sum_{t=1}^T z_{j,k,t} (\B_t-\bar{\B})_{ms} X_{t,nq} -\b{E}\Big( \frac{1}{T} \sum_{t=1}^T z_{j,k,t} (\B_t-\bar{\B})_{ms} X_{t,nq} \Big)\Big| \\
            &= O(c_T d),
		\end{split}
	\end{equation}
	where the second last equality used Assumption (R1) and $\cA_4, \,\cA_{12}, \,\cA_{13}$, while the last used $\cA_4, \,\cA_8, \,\cA_9$. In particular, we used the following as an immediate result of $\cA_{13}$,
	\[
	\max_{m\in[p]} \max_{n\in[l_m] \cup\{0\}} \bigg| \frac{1}{T}\sum_{t=1}^T z_{m,n,t} - \b{E}\Big\{ \frac{1}{T}\sum_{t=1}^T z_{m,n,t}\Big\}\bigg| \leq c_T \vee 0.
	\]
	Lastly for any $j\in[p], k\in[l_j] \cup\{0\}$, similar to (\ref{eqn: rate_I4}), we have on $\cM$ that
	\begin{equation}
		\label{eqn: rate_1_norm_A5}
		\begin{split}
			\big\|\A_{5,j,k} \big\|_1 &= \big\|\V_{\W_j^\top, v}^\top \big(\I_d \otimes \U_{\bepsilon, j,k}\big) \V_{\bPi_t^\ast, r} \big\|_1
			\leq
			\big\|\V_{\W_j^\top, v}^\top \big(\I_d \otimes \U_{\bepsilon, j,k}\big) \big\|_\text{max} \big\|\V_{\bPi_t^\ast, r} \big\|_1 \\
			&\leq
			\big\|\W_j^\top \big\|_1 \max_{m,n\in[d]} \max_{s\in[v]} \Big|\frac{1}{T} \sum_{t=1}^T z_{j,k,t} (\B_t-\bar{\B})_{ms} \epsilon_{t,n}\Big| \cdot \big\|\V_{\bPi_t^\ast, r} \big\|_1 = O(c_T d).
		\end{split}
	\end{equation}

	Consider now (\ref{eqn: rate_H20_error}), we have
	\begin{align*}
		&\hspace{12pt}
		\Big\| \H_{20} -T^{-1/2}d^{a/2} \bXi\Y_W \Big\|_\text{max} = \max_{j,k} \Big\| \A_1\A_2\A_3 (\A_{4,j,k} + \A_{5,j,k}) - \A_1^0\A_2^0\A_3^0 \A_{4,j,k}^0 \Big\|_\text{max} \\
		&\leq
		\max_{j,k} \big\|\A_1\big\|_\text{max} \big\|\A_2\big\|_1 \big\|\A_3\big\|_1 \big\|\A_{5,j,k}\big\|_1 \\
		&\hspace{12pt}
		+ \max_{j,k}\Big\{\big\| \A_1\big\|_\text{max} \big\|\A_2\A_3\A_{4,j,k} -\A_2^0\A_3^0\A_{4,j,k}^0 \big\|_1 + \big\| \A_1 -\A_1^0 \big\|_\text{max} \big\|\A_2^0\A_3^0\A_{4,j,k}^0 \big\|_1 \Big\} ,
		\,\,\, \text{with}\\
		& \max_{j,k}\big\|\A_2\A_3\A_{4,j,k} -\A_2^0\A_3^0\A_{4,j,k}^0 \big\|_1
		\leq \max_{j,k} \Big\{ \big\|\A_2\big\|_1 \big\|\A_3 - \A_3^0\big\|_1 \big\|\A_{4,j,k}\big\|_1 \\
		&\hspace{120pt}
		+ \big\|\A_2\big\|_1 \big\|\A_3^0\big\|_1 \big\|\A_{4,j,k} -\A_{4,j,k}^0 \big\|_1 + \big\|\A_2 -\A_2^0 \big\|_1 \big\|\A_3^0\big\|_1 \big\|\A_{4,j,k}^0 \big\|_1 \Big\}.
	\end{align*}
	Together with all the rates from (\ref{eqn: rate_max_norm_A1_diff}) to (\ref{eqn: rate_1_norm_A5}), we have (\ref{eqn: rate_H20_error}) true on $\cM$.

	For (\ref{eqn: rate_H10_error}), consider for any $j\in[p], k\in[l_j] \cup \{0\}$, we have on $\cM$ that
	\begin{equation}
		\label{eqn: rate_max_norm_H10_part1}
		\begin{split}
			&\hspace{12pt}
			\Big\|\Big[\I_d \otimes (\bgamma^\top \otimes \I_d) \U_{\bepsilon, j,k} \bPi_t^{\ast \top}
			\Big]\Vec{\W_j^\top} \Big\|_\text{max}
			=
			\max_{i\in[d]} \big\|(\bgamma^\top \otimes \I_d) \U_{\bepsilon, j,k} \bPi_t^{\ast \top} \W_{j,i\cdot} \big\|_\text{max} \\
			&\leq
			\max_{i\in[d]} \big\|(\bgamma^\top \otimes \I_d) \U_{\bepsilon, j,k} \big\|_\text{max} \big\| \bPi_t^\ast \big\|_\infty \big\|\W_{j,i\cdot} \big\|_1 \\
			&=
			O(1) \cdot \|\bgamma\|_1 \cdot \max_{m,n\in[d]} \max_{s\in[v]} \Big| \frac{1}{T} \sum_{t=1}^T z_{j,k,t} (\B_t-\bar{\B})_{ms} \epsilon_{t,n} \Big| =O(c_T),
		\end{split}
	\end{equation}
	where the second last equality used Assumption (R1) and the result below (\ref{eqn: def_bPi_ast}), and the last is similar to (\ref{eqn: rate_I4}). In a similar way on $\cM$,
	\begin{equation}
		\label{eqn: rate_max_norm_H10_part2}
		\begin{split}
			&\hspace{12pt}
			\Big\|\Big[\I_d \otimes (\bgamma^\top \otimes \I_d) \big(\U_{\bmu, j,k} -\b{E}(\U_{\bmu, j,k})\big) \bPi_t^{\ast \top} \Big] \Vec{\W_j^\top} \Big\|_\text{max} \\
			&\leq
			\max_{i\in[d]} \big\|(\bgamma^\top \otimes \I_d) \big(\U_{\bmu, j,k} -\b{E}(\U_{\bmu, j,k})\big) \big\|_\text{max} \big\|\bPi_t^\ast \big\|_\infty \big\|\W_{j,i\cdot} \big\|_1 \\
			&=
			O\Big(\|\bgamma\|_1 \|\bmu^\ast \|_\text{max} \Big) \cdot \Big\|\frac{1}{T} \sum_{t=1}^T z_{j,k,t}(\B_t - \bar{\B})^\top -\b{E}\Big\{ \frac{1}{T}\sum_{t=1}^T z_{j,k,t}(\B_t - \bar{\B})^\top \Big\} \Big\|_\text{max} = O(c_T),
		\end{split}
	\end{equation}
	with the last line similar to (\ref{eqn: rate_1_norm_A4_diff}) which is also involved in the last line of the following,
	\begin{equation}
		\label{eqn: rate_max_norm_H10_part3}
		\begin{split}
			&\hspace{12pt}
			\Big\|\Big[\I_d \otimes (\bgamma^\top \otimes \I_d) \big(\U_{\x, j,k} -\b{E}(\U_{\x, j,k})\big) (\bbeta^\ast \otimes \I_d) \bPi_t^{\ast \top} \Big] \Vec{\W_j^\top} \Big\|_\text{max} \\
			&\leq
			\max_{i\in[d]} \big\|(\bgamma^\top \otimes \I_d) \big(\U_{\x, j,k} -\b{E}(\U_{\x, j,k})\big) \big\|_\text{max} \big\|\bbeta^\ast \otimes \I_d \big\|_1 \big\|\bPi_t^\ast \big\|_\infty \big\|\W_{j,i\cdot} \big\|_1 \\
			&=
			O\Big(\|\bgamma\|_1 \Big)\, \max_{m,n\in[d]} \max_{s\in[v]} \max_{q\in[r]} \Big|\frac{1}{T} \sum_{t=1}^T z_{j,k,t} (\B_t-\bar{\B})_{ms} X_{t,nq} -\b{E}\Big( \frac{1}{T} \sum_{t=1}^T z_{j,k,t} (\B_t-\bar{\B})_{ms} X_{t,nq} \Big)\Big| \\
            &= O(c_T).
		\end{split}
	\end{equation}
	Combining (\ref{eqn: rate_max_norm_H10_part1}), (\ref{eqn: rate_max_norm_H10_part2}) and (\ref{eqn: rate_max_norm_H10_part3}), we have (\ref{eqn: rate_H10_error}) true by the following decomposition,
	\begin{align*}
		&\hspace{12pt}
		\Big\| \H_{10}- T^{-1/2}d^{a/2} \B^\top\V \Big\|_\text{max} \\
		&\leq \max_{j,k} \Big\|\Big[\I_d \otimes (\bgamma^\top \otimes \I_d) \U_{\bepsilon, j,k} \bPi_t^{\ast \top}
		\Big]\Vec{\W_j^\top} \Big\|_\text{max} \\
		&\hspace{12pt}
		+ \max_{j,k} \Big\|\Big[\I_d \otimes (\bgamma^\top \otimes \I_d) \big(\U_{\bmu, j,k} -\b{E}(\U_{\bmu, j,k})\big) \bPi_t^{\ast \top} \Big] \Vec{\W_j^\top} \Big\|_\text{max} \\
		&\hspace{12pt}
		+ \max_{j,k} \Big\|\Big[\I_d \otimes (\bgamma^\top \otimes \I_d) \big(\U_{\x, j,k} -\b{E}(\U_{\x, j,k})\big) (\bbeta^\ast \otimes \I_d) \bPi_t^{\ast \top} \Big] \Vec{\W_j^\top} \Big\|_\text{max} .
	\end{align*}

	Next for $F_2$ and $F_3$, we consider first on $\cM$,
	\begin{equation}
		\label{eqn: rate_F2F3_aux}
		\begin{split}
			&\hspace{12pt}
			\Big\| \Big\{\frac{1}{T} \sum_{t=1}^T \Big(\I_d \otimes (\B_t - \bar{\B}) \bgamma \Big) \X_t \Big\} (\X^\top \B^\nu (\B^\nu)^\top \X)^{-1} \X^\top \B^\nu (\B^\nu)^\top \bepsilon^\nu \Big\|_1 \\
			&=
			\bigg\|\A_1 \A_2 \A_3 \,\frac{1}{T} \sum_{t=1}^T (\B_t- \bar{\B})^\top \bepsilon_t \bigg\|_1
			\leq \big\|\A_1 \big\|_1 \big\|\A_2 \big\|_1 \big\|\A_3 \big\|_1 \cdot v \max_{j\in[r]} \bigg| \frac{1}{T}\sum_{t=1}^T \sum_{q=1}^d B_{t,qj}\epsilon_{t,q} \bigg| \\
			&=
			O\big(d^{1+a}\cdot d^{-2}\cdot d \cdot c_T d^{\frac{1}{2} + \frac{1}{2w}} \big)
			= O\big( c_T d^{\frac{1}{2} +\frac{1}{2w} +a} \big) ,
		\end{split}
	\end{equation}
	where the first equality used the fact that $\X_t = \X_t\otimes 1$, the second last used (\ref{eqn: rate_1_norm_A1}), (\ref{eqn: rate_1_norm_A3_all}), (\ref{eqn: rate_1_norm_A2_0}) and $\cA_3$ in Lemma \ref{lemma: LamSouza_theorem_S1}. Then for $F_2$, we have on $\cM$ that
	\begin{equation}
		\label{eqn: rate_F2}
		\begin{split}
			\|F_2\|_1 &\leq \frac{L^{1/2} \cdot L}{C d^{1+a}}\Big( \Big\|(\H_{20} - T^{-1/2}d^{a/2} \bXi\Y_W)\Big\|_\text{max} + \Big\|(T^{-1/2}d^{a/2} \B^\top\V -\H_{10})\Big\|_\text{max} \Big) \\
			&\hspace{12pt}
			\cdot \Big\| \Big\{\frac{1}{T} \sum_{t=1}^T \Big(\I_d \otimes (\B_t - \bar{\B}) \bgamma \Big) \X_t \Big\} (\X^\top \B^\nu (\B^\nu)^\top \X)^{-1} \X^\top \B^\nu (\B^\nu)^\top \bepsilon^\nu \Big\|_1 \\
			&=
			O\big(L^{3/2} \cdot d^{-1-a} \cdot c_T \cdot c_T d^{\frac{1}{2} +\frac{1}{2w} +a} \big) = O\big(c_T^2 L^{3/2} d^{-\frac{1}{2} +\frac{1}{2w}} \big),
		\end{split}
	\end{equation}
	where the last line used (\ref{eqn: bound_inv_H2010}), (\ref{eqn: rate_H20_error}), (\ref{eqn: rate_H10_error}) and (\ref{eqn: rate_F2F3_aux}). Similarly, we have $\|F_3\|_1 = O\big(c_T L^{3/2} d^{-\frac{1}{2} +\frac{1}{2w}} \big)$ on $\cM$, and hence together with $L=O(1)$, (\ref{eqn: rate_F1}) and (\ref{eqn: rate_F2}), it holds on $\cM$ that
	\begin{equation}
		\label{eqn: rate_D2}
		\big\|D_2 \big\|_1 \leq \big\|F_1 \big\|_1 + \big\|F_2 \big\|_1 + \big\|F_3 \big\|_1 = O\big(c_T d^{-\frac{1}{2} +\frac{1}{2w}} \big).
	\end{equation}

	Similar to the way we decompose $D_2$, we can rewrite $D_1 = F_4 + F_5 - F_6$ where
	\begin{align*}
		F_4 &= [(\H_{20} -\H_{10})^\top (\H_{20} -\H_{10})]^{-1} \Big\{(\H_{20} -\H_{10})^\top (\H_{20} -\H_{10}) \\
		&\hspace{12pt}
		- T^{-1}d^a (\B^\top\V -\bXi \Y_W )^\top (\B^\top\V - \bXi\Y_W) \Big\}D_1 , \\
		F_5 &= [(\H_{20} -\H_{10})^\top (\H_{20} -\H_{10})]^{-1} \Big[(T^{-1/2}d^{a/2} \B^\top\V -\H_{10}) -(T^{-1/2}d^{a/2} \bXi \Y_W -\H_{20}) \Big]^\top \\
		&\hspace{12pt}
		\cdot T^{-1/2}d^{a/2} \B^\top\bepsilon ,\\
		F_6 &= [(\H_{20} -\H_{10})^\top (\H_{20} -\H_{10})]^{-1} (\H_{20} -\H_{10})^\top \cdot T^{-1/2}d^{a/2} \B^\top\bepsilon .
	\end{align*}
	From (\ref{eqn: rate_F1}), it is direct that $F_4 = O\big(c_T L^{3/2} d^{1-a} \,\|D_1\|_1 \big)$ on $\cM$. Moreover, (\ref{eqn: rate_H20_error}) and (\ref{eqn: rate_H10_error}) imply that $F_5$ has a smaller rate than that of $F_6$. Given $L=O(1)$, we next construct the asymptotic normality of $\balpha^\top F_6$ for any given non-zero $\balpha \in\b{R}^L$ with $\|\balpha \|_1 \leq c<\infty$.
	
	Denote by $\R_1 := [(\H_{20} -\H_{10})^\top (\H_{20} -\H_{10})]^{-1} (\H_{20} -\H_{10})^\top$, we have
	\begin{align*}
		\balpha^\top F_6 &= \frac{1}{T} \balpha^\top \R_1 \Big(\I_d \otimes \big\{(\B_1 -\bar{\B}, \dots, \B_T -\bar{\B}) (\I_T \otimes \bgamma) \big\} \Big) \vec{(\bepsilon_1, \dots, \bepsilon_T)^\top} \\
		&=
		\frac{1}{T} \balpha^\top \R_1 \Big(\I_d \otimes \big\{(\B_1 -\bar{\B})\bgamma, \dots, (\B_T -\bar{\B})\bgamma \big\} \Big) \sum_{t=1}^T \Big\{ \bepsilon_t \otimes \big(\b{1}_{\{j=t\}} \big)_{j\in[T]} \Big\} \\
		&=
		\frac{1}{T} \sum_{t=1}^T \balpha^\top \R_1 \Big(\bepsilon_t \otimes (\B_t -\bar{\B})\bgamma \Big) = \frac{1}{T} \sum_{t=1}^T \balpha^\top \R_1 \Big(\bepsilon_t \otimes [\B_t -\b{E}(\B_t)] \bgamma (1+ o_P(1))\Big) ,
	\end{align*}
	where the vector $\big( \b{1}_{\{j=t\}} \big)_{j\in[T]} \in\b{R}^T$ has value $\b{1}_{\{j=t\}}$ at each $j$-th entry, and the last equality used $\cA_4$ in Lemma \ref{lemma: LamSouza_theorem_S1}. Hence to apply Theorem 3 (ii) of \cite{Wu2011}, we need to show
	\begin{equation}
		\label{eqn: cond_Wu2011_F6}
		\sum_{t\geq 0}\Big\| P_0\Big\{ \balpha^\top \R_1 \big(\bepsilon_t \otimes [\B_t -\b{E}(\B_t)] \bgamma\big) \Big\} \Big\|_2 <\infty,
	\end{equation}
	where $P_0(\cdot) :=\b{E}_0(\cdot) -\b{E}_{-1}(\cdot)$ and $\b{E}_i(\cdot) := \b{E}(\cdot\mid \sigma(\cG_i, \cH_i))$. Notice that
	\begin{align*}
		&\hspace{12pt}
		\Big\| P_0\Big\{ \balpha^\top \R_1 \big(\bepsilon_t \otimes [\B_t -\b{E}(\B_t)] \bgamma\big)\Big\} \Big\|_2 \\
		&=
		\Big\|\balpha^\top \R_1 \Big\{P_0(\bepsilon_t) \otimes \b{E}_0([\B_t -\b{E}(\B_t)] \bgamma)\Big\} + \balpha^\top \R_1 \Big\{ \b{E}_{-1}(\bepsilon_t) \otimes P_0([\B_t -\b{E}(\B_t)] \bgamma)\Big\} \Big\|_2 \\
		&\leq
		\Big\{2 \balpha^\top \R_1 \Big\{ \b{E}\big( P_0(\bepsilon_t) \,P_0(\bepsilon_t)^\top \big) \otimes \b{E}\big( \b{E}_0([\B_t -\b{E}(\B_t)] \bgamma) \,\b{E}_0(\bgamma^\top [\B_t -\b{E}(\B_t)]^\top) \big)\Big\} \R_1 ^\top \balpha \Big\}^{\frac{1}{2}} \\
		&
		+ \Big(2 \balpha^\top \R_1 \Big\{ \b{E}\big( \b{E}_{-1}(\bepsilon_t) \,\b{E}_{-1}(\bepsilon_t)^\top \big) \otimes \b{E}\big( P_0([\B_t -\b{E}(\B_t)] \bgamma) \,P_0(\bgamma^\top [\B_t -\b{E}(\B_t)]^\top) \big)\Big\} \R_1 ^\top \balpha \Big)^{\frac{1}{2}} \\
		&=
		O\Big(\big\|\balpha \big\|_1 \big\|\R_1 \big\|_\infty \Big) \cdot\Big( \max_{j\in[d]} \big\|P_0(\epsilon_{t,j})\big\|_2 \cdot \max_{j\in[d]} \text{Var}^{1/2}(\B_{t,j\cdot}^\top \bgamma) + \sigma_\text{max} \max_{j\in[d]} \big\| P_0(\B_{t,j\cdot}^\top \bgamma)\big\|_2\Big) \\
		&= O\Big( \max_{j\in[d]} \big\|P_0^\epsilon (\epsilon_{t,j})\big\|_2 + \max_{j\in[d]} \max_{k\in[v]} \big\| P_0^b (B_{t,jk}) \big\|_2\Big),
	\end{align*}
	where the second last equality used $\text{Var}(\cdot) = \text{Var}(\b{E}_i(\cdot)) + \b{E}(\text{Var}_i(\cdot)) \geq \text{Var}(\b{E}_i(\cdot))$, and the last used Assumption (R2) and $\big\|\R_1 \big\|_\infty =O(1)$ which is implied from (\ref{eqn: bound_inv_H2010}), $\big\|\H_{10} \big\|_1 =O(d)$ and $\big\|\H_{20} \big\|_1 =O(d^{1+a})$. With Assumption (R7), (\ref{eqn: cond_Wu2011_F6}) is true. Therefore, with definition
	\begin{equation}
		\label{eqn: def_s0_cov_mat}
		s_1:= \balpha^\top \R_1 \bSigma \R_1^\top \balpha,
		\,\,\,
		\bSigma := \sum_{\tau} \b{E}(\bepsilon_t \bepsilon_{t+\tau}^\top) \otimes \b{E}\big[ (\B_t -\b{E}(\B_t) )\bgamma \bgamma^\top (\B_{t+\tau} -\b{E}(\B_t) )^\top \big],
	\end{equation}
	we have by Theorem 3 (ii) of \cite{Wu2011} that
	\[
	T^{1/2} s_1^{-1/2} \balpha^\top F_6 \xrightarrow{D} \cN(0, 1).
	\]
	Then equivalently we have
	\begin{equation}
		\label{eqn: asymp_F6}
		T^{1/2} (\R_1 \bSigma \R_1^\top)^{-1/2} F_6 \xrightarrow{D} \cN(\0, \I_L),
	\end{equation}
	so that $F_6$ is at least $T^{1/2} d^{(1+a-b)/2}$-convergent which used $\lambda_\text{max}(\R_1 \R_1^\top) =O(d^{-1-a})$ from (\ref{eqn: bound_inv_H2010}), and all eigenvalues of $d^{-b}\bSigma$ uniformly bounded from 0 and infinity by Assumption (R8). Hence, $\big\|D_1 \big\|_1= O\big(\|F_6\|_1 \big)= O\big(T^{-1/2} d^{-(1+a-b)/2} \big)$ on $\cM$, and by (\ref{eqn: rate_D2}) we have
	\[
	\big\|\wt{\bphi}- \bphi^\ast \big\|_1 = O_P\Big(\big\|D_1 \big\|_1 +\big\|D_2 \big\|_1 \Big) = O_P\big(T^{-1/2} d^{-(1+a-b)/2} +c_T d^{-\frac{1}{2} +\frac{1}{2w}}\big) = O_P\big( c_T d^{-\frac{1}{2} +\frac{1}{2w}}\big),
	\]
	where the last equality used Assumption (R10). With the above plugged into (\ref{eqn: beta_consistency}), the proof of Theorem \ref{thm: LASSO_rate} is complete.
	$\square$

	\noindent\textbf{\textit{Proof of Theorem \ref{thm: ada_LASSO_phi_asymp}.}}
	By the KKT condition, $\wh\bphi$ is a solution to the adaptive LASSO problem in (\ref{eqn: simplify_phi_ada_lasso_matrix}) if and only if there exists a subgradient
	\[
	\h = \partial (\u^\top |\wh\bphi|) =\left\{
	\h \in \b{R}^L : \left\{
	\begin{array}{ll}
		h_i = u_i \, \text{sign}(\wh\phi_i), & \hbox{$\wh\phi_i \neq 0$;} \\
		|h_i| \leq u_i, & \hbox{otherwise.}
	\end{array}
	\right.
	\right\} ,
	\]
	such that differentiating the expression in (\ref{eqn: simplify_phi_ada_lasso_matrix}) with respect to $\bphi$, we have
	\[
	T^{-1}(\bXi\Y_W - \B^\top\V)^\top (\bXi\Y_W - \B^\top\V)\bphi + T^{-1}(\bXi\Y_W - \B^\top\V)^\top (\B^\top \y - \bXi\y^\nu) = -\lambda \h .
	\]
	Substituting (\ref{eqn: spatiallag_rewrite1_augmented}) in the above, we arrive at
	\begin{align*}
		-\lambda \h &= T^{-1}(\bXi\Y_W - \B^\top\V)^\top (\bXi\Y_W - \B^\top\V)\bphi \\
		&\hspace{11pt}
		+ T^{-1}(\bXi\Y_W - \B^\top\V)^\top (\B^\top\V \bphi^\ast + \B^\top\X_{\bbeta^\ast} \Vec{\I_d} + \B^\top\bepsilon - \bXi\y^\nu) \\
		&=
		T^{-1}(\bXi\Y_W - \B^\top\V)^\top \B^\top\V (\bphi^\ast - \bphi) + T^{-1}(\bXi\Y_W - \B^\top\V)^\top \B^\top\bepsilon \\
		&\hspace{12pt}
		+ T^{-1}(\bXi\Y_W - \B^\top\V)^\top \B^\top\X_{\bbeta^\ast} \Vec{\I_d} \\
        &\hspace{12pt}
        + T^{-1}(\bXi\Y_W - \B^\top\V)^\top (\bXi\Y_W\bphi^\ast -\bXi\y^\nu + \bXi\Y_W (\bphi-\bphi^\ast) ) \\
		&=
		T^{-1}(\bXi\Y_W - \B^\top\V)^\top (\bXi\Y_W - \B^\top\V) (\bphi -\bphi^\ast) + T^{-1}(\bXi\Y_W - \B^\top\V)^\top \B^\top\bepsilon \\
		&\hspace{12pt}
		+ T^{-1}(\bXi\Y_W - \B^\top\V)^\top \B^\top\X_{\bbeta^\ast - \bbeta(\bphi^\ast)} \Vec{\I_d} ,
	\end{align*}
	where the last equality used the fact that $\B^\top\X_{\bbeta(\bphi^\ast)} \Vec{\I_d} = \bXi\y^\nu - \bXi\Y_W \bphi^\ast$ from (\ref{eqn: simplify_LASSO}). Then we may conclude that there exists a sign-consistent solution $\wh\bphi$ if and only if
	\begin{equation}
		\label{eqn: ada_lasso_sign_consistent}
		\left\{
		\begin{array}{ll}
			-\lambda \h_{H} = T^{-1}(\bXi\Y_{W,H} - \B^\top\V_H)^\top (\bXi\Y_{W,H} - \B^\top\V_H) (\wh\bphi -\bphi^\ast) \\
			\hspace{40pt}
			+ T^{-1}(\bXi\Y_{W, H}- \B^\top\V_H)^\top \B^\top\X_{\bbeta^\ast - \bbeta(\bphi^\ast)} \Vec{\I_d}
			+ T^{-1}(\bXi\Y_{W, H} - \B^\top\V_H)^\top \B^\top\bepsilon , \\
			\lambda \u_{H^c} \geq \Big| T^{-1}(\bXi\Y_{W, H^c} - \B^\top\V_{H^c})^\top \B^\top\X_{\bbeta^\ast - \bbeta(\bphi^\ast)} \Vec{\I_d} + T^{-1}(\bXi\Y_{W, H^c} - \B^\top\V_{H^c})^\top \B^\top\bepsilon \Big| ,
		\end{array}
		\right.
	\end{equation}
	where $\A_H$ and $\a_H$ denote the corresponding submatrix $\A$ with columns restricted on the set $H$ and subvector $\a$ with entries restricted on the set $H$, respectively. Similarly $(\cdot)_{H^c}$ is defined. Consider the first equation in (\ref{eqn: ada_lasso_sign_consistent}), similar to how $D_2$ is decomposed in the proof of Theorem \ref{thm: LASSO_rate}, we write $\wh\bphi -\bphi^\ast = \sum_{j=1}^4 I_{\phi,j}$ where
	\begin{align*}
		I_{\phi,1} &= [(\H_{20} -\H_{10})_H^\top (\H_{20} -\H_{10})_H]^{-1} \Big\{(\H_{20} -\H_{10})_H^\top (\H_{20} -\H_{10})_H \\
		&\hspace{12pt}
		- T^{-1}d^a (\B^\top\V_H -\bXi \Y_{W,H} )^\top (\B^\top\V_H - \bXi\Y_{W,H}) \Big\}(\wh\bphi -\bphi^\ast) , \\
		I_{\phi,2} &= - [(\H_{20} -\H_{10})_H^\top (\H_{20} -\H_{10})_H]^{-1} d^a\lambda \h_H ,\\
		I_{\phi,3} &= [(\H_{20} -\H_{10})_H^\top (\H_{20} -\H_{10})_H]^{-1} T^{-1}d^a (\bXi\Y_{W, H}- \B^\top\V_H)^\top \B^\top\X_{\bbeta(\bphi^\ast) -\bbeta^\ast} \Vec{\I_d} ,\\
		I_{\phi,4} &= [(\H_{20} -\H_{10})_H^\top (\H_{20} -\H_{10})_H]^{-1} T^{-1}d^a (\B^\top\V_H -\bXi\Y_{W, H})^\top \B^\top\bepsilon.
	\end{align*}

	Similar to $F_1$ in the proof of Theorem \ref{thm: LASSO_rate}, we may derive that
	\begin{equation}
		\label{eqn: rate_I_phi1}
		\big\|I_{\phi,1} \big\|_\text{max} = O_P\Big(c_T d^{1-a} \big\|\wh\bphi -\bphi^\ast \big\|_\text{max} \Big)
		= o_P\Big( \big\|\wh\bphi -\bphi^\ast \big\|_\text{max} \Big),
	\end{equation}
	where the first equality used the fact that (R1) implies for a positive constant $u$ that $\sigma_{|H|}^2\{ (\D_W)_H\} \geq du>0$ uniformly as $d\to \infty$, and the conditions in the statement of Theorem \ref{thm: ada_LASSO_phi_asymp}, and the second used (R10). Similarly, with Assumption (R9) we have
	\begin{equation}
		\label{eqn: rate_I_phi2}
		\big\|I_{\phi,2} \big\|_\text{max} = O_P\big(d^{-1-a} \cdot d^a \cdot \lambda \big) =O_P\big(c_T d^{-1} \big).
	\end{equation}
	
	For $I_{\phi,4}$, we may decompose it as the following with the second term dominating the first term similarly to $F_5$ and $F_6$ in the proof of Theorem~\ref{thm: LASSO_rate},
	\begin{align*}
		I_{\phi,4} &=
		\Big( [(\H_{20} -\H_{10})_H^\top (\H_{20} -\H_{10})_H]^{-1} \\
        &\hspace{12pt}
        \cdot \Big[(T^{-1/2}d^{a/2} \B^\top\V_H -\H_{10,H}) -(T^{-1/2}d^{a/2} \bXi \Y_{W,H} -\H_{20,H}) \Big]^\top
		\cdot T^{-1/2}d^{a/2} \B^\top\bepsilon \Big) \\
        &\hspace{12pt}
		- \Big( [(\H_{20} -\H_{10})_H^\top (\H_{20} -\H_{10})_H]^{-1} (\H_{20} -\H_{10})_H^\top \cdot T^{-1/2}d^{a/2} \B^\top\bepsilon \Big).
	\end{align*}
	The second term in the above has rate $T^{-1/2} d^{-(1+a-b)/2}$ by exactly the same way to construct asymptotic normality of $F_6$ in (\ref{eqn: asymp_F6}), except for the restriction to the set $H$ here (proof omitted). Thus,
	\begin{equation}
		\label{eqn: rate_I_phi4}
		\big\|I_{\phi,4} \big\|_\text{max} = O_P\big(T^{-1/2} d^{-(1+a-b)/2} \big) .
	\end{equation}
	
	We next construct the asymptotic normality of $I_{\phi,3}$ and show its rate is of order $T^{-1/2} d^{-(1-b)/2}$ which is dominating over those of $I_{\phi,1}$, $I_{\phi,2}$ and $I_{\phi,4}$ by Assumption (R10). Recall $\R_H =[(\H_{20} -\H_{10})_H^\top (\H_{20} -\H_{10})_H]^{-1} (\H_{20} -\H_{10})_H^\top$, and let non-zero $\balpha \in\b{R}^{|H|}$ such that $\| \balpha\|_1 \leq c<\infty$. Then we have
	\begin{align*}
		&\hspace{12pt}
		\balpha^\top I_{\phi,3} = \balpha^\top \R_H T^{-1} (\B_{\bgamma}- \bar{\B}_{\bgamma})^\top \X_{\bbeta(\bphi^\ast) -\bbeta^\ast} \vec{\I_d} (1+ o_P(1)) \\
		&=
		\balpha^\top \R_H T^{-1} \Big(\I_d \otimes \big\{(\B_1 - \bar{\B}, \dots, \B_T - \bar{\B}) \\
        &\hspace{12pt}
        \cdot (\I_T \otimes \bgamma(\bbeta(\bphi^\ast) -\bbeta^\ast)^\top ) (\X_1, \dots, \X_T)^\top \big\}\Big) \vec{\I_d} (1+ o_P(1)) \\
		&=
		\balpha^\top \R_H \, \frac{1}{T} \sum_{t=1}^T \vec{ (\B_t - \b{E}(\B_t)) \, \bgamma\, (\bbeta( \bphi^\ast) -\bbeta^\ast)^\top \X_t^\top} (1+ o_P(1)) \\
		&=
		\balpha^\top \R_H \, \text{vec}\Big( \frac{1}{T} \sum_{t=1}^T \Big[ \bgamma^\top (\B_{t,i\cdot} -\b{E}(\B_{t,i\cdot}) ) \X_{t,j\cdot}^\top (\bbeta( \bphi^\ast) -\bbeta^\ast) \Big]_{i,j\in[d]} \Big) (1+ o_P(1)) \\
		&=
		\balpha^\top \R_H \S_{\bgamma} (\bbeta( \bphi^\ast) -\bbeta^\ast) (1+ o_P(1)),
	\end{align*}
	where the third last equality used $\cA_4$ in Lemma \ref{lemma: LamSouza_theorem_S1} and the last used $\cA_1$. From (\ref{eqn: rate_diff_U_U0}) and Lemma \ref{lemma: asymp_I2}, we have
	\begin{align*}
		&\hspace{12pt}
		T^{1/2} (\R_{\beta} \bSigma_{\beta} \R_{\beta}^\top)^{-1/2}(\bbeta( \bphi^\ast) -\bbeta^\ast) \\
        &=
		T^{\frac{1}{2}} (\R_{\beta} \bSigma_{\beta} \R_{\beta}^\top)^{-\frac{1}{2}} \Big\{ \big(\X^\top \B^\nu (\B^\nu)^\top \X \big)^{-1} \X^\top \B^\nu (\B^\nu)^\top \bepsilon^\nu \Big\} \\
		&=
		T^{\frac{1}{2}} (\R_{\beta} \bSigma_{\beta} \R_{\beta}^\top)^{-\frac{1}{2}} \Big\{ \big[\b{E}(\X_t^\top \B_t) \b{E}(\B_t^\top \X_t)\big]^{\text{-}1} \frac{1}{T^2} \X^\top \B^\nu (\B^\nu)^\top \bepsilon^\nu (1+o(1)) \Big\}
		\xrightarrow{D} \cN(\0, \I_r) .
	\end{align*}
	Define $s_3:= \balpha^\top \R_H \S_{\bgamma} \R_{\beta} \bSigma_{\beta} \R_{\beta}^\top \S_{\bgamma}^\top \R_H^\top \balpha$, we hence have $T^{1/2} s_3^{-1/2} \balpha^\top I_{\phi,3} \xrightarrow{D} \cN(0, 1)$ and equivalently
	\[
	T^{1/2} (\R_H \S_{\bgamma} \R_{\beta} \bSigma_{\beta} \R_{\beta}^\top \S_{\bgamma}^\top \R_H^\top)^{-1/2} I_{\phi,3} \xrightarrow{D} \cN(\0, \I_{|H|}).
	\]
	
	As shown in the proof of Lemma \ref{lemma: asymp_I2}, the eigenvalues of $\R_{\beta} \bSigma_{\beta} \R_{\beta}^\top$ are of order $d^{b-1}$. Similar to $\R_1$ in the proof of Theorem \ref{thm: LASSO_rate}, $\lambda_\text{max}(\R_H \R_H^\top) =O(d^{-1-a})$. We also have $\lambda_\text{max}(\S_{\bgamma}^\top \S_{\bgamma}) =O(d^{1+a})$ by Assumption (R5). Combining them, we have
	\begin{align*}
	&\hspace{12pt}
    \|\balpha \|_1^2 \lambda_\text{min}(\R_H \S_{\bgamma} \S_{\bgamma}^\top \R_H^\top) \lambda_\text{min}( \R_{\beta} \bSigma_{\beta} \R_{\beta}^\top) \\
    &\leq 
    s_3 \leq \|\balpha \|_1^2 \lambda_\text{max}(\R_H \R_H^\top) \lambda_\text{max}(\S_{\bgamma}^\top \S_{\bgamma}) \lambda_\text{max}( \R_{\beta} \bSigma_{\beta} \R_{\beta}^\top) ,
	\end{align*}
	with the right hand side of order $d^{b-1}$. The left hand side is of the same order by the assumption in the statement of Theorem \ref{thm: ada_LASSO_phi_asymp} that $\R_H \S_{\bgamma} \S_{\bgamma}^\top \R_H^\top$ has the smallest eigenvalue of constant order. Thus, $s_3$ is of order exactly $d^{b-1}$ and hence $\balpha^\top I_{\phi,3}$ has order exactly $T^{-1/2} d^{-(1-b)/2}$. It implies $I_{\phi,3}$ is the leading term in $\wh\bphi -\bphi^\ast$ whose asymptotic normality therefore holds.
	
	As $I_{\phi,1}$ to $I_{\phi,4}$ are all $o_P(1)$, we conclude $\text{sign}(\wh\bphi_H) = \text{sign}(\bphi_H^\ast)$. It remains to show the second part in (\ref{eqn: ada_lasso_sign_consistent}) for the zero consistency of $\wh{\bphi}_{H^c}$.
	
	To this end, notice similar to $I_{\phi,3}$ but with restriction on the set $H^c$, we have
	\begin{align*}
		&\hspace{12pt}
		\big\| T^{-1}(\bXi\Y_{W, H^c} - \B^\top\V_{H^c})^\top \B^\top\X_{\bbeta^\ast - \bbeta(\bphi^\ast)} \Vec{\I_d} \big\|_\text{max} \\
		&= O_P\big( T^{-1/2} d^{-(1-b)/2} \cdot d^{-a} \cdot d^{1+2a} \big) = O_P\big(T^{-1/2} d^{\frac{1}{2} +\frac{b}{2} +a}\big),
	\end{align*}
	which used $\big\|(\H_{20} -\H_{10})^\top (\H_{20} -\H_{10})\big\|_\text{max} \leq \sigma_1^2(\H_{20} - \H_{10}) =O(d^{1+2a})$ similarly to the steps above (\ref{eqn: bound_inv_H2010}). In the same manner, we also have from $I_{\phi,4}$ that
	\[
	\big\| T^{-1}(\bXi\Y_{W, H^c} -\B^\top\V_{H^c})^\top \B^\top\bepsilon \big\|_\text{max}
	=O_P\big(T^{-1/2} d^{\frac{1}{2} +\frac{b}{2} +\frac{a}{2}}\big).
	\]
	
	The left hand side of the second inequality in (\ref{eqn: ada_lasso_sign_consistent}) has minimum value of
	\[
	\frac{\lambda}{\big\| \wt{\bphi}_{H^c} \big\|_\text{max}} \geq \frac{\lambda}{\big\| \wt{\bphi}_{H^c} -\bphi_{H^c}^\ast \big\|_\text{max}} \geq \frac{\lambda}{\big\| \wt{\bphi} -\bphi^\ast \big\|_\text{max}},
	\]
	so it suffices to show
	\[
	\big(T^{-1/2} d^{\frac{1}{2} +\frac{b}{2} +a}\big)\cdot \big\|\wt{\bphi} -\bphi^\ast \big\|_\text{max} =o_P(c_T),
	\]
	which is true by Assumption (R10) and Theorem \ref{thm: LASSO_rate} in which each entry of $F_6$ can be shown to be asymptotically normal. This completes the proof of Theorem \ref{thm: ada_LASSO_phi_asymp}.
	$\square$

	\noindent\textbf{\textit{Proof of Theorem \ref{thm: spatial_weight_rate}.}}
	We have
	\begin{equation}
		\label{eqn: rate_inf_norm_W_diff}
		\begin{split}
			\big\|\wh\W_t - \W_t^\ast \big\|_\infty &= \bigg\|\sum_{j=1}^{p} \Big[ (\wh{\phi}_{j,0} - \phi_{j,0}^\ast) + \sum_{k=1}^{l_j} (\wh{\phi}_{j,k} - \phi_{j,k}^\ast) z_{j,k,t} \Big] \W_j \bigg\|_\infty \\
			&=
			O_P\Big( \big\|\wh\bphi - \bphi^\ast\big\|_1 \cdot \max_j \|\W_j \|_\infty\Big) = O_P\big( T^{-1/2} d^{-(1-b)/2} \big),
		\end{split}
	\end{equation}
	where the last equality used Theorem \ref{thm: ada_LASSO_phi_asymp}, Assumptions (M2) (or (M2')) and (R1). Observe that we have similarly $\big\|\wh\W_t - \W_t^\ast \big\|_1 = O_P\big( T^{-1/2} d^{-(1-b)/2} \big)$ by Assumption (R1), and hence
	\[
	\big\|\wh\W_t - \W_t^\ast \big\| \leq \Big(\big\|\wh\W_t - \W_t^\ast \big\|_1 \, \big\|\wh\W_t - \W_t^\ast \big\|_\infty\Big)^{1/2} = O_P\big( T^{-1/2} d^{-(1-b)/2} \big).
	\]

	With $\bPi_t^\ast$ defined in (\ref{eqn: def_bPi_ast}), we can decompose
	\begin{align*}
		\wh\bmu - \bmu^\ast &=
		\frac{1}{T} \sum_{t=1}^T \Big\{ (\I_d - \bLambda_t \wh\bPhi) \y_t - \X_t \wh\bbeta \Big\} - \frac{1}{T} \sum_{t=1}^T \Big\{(\I_d -\bLambda_t \bPhi^\ast) \y_t - \X_t \bbeta^\ast - \bepsilon_t \Big\} \\
		&=
		\frac{1}{T} \sum_{t=1}^T \Big\{
		\big(\W_t^\ast - \wh\W_t \big) \big(\bPi_t^\ast \bmu^\ast + \bPi_t^\ast\X_t \bbeta^\ast + \bPi_t^\ast\bepsilon_t \big) \Big\}
		+ \bar{\X}(\bbeta^\ast - \wh\bbeta) + \bar{\bepsilon} ,
	\end{align*}
	so that combining (\ref{eqn: rate_inf_norm_W_diff}), Lemma \ref{lemma: LamSouza_theorem_S1} and Theorem \ref{thm: LASSO_rate}, we have
	\begin{align*}
		\big\|\wh\bmu - \bmu^\ast\big\|_\text{max} &=
		O_P\Big\{\max_t \big\|\wh\W_t - \W_t^\ast \big\|_\infty \Big(\big\|\bmu^\ast \big\|_\text{max} +c_T \big\|\bbeta^\ast \big\|_\text{max} +c_T\Big) \\
        &\hspace{125pt}
        + c_T \big\|\bbeta^\ast - \wh\bbeta \big\|_1 + c_T \Big\} = O_P(c_T).
	\end{align*}
	This completes the proof of Theorem~\ref{thm: spatial_weight_rate}. $\square$

\end{document}